\documentclass[10pt, a4paper]{article}

\usepackage[british]{babel}

\usepackage[a4paper,top=2cm,bottom=2cm,left=2.5cm,right=2.5cm,marginparwidth=1.75cm]{geometry}

\usepackage[backend=bibtex, style=numeric-comp, sorting=none]{biblatex}
\usepackage{multirow}
\usepackage{mathrsfs}
\usepackage{amsmath}
\usepackage{graphicx}
\usepackage{adjustbox}
\usepackage[table]{xcolor}
\usepackage[colorlinks=true, allcolors=black]{hyperref}
\usepackage{hyperref}
\usepackage{orcidlink}
\usepackage[title]{appendix}
\usepackage{mathrsfs}
\usepackage{amsfonts}
\usepackage{mathtools}
\usepackage{booktabs} 
\usepackage{caption}  
\usepackage{comment}
\usepackage{threeparttable} 
\usepackage{algorithm}
\usepackage{algorithmicx}
\usepackage{algpseudocode}
\usepackage{listings}
\usepackage{enumitem}
\usepackage{chngcntr}
\usepackage{booktabs}
\usepackage{lipsum}
\usepackage{subcaption}
\usepackage{authblk}
\usepackage[T1]{fontenc}    
\usepackage{csquotes}       
\usepackage{diagbox}
\usepackage{physics}
\usepackage{amssymb}
\usepackage{tikz}
\usepackage{array}
\usepackage{orcidlink}

\usepackage{titlesec}
\titleformat{\section} 
  {\normalfont\Large\bfseries}{\thesection.}{1em}{}
  
\usepackage{lineno} 

\rightlinenumbers 

\usepackage{float}   
\usepackage{caption} 
\usepackage{textcomp}

\usepackage[acronym]{glossaries}
\newacronym{ape}{APE}{absolute percentage error}
\newacronym{cad}{CAD}{computer-aided design}
\newacronym{dft}{DFT}{discrete Fourier transform}
\newacronym{fnn}{FNN}{feedforward neural network}
\newacronym[longplural={Gaussian processes}]{gp}{GP}{Gaussian process}
\newacronym{iga}{IGA}{isogeometric analysis}
\newacronym{lar}{LAR}{least angle regression}
\newacronym{mae}{MAE}{mean absolute error}
\newacronym{mape}{MAPE}{mean absolute percentage error}
\newacronym{ml}{ML}{machine learning}
\newacronym{mse}{MSE}{mean square error}
\newacronym[\glsshortpluralkey=NURBS]{nurbs}{NURBS}{non-uniform rational B-spline}
\newacronym{pca}{PCA}{principal component analysis}
\newacronym{pce}{PCE}{polynomial chaos expansion}
\newacronym{pdf}{PDF}{probability density function}
\newacronym{pmsm}{PMSM}{permanent magnet synchronous machine}
\newacronym[longplural={quantities of interest}]{qoi}{QoI}{quantity of interest}
\newacronym{rom}{ROM}{reduced-order model}
\newacronym{rsm}{RSM}{response surface model}
\newacronym{rv}{RV}{random variable}
\newacronym{sdape}{SDAPE}{standard deviation of absolute percentage error}
\newacronym{uq}{UQ}{uncertainty quantification}

\usepackage{setspace}
\title{Fourier-enhanced reduced-order surrogate modeling for uncertainty quantification in electric machine design}

\author[1,2,*]{Aylar Partovizadeh}
\author[1,2,3]{Sebastian Sch\"ops \orcidlink{0000-0001-9150-0219}}
\author[2,3]{Dimitrios Loukrezis \orcidlink{0000-0003-1264-1182}}
\affil[1]{\small{Computational Electromagnetics Group, Technische Universit\"at Darmstadt Schlo{\ss}gartenstr.~8,~64289~Darmstadt,~Germany}}
\affil[2]{\small{Institute for Accelerator Science and Electromagnetic Fields, Technische Universit\"at Darmstadt Schlo{\ss}gartenstr.~8,~64289~Darmstadt,~Germany}}
\affil[3]{\small{Graduate School of Computational Engineering, Technische Universit\"at Darmstadt Dolivostr.~15,~64293~Darmstadt,~Germany}}

\affil[*]{\small{Corresponding author: \texttt{aylar.partovizadeh@tu-darmstadt.de}}}
\affil[ ]{\small{Contributing authors: \texttt{sebastian.schoeps@tu-darmstadt.de, dimitrios.loukrezis@tu-darmstadt.de}}}

\date{}  

\begin{document}
\maketitle

\begin{abstract}
This work proposes a data-driven surrogate modeling framework for cost-effectively inferring the torque of a permanent magnet synchronous machine under geometric design variations. 
The framework is separated into a reduced-order modeling and an inference part.
Given a dataset of torque signals, each corresponding to a different set of design parameters, torque dimension is first reduced by post-processing a discrete Fourier transform and keeping a reduced number of frequency components. 
This allows to take advantage of torque periodicity and preserve physical information contained in the frequency components.
Next, a response surface model is computed by means of machine learning regression, which maps the design parameters to the reduced frequency components. 
The response surface models of choice are polynomial chaos expansions, feedforward neural networks, and Gaussian processes. 
Torque inference is performed by evaluating the response surface model for new design parameters and then inverting the dimension reduction. 
Numerical results show that the resulting surrogate models lead to sufficiently accurate torque predictions for previously unseen design configurations.
The framework is found to be significantly advantageous compared to approximating the original (not reduced) torque signal directly, as well as slightly advantageous compared to using principal component analysis for dimension reduction. 
The combination of discrete Fourier transform-based dimension reduction with Gaussian process-based response surfaces yields the best-in-class surrogate model for this use case.
The surrogate models replace the original, high-fidelity model in Monte Carlo-based uncertainty quantification studies, where they provide accurate torque statistics estimates at significantly reduced computational cost. 
\end{abstract}


\section{Introduction}
\label{sec:intro}
Several sectors of major societal and economic interest, such as industry, infrastructure, and transportation, currently undergo transformative changes towards their electrification \cite{nadel2019electrification}. These changes drive an ever increasing usage of electric machines, which in turn imposes high demands on design quality and reliability. 
Considering \gls{cad} in particular, discrepancies between computer models and manufactured machines are all but inevitable, often affecting crucial \glspl{qoi} \cite{bramedorfer2020effect}. 
To a large extent, these discrepancies can be attributed to \emph{epistemic} uncertainty, referring to systematic modeling errors due to inexactly known physical mechanisms, or \emph{aleatory} uncertainty, referring to inherently random variations in the design parameters \cite{jyrkama2016separation}.
This work focuses on aleatory uncertainty only.
In this context, \gls{uq} aims at assessing the impact of random design parameters upon \glspl{qoi} \cite{smith2024uncertainty}, such that more robust and reliable designs may be developed.
However, workhorse \gls{uq} methods such as Monte Carlo sampling \cite{zhang2021modern} can incur severe computational costs, due to their need for numerous evaluations of possibly time and resource-demanding numerical models and simulations, as is also the case in electric machine design. 

A common way to mitigate the computational cost of \gls{uq} - or other studies that require repetitive model evaluations for varying parameter values, such as optimization or design exploration - is to replace the computationally demanding, high-fidelity model with an inexpensive albeit sufficiently accurate \emph{surrogate} model \cite{alizadeh2020managing}.
A surrogate modeling approach that dates back to the work of Box and Wilson in the early 1950s, is to use a dataset of model parameters and corresponding \gls{qoi} values to establish a so-called \gls{rsm} that mimics the functional relation between parameters and \gls{qoi} \cite{kleijnen2015response}.
Among other options, \glspl{pce}, \glspl{fnn}, and \glspl{gp} have been routinely used as \glspl{rsm} \cite{hosder2012stochastic, thakre2022uncertainty, beheshti2019new, xu2022artificial, costa2016gaussian, chen2023modeling}.
An alternative approach is to decrease the cost of a computational model by reducing the dimensionality of the algebraic objects encountered in its numerical solution, resulting in a so-called \gls{rom}. 
Depending on application and \gls{rom} method of choice, dimension reduction can be applied to the input parameters, the \gls{qoi}, the system matrices of the numerical model, or combinations thereof.
This typically entails computing an approximate, low-dimensional representation of the original, high-dimensional algebraic object, using methods such as balanced truncation, proper orthogonal decomposition, Krylov subspaces, tensor decompositions, or dynamic mode decomposition, to name but a few relevant options \cite{benner2015survey, chinesta2016model, alla2017nonlinear, nouy2017low}.
As a natural extension, surrogate modeling methods that combine \glspl{rom} with \glspl{rsm} have also been developed \cite{cicci2023uncertainty, drakoulas2023fastsvd, conti2024multi}.
Note that the literature does not always clearly distinguish between \glspl{rom} and surrogate models, even using the terms interchangeably on occasion.
In the context of this work, these terms will be used as described above.

The present work focuses on the design of a \gls{pmsm} under geometric variations, possibly random ones.
Different \gls{pmsm} geometries are obtained by varying $20$ parameters of a \gls{cad} model \cite{wiesheu2024combined}, which take values within predefined ranges. 
The considered \gls{qoi} is the electromagnetic torque developed during the machine's rotation, which is computed using a numerical model based on \gls{iga} \cite{hughes2005isogeometric}.
For the torque signal to have sufficient resolution, e.g., to account for so-called torque ripples, the torque must be evaluated for a large number of rotation angles, thus resulting in a high-dimensional \gls{qoi}.
The high dimensionality of the \gls{qoi} can hinder the application of direct surrogate modeling approaches, such as \glspl{rsm} mapping the design parameters to the \gls{qoi}, by increasing computation time and failing to reach sufficient approximation accuracy. 
An obvious solution is to use a dimension reduction method upon the torque signal.
However, common methods do not take into account important properties of the torque signal, such as periodicity (for steady-state operation) and frequency content. 

To compute a sufficiently accurate surrogate model that predicts the torque of the \gls{pmsm} given different geometric designs and thus enable \gls{uq} studies, this work suggests a data-driven surrogate modeling framework utilizing \gls{dft} for dimension reduction in combination with \glspl{rsm}.
The framework is separated into a reduced-order modeling part and an inference part.
For reduced-order modeling, \gls{dft} is first applied to a dataset of torque signals, each obtained for different design parameters. 
Next, critical frequency components are retained while the rest are omitted, resulting in a reduced \gls{qoi} that preserves the most important frequency content of the signal.
Then, \gls{ml} regression is used to compute an \gls{rsm} that maps the design parameters to the reduced frequency components. 
To infer the torque given a new geometric design of the \gls{pmsm}, the \gls{rsm} is first evaluated, followed by inverting the \gls{dft}, thus obtaining the full torque signal.

From a methodological perspective, the use of \gls{dft} for the purpose of dimension reduction and its utilization in a surrogate modeling context is the main novelty of the suggested framework.
The review of Hou and Behdinan \cite{hou2022dimensionality} lists several prior works which have pursued similar ideas, wherein \gls{dft}-based dimension reduction is not included.
For this particular use case, \gls{dft} ensures that important physical information contained in the frequency components is retained, thus leading to superior results in terms of surrogate modeling accuracy, as also supported by the numerical studies available in this work.
The use of a reduced-order surrogate model, i.e., one that integrates dimension reduction of the \gls{qoi} in combination with an \gls{rsm}, is also a novel contribution concerning surrogate-assisted electric machine design, see for example the review of Cheng et al. \cite{cheng2024review} and the references therein.
Last, contrary to most works encountered in the literature, the suggested framework is not restricted to a single \gls{rsm}. 
Instead, different \gls{rsm} options are evaluated, namely, \gls{pce}, \gls{fnn}, and \gls{gp}, in an attempt to identify the best-in-class surrogate model for this use case.

The remaining of this paper is organized as follows. Section~\ref{sec:problem-setting} presents the \gls{pmsm} model, its numerical approximation, and the considered geometric variations.
Next, the data-driven reduced-order surrogate modeling framework proposed in this work is detailed in section~\ref{sec:rom-framework}. 
Numerical studies regarding the performance of surrogate models obtained with the suggested framework are presented in section~\ref{sec:num-results}.
Last, section~\ref{sec:conclusions} contains concluding remarks.

\section{Problem setting: PMSM with geometry variations}
\label{sec:problem-setting}

\subsection{Geometry representation in CAD}
\label{sec:geometry-representation}

\begin{figure}[t!]
    \centering
    \includegraphics[width=1.0\textwidth]{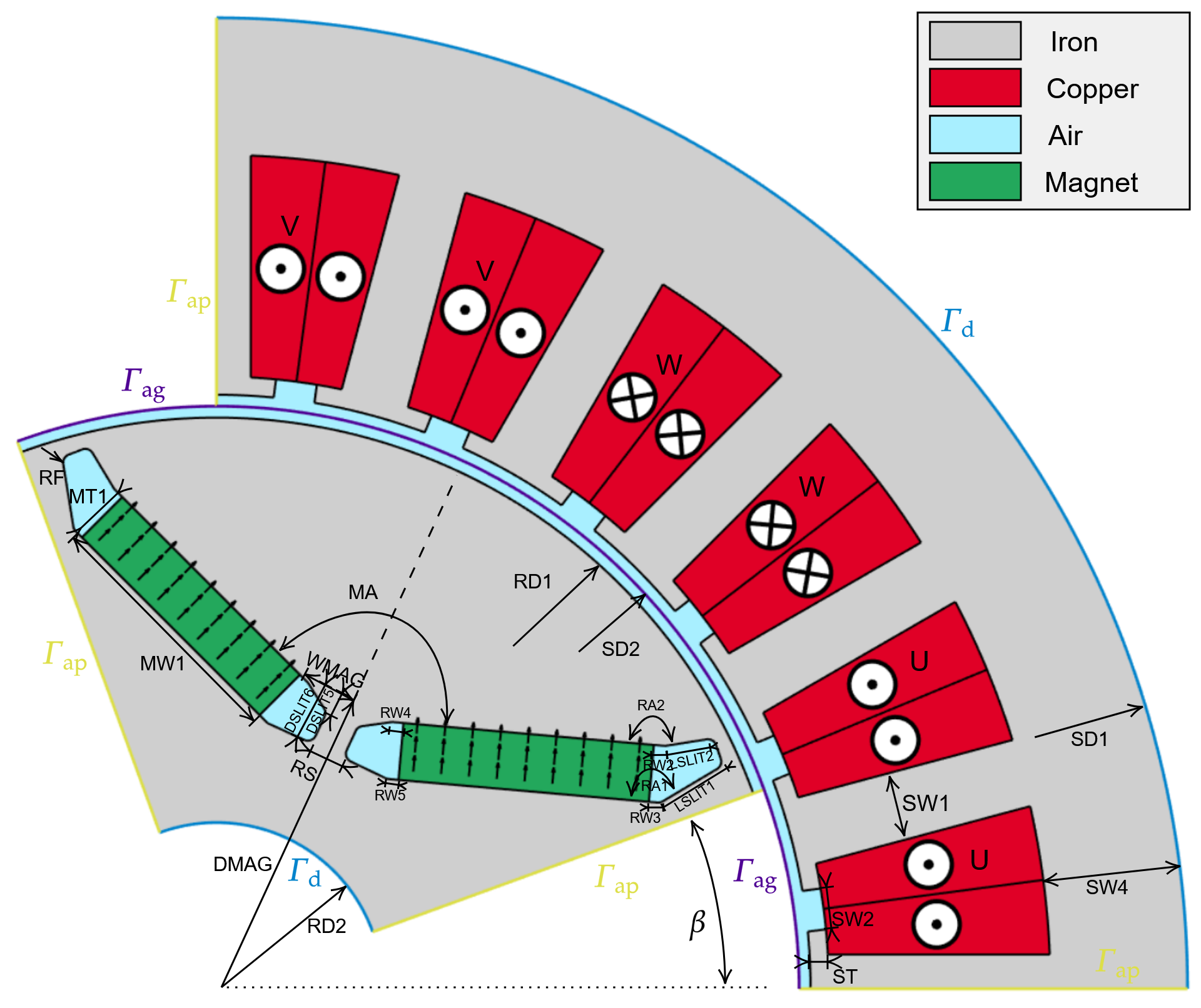}
    \caption{One-fourth of the \gls{pmsm}'s geometry in two dimensions, including material distribution, boundary conditions, and geometric design parameters (see Table~\ref{tab:pmsm-parameters}). The  phase and current direction in the copper coils is also shown. Figure adapted from \cite{wiesheu2024combined}.}
    \label{fig:pmsm}
\end{figure}

We consider a \gls{pmsm} similar to the one depicted in Figure~\ref{fig:pmsm}, where only one-fourth of a two-dimensional cross-section of its geometry is shown, taking into account rotational symmetry and translational invariance.
The \gls{pmsm} consists of a stator and a rotor, separated by an air gap. 
The rotor is composed of an iron part, permanent magnets, and air slots. 
The stator comprises an iron part and six slots where the copper windings carrying the source current are placed. 
For computational purposes, the air gap is divided into a rotor part and a stator part, separated by the interface $\Gamma_{\text{ag}}$.
This is a benchmark \gls{pmsm} geometry obtained from the commercial software \texttt{JMAG} \cite{jsol_library_2024aa}. 
For the purposes of the present study, the geometry is rebuilt using the \texttt{GeoPDEs} software \cite{Wiesheu_2023ag, vazquez2016new}.

As is common in \gls{cad}, the \gls{pmsm}'s geometry is parametrized using free-form curves, such that the geometry is defined through a projection map from a reference domain $\left[0,1\right]^d$ to the physical domain $\Omega \subset \mathbb{R}^r$, $r \geq d$.
The standard tools for this process are B-splines and \glspl{nurbs} \cite{piegl2012nurbs}.
Given a knot vector $\Xi = \left\{\xi_1, \dots, \xi_{n+p+1}\right\} \subset \left[0,1\right]$ such that $\xi_1 \leq \xi_2 \leq \cdots \leq \xi_{n+p+1}$, a basis of $n$ univariate B-splines of degree $p$ can be defined using the Cox-de-Boor recursion formula
\begin{subequations}
\label{eq:cox-de-boor}
\begin{align}
B_i^p\left(\xi\right) &= \frac{\xi - \xi_i}{\xi_{i+p} - \xi_i} B_i^{p-1}\left(\xi\right) + \frac{\xi_{i+p+1} - \xi}{\xi_{i+p+1} - \xi_{i+1}} B_{i+1}^{p-1}\left(\xi\right), \quad i=1,\dots,n, \\
B_i^0\left(\xi\right) &= 
\begin{cases}
1, &\xi_1 \leq \xi < \xi_{i+1}, \\
0, &\text{otherwise}.
\end{cases}
\end{align}
\end{subequations}
Univariate \gls{nurbs} basis functions are defined as
\begin{equation}
\label{eq:nurbs-basis}
N_i^p\left(\xi\right) = \frac{w_i B_i^p\left(\xi\right)}{\sum_{j=1}^n w_j B_j^p\left(\xi\right)},
\end{equation}
where $w_i$ are positive weights.
Using $n$ \glspl{nurbs} basis functions along with $n$ control points $\left\{\mathbf{P}_i\right\}_{i=1}^n \subset \Omega$, a \gls{nurbs} curve can be defined as 
\begin{equation}
\label{eq:nurbs-curve}
\mathbf{C}\left(\xi\right) = \sum_{i=1}^n \mathbf{P}_i N_i^p\left(\xi\right).
\end{equation}
\Gls{nurbs} surfaces or higher-dimensional objects can be created using tensor products of \gls{nurbs} curves.
Using these constructions, it is possible to define a map $F: \left[0,1\right]^d \rightarrow \Omega$ that parametrizes the physical domain such that $\mathbf{x} = F(\boldsymbol{\xi})$, $\mathbf{x} \in \Omega$, $\boldsymbol{\xi} \in \left[0,1\right]^d$. 

Not that, in most practical applications, the geometry cannot be parametrized using a single projection map from the reference to the physical domain. 
This is also the case for the considered \gls{pmsm}, which features a complex geometry with different material subdomains.
In such cases, a multi-patch parametrization is used, where the physical domain is decomposed into a collection of subdomains, each with a corresponding projection map, to be appropriately combined \cite{buffa2015approximation}.

\subsection{Magnetostatic model}
\label{sec:magnetostatic-model}

The magnetic field distribution on the PMSM is obtained by the magnetostatic formulation, which for a magnetic vector potential $\mathbf{A}$ reads
\begin{equation}
\label{eq:magnetostatic}
\nabla \times (\nu \nabla \times \mathbf{A}) = \mathbf{J}_{\text{src}} + \nabla \times (\nu \mathbf{B}_\text{r}),
\end{equation}
where $\mathbf{B} = \nabla \times \mathbf{A}$ denotes the magnetic flux density, $\nu = \nu(\left|\mathbf{B}\right|)$ the (nonlinear) reluctivity, $\mathbf{J}_{\text{src}}$ the source current density, and $\mathbf{B}_\text{r}$ the remanent flux density of the permanent magnets. 
Note that this formulation neglects displacement and eddy currents, which is common practice for laminated \glspl{pmsm}.
Moreover, assuming invariance along the $z$-axis and neglecting three-dimensional effects, e.g., due to end windings, the two-dimensional problem is considered, simplifying \eqref{eq:magnetostatic} to the Poisson problem
\begin{subequations}
\label{eq:poisson}
\begin{align}
\nabla \cdot (\nu \nabla A_{z,\text{rt}}) &= \nu \nabla \cdot \mathbf{B}_{\text{r}}, && \text{in } \Omega_{\text{rt}}, \\
\nabla \cdot (\nu \nabla A_{z,\text{st}}) &= -J_{z,\text{src}}, && \text{in } \Omega_{\text{st}}, 
\end{align}
\end{subequations}
where  $\Omega_{\text{rt}}$ and $\Omega_{\text{st}}$ denote the rotor and stator domains, respectively.
Note that \eqref{eq:poisson} requires only the $z$-components of the magnetic vector potential and the source current density, i.e., $\mathbf{A} = \left(0,0,A_z\right)$ and $\mathbf{J}_{\text{src}} = \left(0,0,J_{z,\text{src}}\right)$.
Problem \eqref{eq:poisson} is complemented with homogeneous Dirichlet and anti-periodic boundary conditions, respectively applied to the boundaries denoted with $\Gamma_{\text{d}}$ and $\Gamma_{\text{ap}}$ in Figure~\ref{fig:pmsm}.
Additionally, field continuity  at the air gap interface $\Gamma_{\text{ag}}$ is ensured by the coupling conditions
\begin{subequations}
\label{eq:air-gap-continuity}
\begin{align}
A_{z,\text{st}}(\theta) &= A_{z,\text{rt}}(\theta - \beta), && \text{on } \Gamma_{\text{ag}}, \\
H_{\theta,\text{st}}(\theta) &= H_{\theta,\text{rt}}(\theta - \beta), && \text{on } \Gamma_{\text{ag}},
\end{align}
\end{subequations}
where $\mathbf{H} = \nu \mathbf{B}$ is the magnetic field strength and $A_z$, $H_{\theta}$ are evaluated in local, rotor- or stator-specific coordinate systems that depend on the rotation angle $\beta$ \cite{egger2022torque}.
The remanent flux density $\mathbf{B}_{\text{r}}$ is given as $\mathbf{B}_{\text{r}} = B_\text{r} \left( -\sin\left(\alpha\right), \cos\left(\alpha\right) \right)$, where $\alpha$ is the magnets' direction angle.
Last, the source current density is given as $J_{z,\text{src}} = \sum_{k} J_{z,\text{src}}^{(k)}$, $k\in\left\{1,2,3\right\}$, and the $k$-th phase current is given as 
\begin{equation}
\label{eq:source-current}
J^{(k)}_{z,\text{src}} = J_0 \sin\left( n_{\text{pp}} \beta + \phi_0 +\frac{2 \pi}{3}k \right),
\end{equation}
where the magnitude $J_0$ depends on the application current and the coils' characteristics, $\phi_0$ is the electric phase offset, and $n_{\text{pp}}$ the number of pole pairs. 
A uniform distribution of $J_{z,\text{src}}$ within the coils is assumed.

\subsection{Numerical approximation}
\label{sec:numerical-approx}
The \gls{iga} method is used to solve problem \eqref{eq:poisson} numerically, such that the basis functions of the \gls{cad} geometry representation are also used as basis functions for the numerical approximation of the solution \cite{hughes2005isogeometric}.
Additionally, using harmonic mortaring for rotor-stator coupling \cite{bontinck2018isogeometric} results in the matrix system
\begin{equation}
\underbrace{
\begin{pmatrix}
\mathbf{K}_\text{rt} & \mathbf{0} & -\mathbf{G}_\text{rt} \\
\mathbf{0} & \mathbf{K}_\text{st} & \mathbf{G}_\text{st} \mathbf{R}_{\beta} \\
-\mathbf{G}_\text{rt}^\top & \mathbf{R}_{\beta}^\top \mathbf{G}_\text{st}^\top & \mathbf{0} 
\end{pmatrix}
}_{\eqqcolon \mathbf{K}}
\underbrace{
\begin{pmatrix}
\mathbf{u}_\text{rt} \\
\mathbf{u}_\text{st} \\
\boldsymbol{\lambda} 
\end{pmatrix}
}_{\eqqcolon \mathbf{u}}
=
\underbrace{
\begin{pmatrix}
\mathbf{b}_\text{rt} \\
\mathbf{b}_\text{st} \\
\mathbf{0} 
\end{pmatrix}
}_{\eqqcolon \mathbf{b}},
\label{eq:system}
\end{equation}
where $\mathbf{K}_\text{rt}$, $\mathbf{K}_\text{st}$ are stiffness matrices and $\mathbf{G}_\text{rt}$, $\mathbf{G}_\text{st}$ coupling matrices for the rotor and stator, respectively, $\mathbf{R}_{\beta} $ is the rotation matrix for a given rotation angle $\beta$, $\boldsymbol{\lambda}$ is the vector of Lagrange multipliers, and $\mathbf{b}_\text{rt} $, $\mathbf{b}_\text{st} $ are the rotor- and stator-specific right-hand side vectors.  
For more details on constructing the matrix system \eqref{eq:system} the reader is referred to \cite{wiesheu2024combined}.
Note that the system matrix $\mathbf{K}$ actually depends on the solution vector $\mathbf{u}$ due to the nonlinear constitutive law.
Based on energy conservation principles \cite{egger2022torque}, the electromagnetic torque, denoted as $\tau_{\beta}$ for a given rotation angle $\beta$, is computed as
\begin{equation}
\tau_{\beta}(\mathbf{u}) = -\mathbf{u}_\text{st}^\top \mathbf{G}_\text{st} \frac{\mathrm{d} \mathbf{R}_\beta}{\mathrm{d} \beta} \boldsymbol{\lambda} L,
\end{equation}
where $L$ is the axial length of the \gls{pmsm}. 
The latter must be explicitly taken into account since the numerical solution corresponds to the two-dimensional problem.

\subsection{PMSM model parameters and design variations}
\label{sec:design-variations}

Following Wiesheu et al.~\cite{wiesheu2024combined}, a \gls{pmsm} model with the standard M330-50A material is utilised. 
The equivalent M27 data from the FEMM software \cite{meeker2019finite} are used to define the nonlinear relative permeability, i.e., the inverse of the magnetic reluctivity (see section~\ref{sec:magnetostatic-model}) of the iron parts. 
Nd-Fe-B (linear) magnets with remanence $B_{\text{r}} = 1.0$~T and relative permeability $\mu_{\text{r}} = 1.05$ are used. 
The \gls{pmsm} operates with an applied current $I_{\text{app}}=3$~A and its windings have $n_{\text{wind}} = 35$ turns. 
The \gls{iga} discretization of the \gls{pmsm}'s geometry utilises $32$ patches with $444$ control points for the rotor domain and $144$ patches with $365$ control points for the stator domain.  
The geometry of the \gls{pmsm} depends on $P = 20$ design parameters, which are shown in Figure~\ref{fig:pmsm} and listed in Table~\ref{tab:pmsm-parameters}.
These parameters take values within predefined ranges, also listed in Table~\ref{tab:pmsm-parameters}.
The different geometric design configurations arising from the possible parameter combinations have a significant impact on the computational domain of the \gls{pmsm} model, its numerical solution, and finally on \glspl{qoi} estimated by post-processing the solutions, such as the electromagnetic torque. 
The latter is the main \gls{qoi} in this work. 
Collecting all geometric design parameters in a vector $\mathbf{p} \in \mathbb{R}^P$, system \eqref{eq:system} becomes parameter-dependent with system matrix $\mathbf{K} = \mathbf{K}\left(\mathbf{p}\right)$, right-hand side $\mathbf{b} = \mathbf{b}\left(\mathbf{p}\right)$, and solution $\mathbf{u} = \mathbf{u}\left(\mathbf{p}\right)$. Accordingly, the electromagnetic torque is parameter-dependent as well. We write, by abuse of notation, $\tau_{\beta}\left(\mathbf{u}\right) = \tau_{\beta}\left(\mathbf{u}\left(\mathbf{p}\right)\right) = \tau_{\beta}\left(\mathbf{p}\right)$. 

\begin{table*}[b!]
\centering
\caption{Geometric design parameters of the \gls{pmsm} and their value ranges. The lower and upper bounds of the parameter values are given in mm.}
\label{tab:pmsm-parameters}
\begin{threeparttable}
\begin{tabular*}{\textwidth}{@{\extracolsep{\fill}}cccc@{\extracolsep{\fill}}}
\toprule
\# & Parameter & Lower bound & Upper bound \\ [0.5ex]
\toprule
1  & LSLIT1 & 6.1      & 6.7 \\
2  & LSLIT2 & 4.1      & 4.5 \\
3  & DSLIT5 & 0.9      & 1.1 \\
4  & DSLIT6 & 1.9      & 2.1 \\
5  & MA     & 142.5    & 157.5 \\
6  & MT1    & 3.8      & 4.2 \\
7  & MW1    & 20.9     & 23.1 \\
8  & RA1    & 136.8    & 151.2 \\
9  & RA2    & 157.7    & 174.3 \\
10 & RS     & 0.9      & 1.1 \\
11 & RW2    & 0.9      & 1.1 \\
12 & RW3    & 0.9      & 1.1 \\
13 & RW4    & 0.9      & 1.1 \\
14 & RW5    & 0.9      & 1.1 \\
15 & WMAG   & 3.8      & 4.2 \\
16 & DMAG   & 28.5     & 31.5 \\
17 & ST     & 1.5      & 1.7 \\
18 & SW1    & 5.7      & 6.3 \\
19 & SW2    & 3.4      & 3.8 \\
20 & SW4    & 24.2     & 26.8 \\
\bottomrule
\end{tabular*}
\end{threeparttable}
\end{table*}

\begin{figure}[t!]
\captionsetup{singlelinecheck=off}
\captionsetup[subfigure]{justification=centering}
\begin{subfigure}[b]{0.32\textwidth}
\centering
\includegraphics[width=\textwidth]{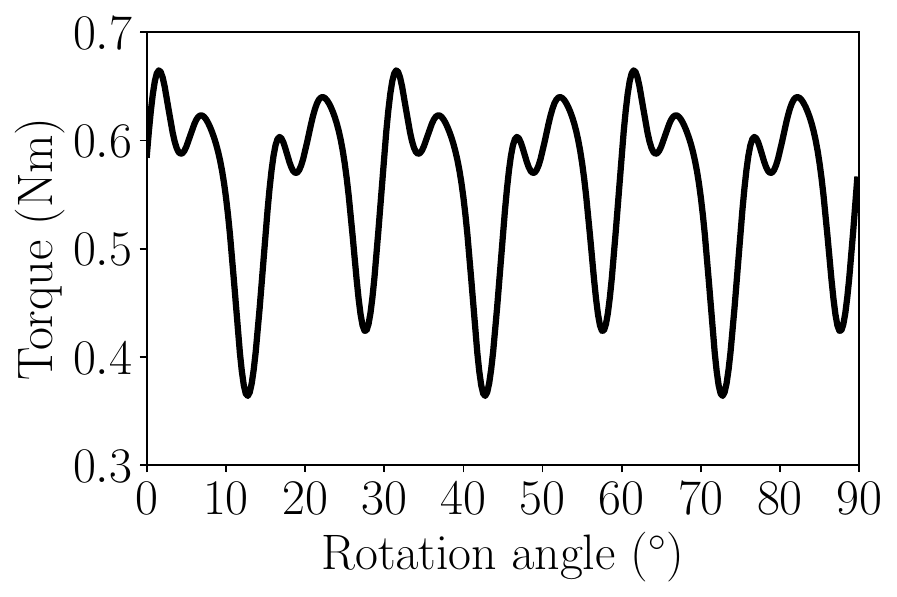}
\caption{}
\end{subfigure}
\begin{subfigure}[b]{0.32\textwidth}
 \centering
 \includegraphics[width=\textwidth]{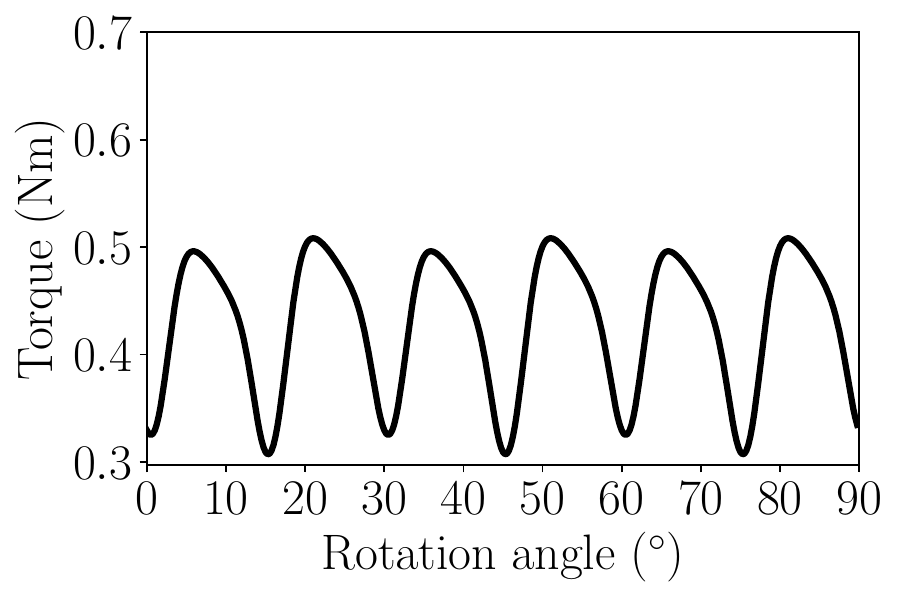}
\caption{}
\end{subfigure}
\begin{subfigure}[b]{0.32\textwidth}
 \centering
 \includegraphics[width=\textwidth]{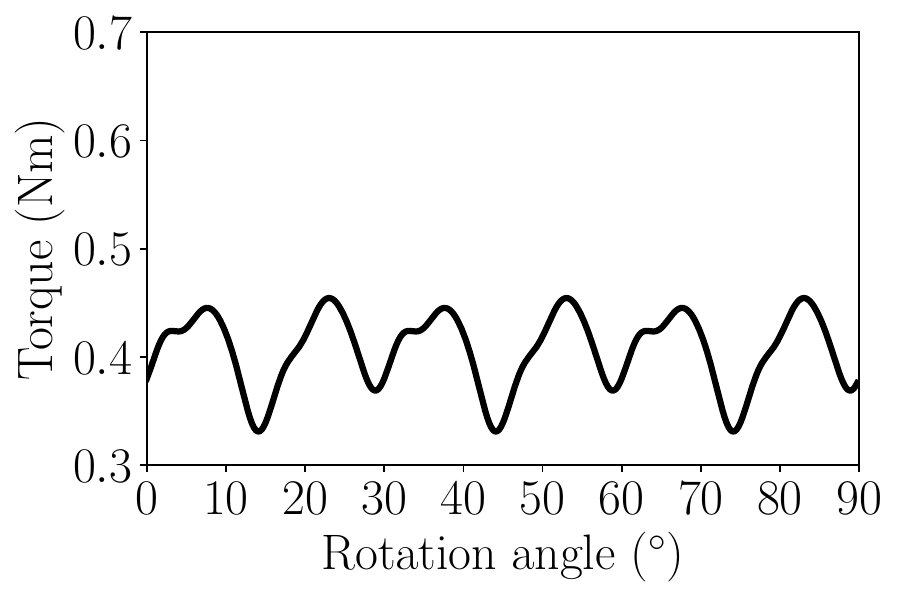}
 \caption{}
\end{subfigure}
\\
\begin{subfigure}[b]{0.32\textwidth}
 \centering
 \includegraphics[width=\textwidth]{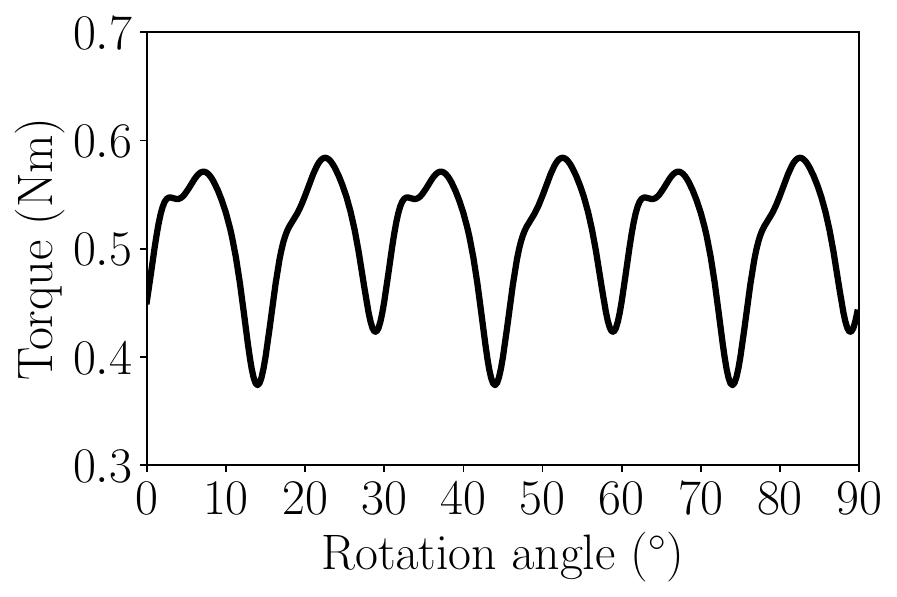}
 \caption{}
\end{subfigure}
\begin{subfigure}[b]{0.32\textwidth}
 \centering
 \includegraphics[width=\textwidth]{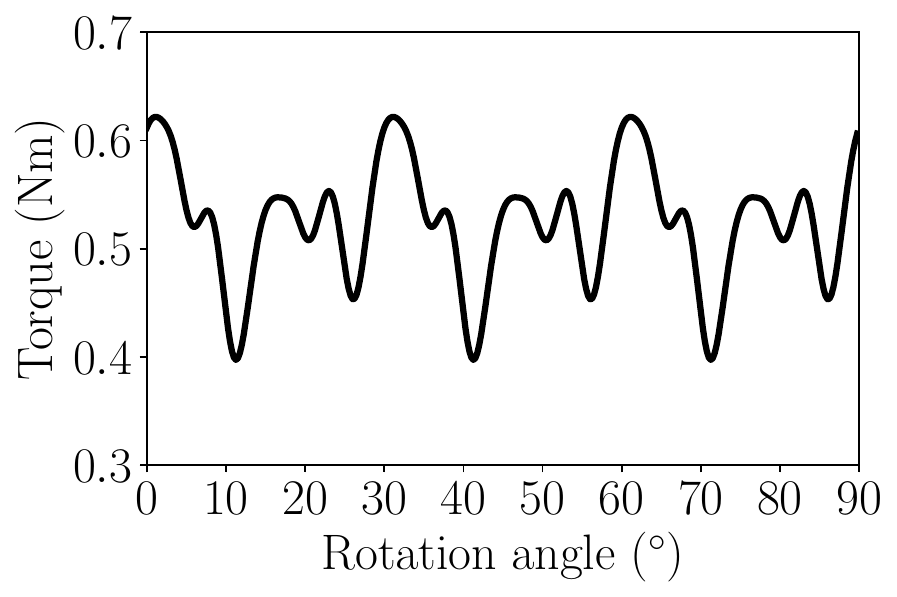}
 \caption{}
\end{subfigure}
\begin{subfigure}[b]{0.32\textwidth}
 \centering
 \includegraphics[width=\textwidth]{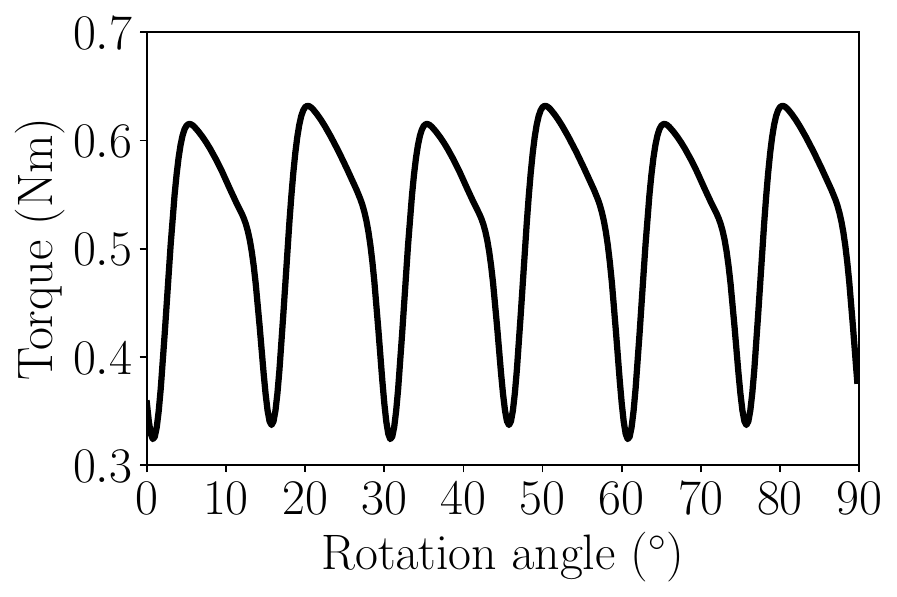}
 \caption{}
\end{subfigure}
\\
\begin{subfigure}[b]{0.32\textwidth}
 \centering
 \includegraphics[width=\textwidth]{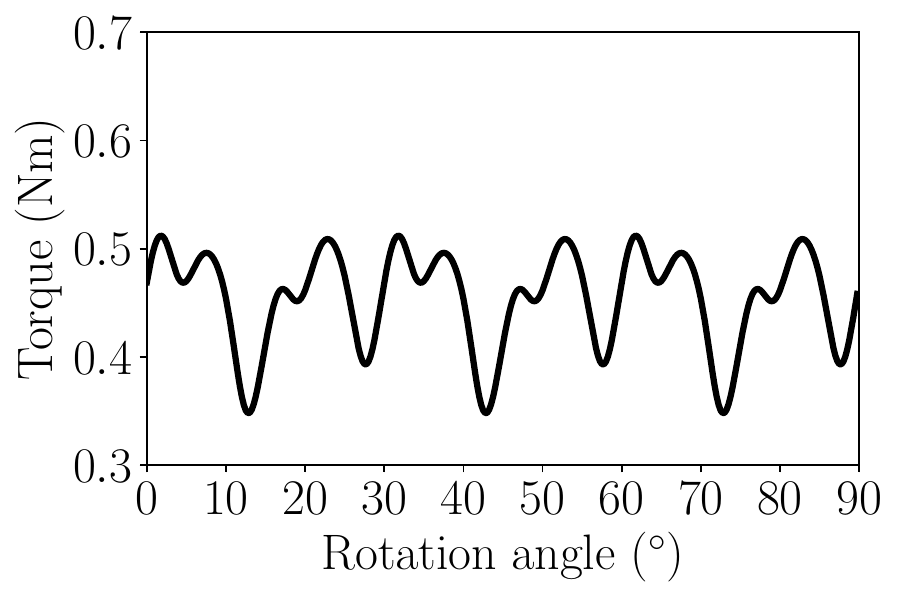}
 \caption{}
\end{subfigure}
\begin{subfigure}[b]{0.32\textwidth}
 \centering
 \includegraphics[width=\textwidth]{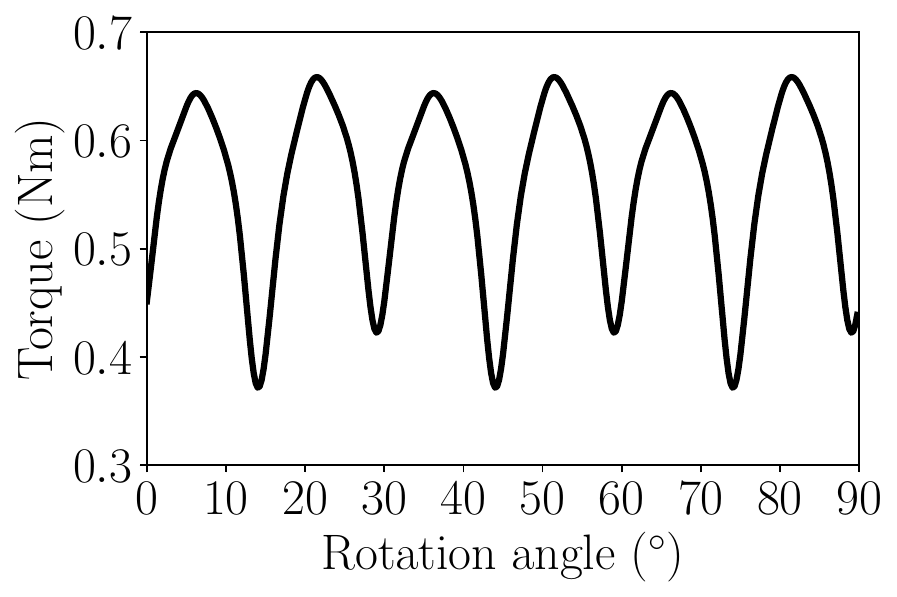}
 \caption{}
\end{subfigure}
\begin{subfigure}[b]{0.32\textwidth}
 \centering
 \includegraphics[width=\textwidth]{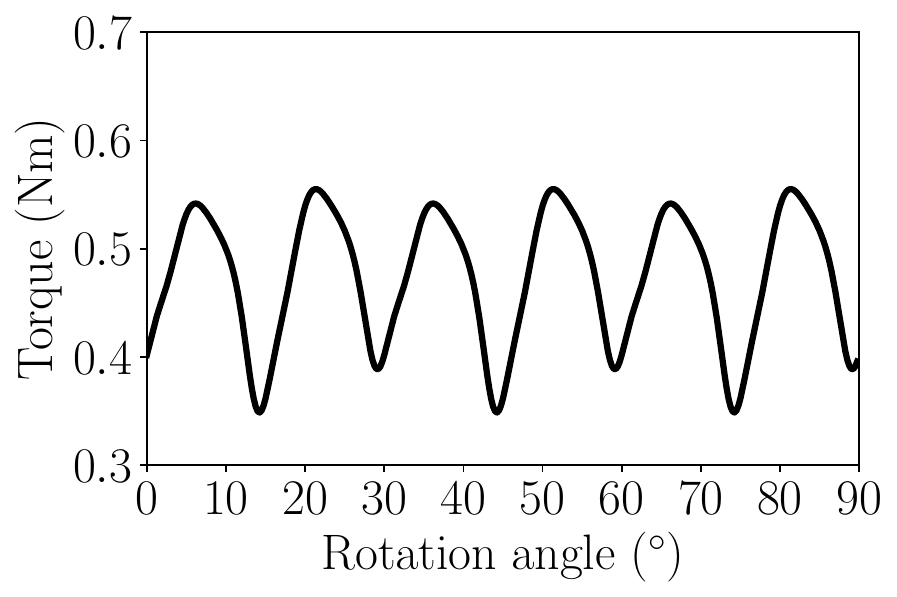}
 \caption{}
\end{subfigure}
\caption{Torque signals for different geometric design configurations of the \gls{pmsm}.}
\label{fig:torque-samples}
\end{figure}

To illustrate the impact of geometric design variations upon the torque of the \gls{pmsm}, Figure~\ref{fig:torque-samples} shows samples of torque signals, each corresponding to a different set of design parameter values given in Table~\ref{tab:figure2-parameters}. 
Note that the torque repeats every $30^\circ$ of rotation due to the symmetry in the stator's winding pattern.
As can be observed, geometric design variations lead to significant changes in the torque signal.
Also note that generating the torque signal for each geometric design variation entails running a computationally expensive numerical simulation for the solution of problem \eqref{eq:poisson}. 
A computationally inexpensive estimation of the torque would be preferable, especially considering \gls{uq} \cite{bontinck2015response, offermann2015uncertainty, galetzka2019multilevel, beltran2020uncertainty} or related studies like robust or reliability-based optimization \cite{ion2018robust, lei2020robust, bramerdorfer2020multiobjective}, that necessitate repetitive and possibly numerous model evaluations for varying \gls{pmsm} designs.

\begin{table}[b!]
\centering
\caption{Geometric design parameters values (in mm) corresponding to the torque signals shown in Figure~\ref{fig:torque-samples}.}
\label{tab:figure2-parameters}
\begin{tabular}{c c c c c c c c c c c c}
\toprule
\# & Parameter & \ref{fig:torque-samples}(a) & \ref{fig:torque-samples}(b) & \ref{fig:torque-samples}(c) & \ref{fig:torque-samples}(d) & \ref{fig:torque-samples}(e) & \ref{fig:torque-samples}(f) & \ref{fig:torque-samples}(g) & \ref{fig:torque-samples}(h) & \ref{fig:torque-samples}(i) \\ [0.5ex]
\toprule
1  & LSLIT1 & 6.27 & 6.26 & 6.51 & 6.25 & 6.54 & 6.18 & 6.53 & 6.29 & 6.39\\
2  & LSLIT2 & 4.26 & 4.09 & 4.26 & 4.20 & 4.10 & 4.17 & 4.38 & 4.40 & 4.45\\
3  & DSLIT5 & 1.02 & 0.96 & 0.96 & 0.98 & 0.97 & 1.04 & 0.98 & 0.99 & 1.02\\
4  & DSLIT6 & 1.96 & 2.00 & 1.93 & 1.94 & 2.07 & 2.01 & 1.99 & 1.98 & 2.06\\
5  & MA     & 156.89 & 147.96 & 155.08 & 155.12 & 153.99 & 149.09 & 156.30 & 149.13 & 151.70\\
6  & MT1    & 3.80 & 3.92 & 3.82 & 4.09 & 3.84 & 3.86 & 4.16 & 3.91 & 4.03\\
7  & MW1    & 22.99 & 20.93 & 20.98 & 21.64 & 22.94 & 20.98 & 21.68 & 22.11 & 21.18\\
8  & RA1    & 138.02 & 136.84 & 145.08 & 150.34 & 148.50 & 138.26 & 139.43 & 146.07 & 143.15\\
9  & RA2    & 158.11 & 172.73 & 157.91 & 163.86 & 173.99 & 162.99 & 169.16 & 159.35 & 173.63\\
10 & RS     & 0.97 & 1.01 & 1.00 & 1.02 & 0.99 & 1.05 & 0.99 & 0.96 & 0.99\\
11 & RW2    & 1.01 & 0.99 & 0.98 & 0.96 & 0.96 & 0.97 & 0.96 & 1.04 & 0.98\\
12 & RW3    & 0.99 & 0.97 & 0.99 & 1.00 & 0.97 & 1.02 & 0.97 & 0.96 & 0.95\\
13 & RW4    & 1.02 & 0.96 & 0.99 & 1.02 & 1.02 & 1.04 & 1.02 & 0.96 & 0.96\\
14 & RW5    & 1.00 & 1.05 & 1.00 & 1.00 & 1.01 & 0.98 & 1.02 & 1.01 & 0.98\\
15 & WMAG   & 4.00 & 3.99 & 3.86 & 4.02 & 3.87 & 4.05 & 3.98 & 4.19 & 3.85\\
16 & DMAG   & 30.01 & 28.97 & 28.52 & 30.25 & 29.06 & 30.59 & 29.30 & 29.27 & 29.84\\
17 & ST     & 1.65 & 1.67 & 1.63 & 1.63 & 1.62 & 1.56 & 1.65 & 1.54 & 1.63\\
18 & SW1    & 6.01 & 5.85 & 6.13 & 6.09 & 5.90 & 5.99 & 5.90 & 5.90 & 5.85\\
19 & SW2    & 3.74 & 3.61 & 3.52 & 3.56 & 3.48 & 3.48 & 3.61 & 3.68 & 3.55\\
20 & SW4    & 25.48 & 25.41 & 26.66 & 24.56 & 26.13 & 25.85 & 26.71 & 24.41 & 26.07\\
\bottomrule
\end{tabular}
\end{table}

\section{Reduced-order surrogate modeling framework}
\label{sec:rom-framework}
To address the need for computationally inexpensive torque estimates given different design configurations of the \gls{pmsm}, we propose a data-driven framework for computing a surrogate model that can reliably replace the original, high-fidelity model in parameter studies such as \gls{uq}. 
The framework consists of a reduced-order modeling part and an inference part. 

The reduced-order modeling part of the framework consists of the following steps:
\begin{enumerate}
\setcounter{enumi}{-1}
\item \emph{Data acquisition:} We assume that a dataset $\mathcal{D} = \left\{\mathbf{p}^{(m)}, \boldsymbol{\tau}^{(m)} = \boldsymbol{\tau}\left(\mathbf{p}^{(m)}\right) \right\}_{m=1}^M$ is made available, 
containing different parameter realizations $\mathbf{p}^{(m)} \in \mathbb{R}^P$ and the corresponding torque signals $\boldsymbol{\tau}^{(m)} \in \mathbb{R}^{N}$, where $N$ is the signal's dimension, defined by the number of rotation angles $\beta_n$ for which the corresponding torque values $\tau_{\beta_n}$ are obtained, such that $\boldsymbol{\tau}^{(m)}= \left\{\tau_{\beta_n}^{(m)}\right\}_{n=0}^{N-1}$.
Continuous torque signals, e.g., as in Figure~\ref{fig:torque-samples}, are obtained by means of linear interpolation.
In this work, such a dataset is obtained by evaluating the \gls{pmsm}'s numerical model for different design configurations. However, it is possible to use experimental data as well, e.g., torque signals measured for different machine configurations, possibly even mixing simulation-generated and measured data.
\item \emph{Dimension reduction of the \gls{qoi}:} The torque signals $\boldsymbol{\tau}^{(m)}$, $m=1,\dots,M$, are processed such that the original dimension of the \gls{qoi} is significantly reduced. 
In this work, dimension reduction is performed by applying \gls{dft} upon the torque signals and retaining the most important frequency components. 
Essentially, the dimension reduction step creates a map $\mathcal{R} \colon \boldsymbol{\tau} \rightarrow \mathbf{r}$, where $\mathbf{r}$ denotes the reduced \gls{qoi}, equivalently, the representation of the original \gls{qoi} in a reduced frequency space. 

\item \emph{\Gls{ml} regression:} \Glspl{rsm} are trained by means of \gls{ml} regression, as to map the design parameters to the reduced frequency components of the torque signals. 
We denote this map as $\mathcal{S} \colon \mathbf{p} \rightarrow \mathbf{r}$. 
Three \glspl{rsm} are employed for that purpose, namely, \gls{pce}, \gls{fnn}, and \gls{gp}. 
\end{enumerate}

The inference part of the framework consists of the following steps:
\begin{enumerate}
\item \emph{Response surface evaluation:} Given a new design configuration and the corresponding design parameter vector $\mathbf{p}^*$, the \gls{rsm} is evaluated as to yield an estimate of the reduced \gls{qoi}, such that $\mathbf{r}^* = \mathcal{S}\left(\mathbf{p}^*\right)$. 
\item \emph{Dimension reduction inversion:} The torque estimate for the new design configuration is obtained by inverting the dimension reduction procedure used in the reduced-order modeling part, such that $\boldsymbol{\tau}^* = \mathcal{R}^{-1}\left(\mathbf{r}^*\right)$. 
\end{enumerate}

\begin{figure}[t!]
    \centering
    \includegraphics[width=1.0\textwidth]{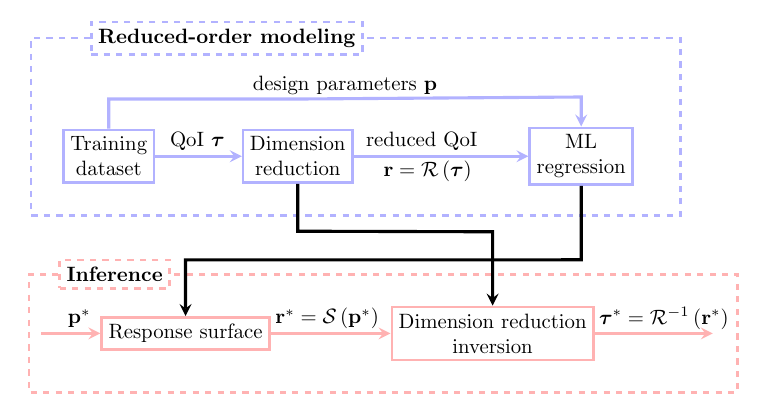}
    \caption{Schematic illustration of the surrogate modeling framework, separated into reduced-order modeling and inference parts.}
    \label{fig:rom-framework}
\end{figure}

A visualization of the complete framework is provided in Figure~\ref{fig:rom-framework}.
For \gls{uq} purposes, the resulting surrogate model can replace the high-fidelity model within a Monte Carlo sampling algorithm to obtain estimates of the torque's statistics (e.g., expected value, variance) cost-effectively.
Details regarding the \gls{dft}-based dimension reduction approach are given in section~\ref{sec:dimension-reduction}. 
The computation of \glspl{rsm} by means of \gls{ml} regression is presented in section~\ref{sec:ml}.

\paragraph{Remark on computational cost} The computational cost of the suggested surrogate modeling framework concerns mainly the reduced-order modeling part, in particular data acquisition and \gls{ml} regression, collectively called ``offline'' costs.
In this work, data acquisition is the main computational cost to be considered, since datasets are generated by running computationally expensive numerical model evaluations. 
In other cases, e.g., for datasets that are readily available, this cost can become insignificant. 
ML regression costs may range from negligible to quite significant, depending on choices regarding the type of ML model, training/optimization algorithm, and size of training dataset.
In this work, \gls{ml} regression costs can be neglected next to data acquisition costs.
Inference (also called ``online'') costs can similarly be neglected, even if one considers numerous surrogate model evaluations, for example, in the context of sampling-based \gls{uq} studies. 
The cost remains negligible due to the computationally inexpensive surrogate model.
A detailed discussion on the computational costs and gains in the context of the present work is provided in section~\ref{sec:costs-and-gains}.

\subsection{Dimension reduction via discrete Fourier transform}
\label{sec:dimension-reduction}

In the following, we assume that $M$ torque signals $\boldsymbol{\tau}^{(m)}$, $m=1,\dots,M$, are available, consisting of $N$ torque values within one rotation period. 
Longer torque signals can always be reduced to a single period without loss of information.
Due to the fact that the dimensionality of the torque signal can be prohibitively large for the subsequent steps of the surrogate modeling framework, we suggest an approach based on the \gls{dft} to reduce the torque's dimension to its minimum necessary frequency components. 

The \gls{dft} is first used to convert a finite sequence of equally-spaced torque samples (in time, equivalently, with respect to rotation angle) into a sequence of the same length, where the latter contains information about the underlying frequency components that contribute to the given torque signal. 
Given the torque signal as a sequence of torque values per rotation angle, i.e., $\boldsymbol{\tau}^{(m)}= \left\{\tau_{\beta_n}^{(m)}\right\}_{n=0}^{N-1}$, the \gls{dft} yields the sequence of complex values $\mathbf{c}^{(m)} = \left\{c_k^{(m)}\right\}_{k=0}^{N-1}$, such that
\begin{equation}
\label{eq:dft}
c_k^{(m)} = \sum_{n=0}^{N-1} \tau_{\beta_n}^{(m)} \exp\left(-\frac{\imath 2 \pi kn}{N}\right) = \sum_{n=0}^{N-1} \tau_{\beta_n}^{(m)} \left( \cos{\left(\frac{2 \pi k n}{N}\right)} - \imath \sin{\left(\frac{2 \pi k n}{N}\right)} \right),
\end{equation}
where $c_0^{(m)}\in\mathbb{R}$ corresponds to the average value of the original torque signal $\boldsymbol{\tau}^{(m)}$ across all rotation-angle samples. 
Since the torque signal is real-valued, its average value is real-valued as well. 
The \gls{dft} can be inverted to reconstruct the original signal, such that 
\begin{equation} 
\label{eq:inverted-dft}
\tau_{\beta_n}^{(m)} = \frac{1}{N}\sum_{k=0}^{N-1} c_k^{(m)} \exp\left(\frac{\imath 2 \pi k n}{N} \right).
\end{equation}

For the purpose of dimension reduction, we take advantage of the fact that the power spectrum of a given torque signal $\boldsymbol{\tau}^{(m)}$ can be given in terms of its frequency components as
\begin{equation}
\label{eq:power-spectrum}
\mathrm{PS}_k^{(m)} = \frac{1}{N} \left|c_k^{(m)}\right|^2,
\end{equation} 
meaning that each frequency component has a distinct contribution to the power spectrum, quantified by the value $\eta_k^{(m)} = \left|c_k^{(m)}\right|^2$.
Omitting components with negligible contributions to the power spectrum allows to reduce the dimensionality of the output of the \gls{dft} without significantly affecting the value of the power spectrum, but also without compromising the accuracy of the torque's reconstruction given in \eqref{eq:inverted-dft} due to the small values of the omitted constants $c_k^{(m)}$.

However, for varying design configurations, each corresponding to a different torque signal, it is to be expected that the \gls{dft} output will differ from one signal to another.
In turn, the classification of frequency components to negligible or retainable shall differ among the torque signals.
To address this issue, we suggest to retain the frequency components that remain important over the possible design configurations in the sense of  expected contribution to the power spectrum. 
To that end, the contribution of each frequency component to the power spectrum over all given torque signals $\boldsymbol{\tau}^{(m)}$, $m=1,\dots,M$, is averaged as to obtain a single contribution per component, computed as
\begin{equation}
\overline{\eta}_k 
= \frac{1}{M} \sum_{m=1}^M \eta_k^{(m)}
= \frac{1}{M} \sum_{m=1}^M \left|c_k^{(m)}\right|^2.
\end{equation}
Subsequently, the frequency components are sorted based on their averaged power spectrum contributions. 
The dimension reduction of the \gls{dft}'s output proceeds by keeping the frequency components corresponding to significant averaged power spectrum contributions $\overline{\eta}_k$ and omitting the rest.
One way of distinguishing between frequency components being negligible or not is presented in section~\ref{sec:results-dimension-reduction}, based on the accuracy of the reconstructed torque signal.

This procedure results in a reduced \gls{qoi} $\mathbf{r} \in \mathbb{C}^R$, $R<N$, with $\mathbf{r} = \left\{c_{k'}\right\}_{k' \in \mathcal{I}}$, where $\mathcal{I} \subset \left\{k\right\}_{k=0}^{N-1}$ is an index set with cardinality equal to $R$ that contains the indices of the retained frequency components. 
Put in a \gls{rom} context, $\mathbf{r} \in \mathbb{C}^R$ is the reduced representation (here, in frequency space) of $\boldsymbol{\tau} \in \mathbb{R}^N$ and $\left\{c_{k'}\right\}_{k' \in \mathcal{I}}$ is the corresponding reduced basis. 
Accordingly, the realizations of the reduced \gls{qoi} are given as $\mathbf{r}^{(m)} = \left\{c_{k'}^{(m)}\right\}_{k' \in \mathcal{I}}$, $m=1.\dots,M$. 
Note that the reduced \gls{qoi} can be given equivalently as a real vector $\mathbf{r} \in \mathbb{R}^{2R-1}$, by considering separately the real and imaginary parts of the complex elements $c_{k'}$, $k' \neq 0$, where $c_0 \in \mathbb{R}$. 
We opt for the more concise complex notation. 

\subsection{Regression-based response surface models}
\label{sec:ml}

Within the suggested surrogate modeling framework, data-driven, regression-based \glspl{rsm} are employed to approximate the functional relationship between the geometric parameters of the \gls{pmsm} and the reduced frequency components of the torque signal, thus providing a map $\mathcal{S}: \mathbf{p} \rightarrow \mathbf{r}$. 
For comparison purposes, in the numerical experiments presented in section~\ref{sec:num-results}, the \glspl{rsm} are employed as approximations to full torque signals and to their reduced principal components.
Therefore, in the following, the functional relation is denoted as $\mathbf{y} = f\left(\mathbf{p}\right)$, where $\mathbf{y}$ may refer to either the full or reduced \gls{qoi}.
Without loss of generality, we assume that $\mathbf{y} \in \mathbb{R}^N$.
For the regression-based computation of the \glspl{rsm} , we assume the existence of a dataset $\mathcal{D}_{\text{t}} = \left\{\mathbf{p}^{(m)}, \mathbf{y}^{(m)} \right\}_{m=1}^M$ containing input parameter realizations along with the corresponding values of the \gls{qoi}.
Three \glspl{rsm} are examined, namely, \gls{pce}, \gls{fnn}, and \gls{gp}, which are briefly presented next.

\subsubsection{Polynomial chaos expansion (PCE)}
\label{sec:pce}

For response surface modeling by means of the \gls{pce}, the parameter vector $\mathbf{p}$ is assumed to be a random vector defined on the probability space $(\Theta,\Sigma, \mathcal{P})$, where $\Theta$ denotes the sample space, $\Sigma$ the sigma algebra of events, and $\mathcal{P} \colon \Sigma \to [0,1]$ a probability measure, and with the \gls{pdf} $\varrho(\mathbf{p}) \colon \Xi \to \mathbb{R}_{\geq 0}$, where $\Xi \subset \mathbb{R}^P$ denotes the random vector's support. 
Under these assumptions, a \gls{pce} is a global polynomial approximation of the form
\begin{equation}
\label{eq:spectral_approx}
f(\mathbf{p}) \approx f_{\text{PCE}}(\mathbf{p}) = \sum_{k=1}^K \mathbf{a}_k \Psi_k(\mathbf{p}),
\end{equation}
where $\mathbf{a}_k \in \mathbb{R}^N$ are expansion coefficients and $\Psi_k$ are multivariate polynomials that satisfy the orthogonality property
\begin{equation}
	\label{eq:orthogonality}
	\mathbb{E}\left[\Psi_k \Psi_l\right] = \int_{\Xi}  \Psi_k\left(\mathbf{p}\right) \Psi_l\left(\mathbf{p}\right) \varrho\left(\mathbf{p}\right) \mathrm{d}\mathbf{p} =  \mathbb{E}\left[\Psi_k^2\right] \delta_{kl},
\end{equation}
where $\delta_{kl}$ is the Kronecker delta. 
Depending on the \gls{pdf}, the polynomials are chosen either according to the Wiener-Askey scheme \cite{xiu2002wiener} or are constructed numerically \cite{feinberg2018multivariate}.
To evenly cover the design space, we assume here that all design parameters are uniformly distributed within their value ranges, also see section~\ref{sec:results-dimension-reduction}. Therefore, Legendre polynomials are used.
The polynomial basis of the \gls{pce} can be formed in several ways, which also define the truncation limit $K$. 
Common options include total degree, hyperbolic truncation, sparse, and adaptive bases, and combinations thereof \cite{luethen2021sparse}.

Training the \gls{pce} means fitting its coefficients based on the data $\mathcal{D}_{\text{t}}$.
In particular, the coefficients are computed by solving the least squares regression problem 
\begin{align}
\label{eq:regression}
\underset {\mathbf{a}_1, \dots, \mathbf{a}_K}{\arg\min} \left\{\frac{1}{M}\sum_{m=1}^{M} \left( \mathbf{y}^{(m)} - \sum_{k=1}^K \mathbf{a}_k \Psi_k\left(\mathbf{p}^{(m)}\right) \right)^2\right\}.
\end{align}
For sparse \glspl{pce}, the regression problem \eqref{eq:regression} is complemented with an additional penalty term that induces sparsity in its solution.
In this work, we opt for a sparse and adaptive \gls{pce} method based on \gls{lar} \cite{blatman2011adaptive}, which is implemented with the \texttt{UQLab} software \cite{marelli2014uqlab}.

\subsubsection{Feedforward neural network (FNN)}
\label{sec:nn}
To define an \gls{fnn}, we must first introduce the concepts of neuron and layer.
A neuron  is the smallest unit of an \gls{fnn} and is mathematically described by the function
\begin{equation}
\nu\left(\mathbf{z}\right) = \sigma \left(\mathbf{z}^\top \mathbf{w} + b\right),
\end{equation}
where $\mathbf{z} \in \mathbb{R}^{D}$ is the input to the neuron, $\mathbf{w} \in \mathbb{R}^{D}$ a weight vector, $b \in \mathbb{R}$ a bias term, and $\sigma: \mathbb{R} \rightarrow \mathbb{R}$ a nonlinear function commonly referred to as the activation.
A layer is formed by a set of $K$ neurons that receive the same input $\mathbf{z} \in \mathbb{R}^D$ and is mathematically described by the function
\begin{equation}
\ell\left(\mathbf{z}\right) = \sigma \left(\mathbf{z}^\top \mathbf{W} + \mathbf{b}\right),
\end{equation}
where $\mathbf{W} \in \mathbb{R}^{D \times K}$ is a matrix that contains the weights of the $K$ neurons in the layer, $\mathbf{b} \in \mathbb{R}^K$ is the corresponding bias vector, and the activation function is applied elementwise per neuron.
The \gls{fnn} is formed as a sequence of $L$ layers and is mathematically described by the layer composition formula
\begin{equation}
f_{\text{FNN}}\left(\mathbf{z}\right) = \ell^{L}\left( \ell^{L-1}\left( \cdots \left(\ell^{1}\left(\mathbf{z}\right)\right) \right) \right).
\end{equation}
 
To be used as an \gls{rsm} $f_{\text{FNN}}(\mathbf{p}) \approx f(\mathbf{p})$, the \gls{fnn}'s trainable parameters, i.e., the collection of its weights and biases $\boldsymbol{\theta} = \left\{\mathbf{W}^{(l)}, \mathbf{b}^{(l)}\right\}_{l=1}^L$ must be fitted to the training data $\mathcal{D}_{\text{t}}$.
Here, this is accomplished by minimizing the \gls{mse} loss function
\begin{equation}
\mathcal{L}\left(\boldsymbol{\theta}\right) = \frac{1}{M} \sum_{m=1}^M \left(\mathbf{y}^{(m)} - f_{\text{FNN}}\left(\boldsymbol{\theta}; \mathbf{p}^{(m)}\right)\right)^2,
\end{equation}
resulting in a nonlinear optimization problem, the solution of which is typically computed by means of stochastic gradient descent algorithms \cite{bottou2018optimization} enabled by automatic differentiation \cite{baydin2018automatic}.

In this work we opt for an \gls{fnn} consisting of four fully-connected hidden layers with $45$, $60$, $80$, and $25$ neurons, respectively. The number of layers and neurons per layer has been determined by considering an initial \gls{fnn} with three layers and 20 neurons per layer, and then incrementally increasing the number of layers and neurons per layer until no significant improvement in the approximation error was observed.
Naturally, the number of neurons used in the output layer coincide with the dimensions of the \gls{qoi} (original or reduced).
Rectified linear unit (ReLU) activation functions are used for all neurons.
The \gls{fnn} is trained for $600$ epochs using a standard stochastic gradient descent optimizer with scheduled learning rate. 
The \texttt{Tensorflow} software \cite{abadi2016tensorflow} is used for implementation.

\subsubsection{Gaussian process (GP)}
\label{sec:gp}
A \gls{gp} is formally defined as a collection of random variables, any finite number of which have a joint Gaussian distribution \cite{williams2006gaussian}.
Considering temporarily a scalar \gls{qoi} $y = f(\mathbf{p})$, an approximation by a \gls{gp} interprets the functional relation as a probability distribution in function space, such that
\begin{equation}
\label{eq:gp-prior}
f(\mathbf{p}) \sim \mathcal{GP}\left( m\left(\mathbf{p}\right), k\left(\mathbf{p}, \mathbf{p}'\right) \right), 
\end{equation}
where $m(\mathbf{p})$ is a mean function and $k(\mathbf{p}, \mathbf{p}')$ a covariance function. 
The mean and covariance functions encode prior assumptions regarding the target function, e.g., concerning its expected behavior and regularity.
The prior assumptions are then combined with the data available in the training dataset $\mathcal{D}_{\text{t}}$ as to obtain a new \gls{gp}, called the posterior, with updated mean and covariance functions.
Defining the matrix $\mathbf{K} \in \mathbb{R}^{M \times M}$ with elements $k_{ij} = k\left(\mathbf{p}^{(i)}, \mathbf{p}^{(j)}\right)$, and the vector $\mathbf{m} = \left(m\left(\mathbf{p}^{(1)}\right), \dots, m\left(\mathbf{p}^{(M)}\right)\right) \in \mathbb{R}^M$, the updated (posterior) distribution of the target function, i.e., conditioned on the training dataset $\mathcal{D}_{\text{t}}$, is given as
\begin{equation}
\label{eq:gp-posterior}
f(\mathbf{p}) | \mathcal{D}_{\text{t}} \sim \mathcal{N}\left(\mathbf{m}, \mathbf{K}\right),
\end{equation}
where $\mathcal{N}\left(\cdot, \cdot\right)$ denotes a Gaussian distribution.
For a new data point $\mathbf{p}^* \not \in \mathcal{D}_{\text{t}}$, the predictive distribution is given as
\begin{equation}
\label{eq:gp-predictive}
f(\mathbf{p}) | \mathbf{p}^*, \mathcal{D}_{\text{t}} \sim \mathcal{N}\left( \underbrace{m\left(\mathbf{p}^*\right) + \mathbf{k}\left(\mathbf{p}^*, \mathbf{P}\right) \mathbf{K}^{-1} \left(\mathbf{y} - \mathbf{m}\right)^\top}_{\text{mean}}, \underbrace{k\left(\mathbf{p}^*,\mathbf{p}^*\right) - \mathbf{k}\left(\mathbf{p}^*,\mathbf{P}\right) \mathbf{K}^{-1} \mathbf{k}\left(\mathbf{P},\mathbf{p}^*\right)^\top}_{\text{variance}} \right),
\end{equation}
where $\mathbf{P} = \left(\mathbf{p}^{(m)}\right)_{m=1}^M$, $\mathbf{y} = \left(y^{(m)}\right)_{m=1}^M$, $\mathbf{k}\left(\mathbf{p}^*, \mathbf{P}\right) = \left(k\left(\mathbf{p}^*, \mathbf{p}^{(m)}\right)\right)_{m=1}^M$, and $\mathbf{k}\left(\mathbf{P}, \mathbf{p}^*\right) = \left(k\left(\mathbf{p}^{(m)}, \mathbf{p}^*\right)\right)_{m=1}^M$.
The predictions of the \gls{gp}-based \gls{rsm} are obtained by evaluating the mean function of the predictive distribution upon new data points, equivalently,
\begin{equation}
f\left(\mathbf{p}\right) \approx f_{\text{GP}}\left(\mathbf{p}\right) = m\left(\mathbf{p}\right) + \mathbf{k}\left(\mathbf{p}, \mathbf{P}\right) \mathbf{K}^{-1} \left(\mathbf{y} - \mathbf{m}\right)^\top.
\end{equation}

In this work, the assumed prior has zero mean and the ellipsoidal Gaussian covariance function
\begin{equation}
\label{eq:ellipsoidal-covariance}
k(\mathbf{p}, \mathbf{p}'|\boldsymbol{\theta}) = \exp \left( -\frac{1}{2} \sum_{i=1}^{P} \left(\frac{p_i - p_i'}{\theta_i}\right)^2  \right),
\end{equation}
where $\boldsymbol{\theta} = \left(\theta_i\right)_{i=1}^P$ is a vector of hyperparameters. 
Note that, for ease of notation, the hyperparameter vector $\boldsymbol{\theta}$ has been omitted prior to \eqref{eq:ellipsoidal-covariance}.
Training the \gls{gp}-based \gls{rsm} entails optimizing the hyperparameters such that the likelihood of observing the data contained within the training dataset $\mathcal{D}_{\text{t}}$, denoted as $\mathcal{P}\left(\mathcal{D}_{\text{t}}, \boldsymbol{\theta}\right)$, is maximized \cite{williams2006gaussian}. 
Here, this is accomplished with a covariance matrix adaptation evolution strategy (CMA-ES) optimizer \cite{hansen2001completely}.  
Similar to the \gls{pce}, the implementation of the \gls{gp}-based \gls{rsm} is based on the \texttt{UQLab} software \cite{marelli2014uqlab}.

Note that the presentation above concentrated on a scalar \gls{qoi} and its approximation by a \gls{gp}.
Multi-output \glspl{gp} can be obtained in various ways, for example, using a so-called co-Kriging model or assuming a multi-output covariance \cite{liu2018remarks}.
In this work, we opt for the simplest solution of considering the elements of a vector-valued \gls{qoi} to be independent and treating them separately, which was found to be sufficient.

\section{Numerical studies and results}
\label{sec:num-results}

In the following, we assess several aspects of suggested reduced-order surrogate modeling framework, in particular concerning the dimension reduction achieved for the \gls{qoi}, the performance of the resulting surrogates in terms of prediction accuracy, and the accuracy of statistics estimates when the surrogates are used for \gls{uq}.
The corresponding numerical studies are presented in sections~\ref{sec:results-dimension-reduction}, \ref{sec:results-rom}, and \ref{sec:results-uq}, respectively.

\subsection{Dimension reduction of the QoI}
\label{sec:results-dimension-reduction}

As previously explained in section~\ref{sec:dimension-reduction}, by keeping a reduced number of the frequency components obtained via \gls{dft}, the dimensions of the QoI can be reduced significantly. 
However, the number of reduced frequency components must be determined considering the trade-off between dimension reduction and the error between the reconstructed \gls{qoi}, obtained by inverting the \gls{dft}, and the original. 
Therefore, in this first numerical study, we examine the optimal number of reduced \gls{qoi} dimensions, such that torque signal reconstruction is sufficiently accurate compared to the original signals.

\begin{figure}[t!]
\centering
\begin{subfigure}[t]{0.4\textwidth}
    \hspace*{0.5\textwidth}
    \fbox{\includegraphics[width=\textwidth]{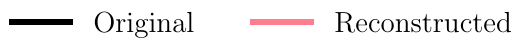}}
 \end{subfigure}
 \\
 \begin{subfigure}[t]{0.32\textwidth}
     \centering
     \includegraphics[width=1\textwidth]{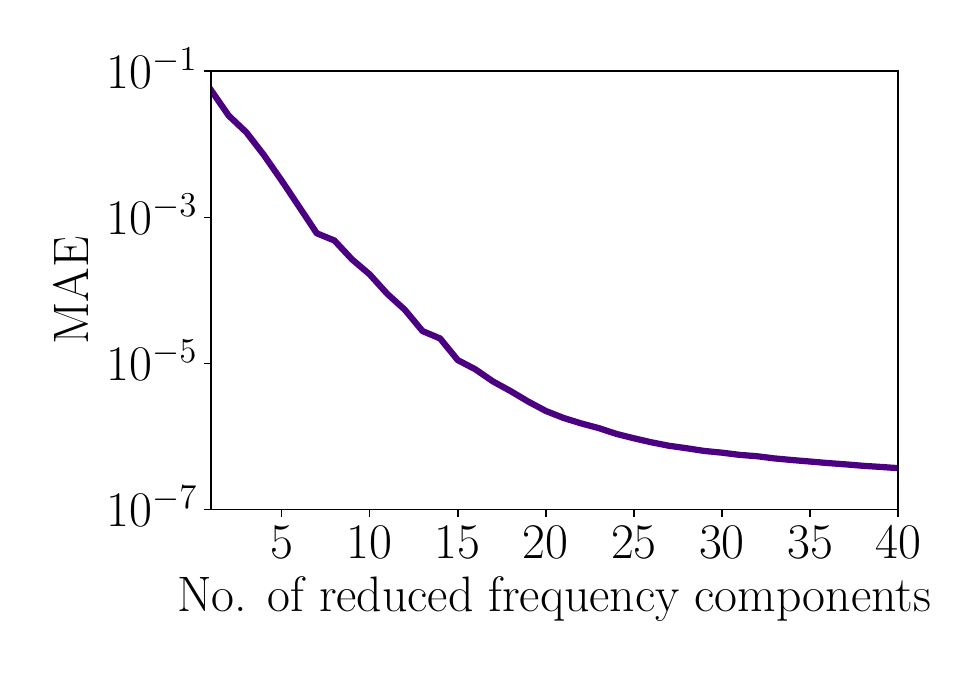}
     \caption{Reconstruction error versus reduced frequency components.}
     \label{fig:3-fft-all}
 \end{subfigure}
 \hfill
 \begin{subfigure}[t]{0.32\textwidth}
     \centering
     \includegraphics[width=1\textwidth]{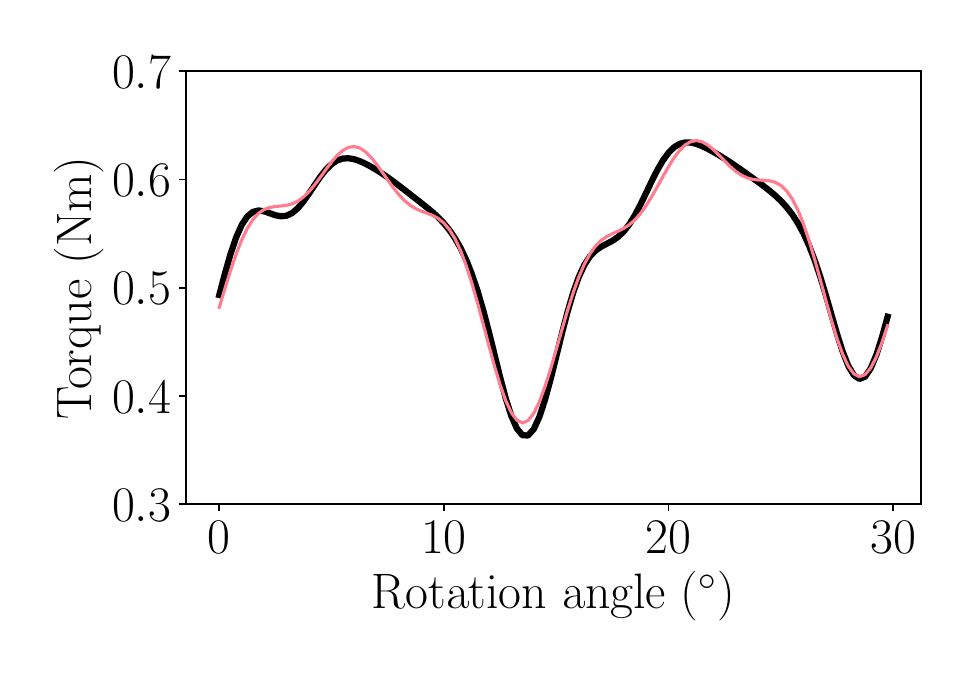}
     \caption{Worst-case reconstruction with $R=5$ frequency components.} 
     \label{fig:3-fft-9}
 \end{subfigure}
 \hfill
 \begin{subfigure}[t]{0.32\textwidth}
     \centering
     \includegraphics[width=1\textwidth]{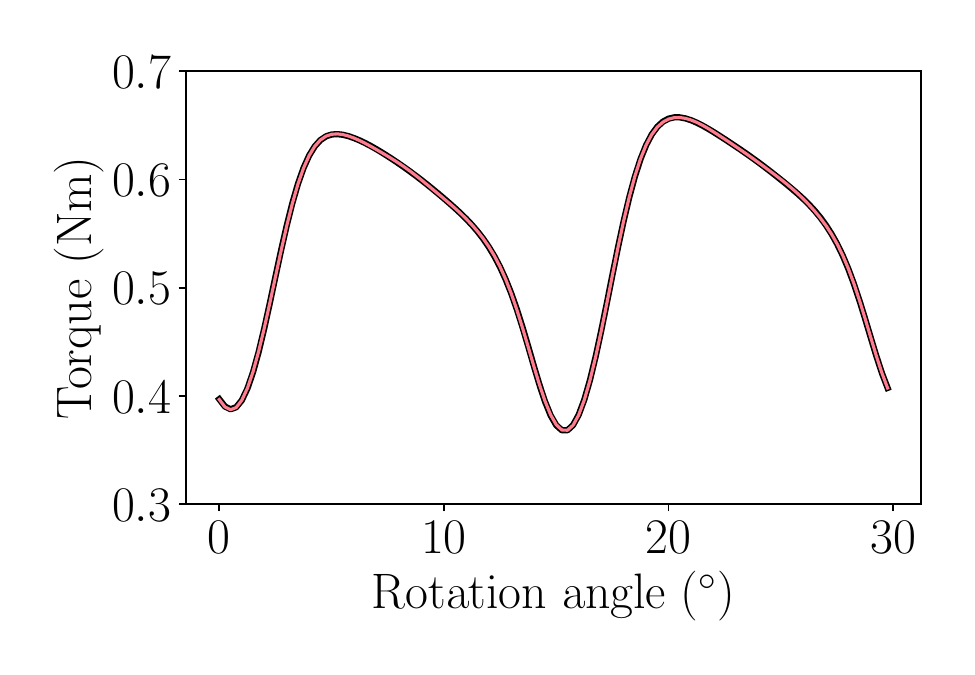}
     \caption{Worst-case reconstruction with $R=11$ frequency components.} 
     \label{fig:3-fft-21}
 \end{subfigure}
    \caption{Reconstruction error versus dimension reduction and worst-case torque signal reconstructions using \gls{dft}-based dimension reduction with $R=5$ and $R=11$ reduced frequency components.}
    \label{fig:3-fft}
\end{figure}

To identify the optimal level of dimension reduction, we first evaluate the \gls{mae} 
\begin{equation}
\label{eq:mae}
\mathrm{MAE} = \frac{1}{M} \sum_{m=1}^M \left\| \boldsymbol{\tau}^{(m)} - \mathcal{R}^{-1}\left(\mathbf{r}^{(m)}\right)\right\|_1,
\end{equation}
where $\mathcal{R} \colon \boldsymbol{\tau} \in \mathbb{R}^N \rightarrow \mathbf{r} \in \mathbb{C}^R$, $R < N$, as discussed in section~\ref{sec:dimension-reduction}.
To that end, we use a sample of $M=2000$ torque signals, each containing $N=120$ torque values $\tau_{\beta_n}$ evaluated at equidistantly distributed rotation angles $\beta_n \in \left[0^\circ, 30^\circ\right)$, $n=0,\dots,N-1$.
The torque signals correspond to uniform random samples of the design space defined in Table~\ref{tab:pmsm-parameters}, i.e., for sampling purposes, each parameter is assumed to be a random variable that is uniformly distributed within its value range, additionally assuming that the parameters are independent from one another.
It should be noted that if $R=N$, the original torque signal can be fully recovered without information loss, equivalently, the reconstruction error should be equal to zero (bearing machine accuracy).
Figure~\ref{fig:3-fft-all} shows the relationship between reduced frequency components and the associated reconstruction error, which may provide an indication as to how many frequency components should be retained to achieve an average reconstruction accuracy.

However, the \gls{mae} is not informative concerning worst-case reconstruction errors. 
This can be verified by the results shown in Figure~\ref{fig:3-fft-9}, which shows the worst-case signal reconstruction for $R=5$ frequency components.
In this case, the $\mathrm{MAE}$ is equal to $3.2 \cdot 10^{-3}$, which should be sufficient for a reconstruction that is visually identical to the original signal, based on the torque values. 
Still, the reconstructed signal deviates significantly from the original, thus indicating that the retained frequency components are insufficient.
The minimum number of frequency components to obtain a worst-case signal reconstruction that is visually identical to the original torque signal is $R = 11$, as can be observed in 
Figure~\ref{fig:3-fft-21}.
In this case, the $\mathrm{MAE}$ equals $9.0\cdot 10^{-5}$.
Based on these observations and numerical results, in the following, $R=11$ frequency components will be retained after the \gls{dft}. 

\paragraph{Remark on the selection of frequency components} In principle, it is possible to determine the set of influential frequency components based on the electric machine's geometry and configuration, e.g., based on the number of phases, poles, and slots. 
However, this requires expert knowledge. 
Moreover, geometric and material uncertainties can cause the number of influential frequency components to rise significantly \cite{gandhi2010recent, handgruber2013evaluation}, in which case the geometrical characteristics of the machine alone do not suffice.
The suggested approach allows to select the important frequency components in an automatic manner, without resorting to expert knowledge.

\subsection{Surrogate model accuracy}
\label{sec:results-rom} 
In this second numerical study, we assess the performance of surrogates obtained with the suggested framework in terms of prediction accuracy. 
Moreover, comparisons against alternative surrogates are performed, which use the same \glspl{rsm} but either employ \gls{pca} \cite{abdi2010principal} instead of \gls{dft} for dimension reduction or omit dimension reduction altogether.
In the former case, $21$ principal components are kept, resulting in an \gls{mae} equal to $2.8 \cdot 10^{-5}$ with respect to torque signal reconstruction. 
Note that the chosen number of principal components coincides with the number of reduced components obtained if the \gls{dft}-based, complex-valued, reduced \gls{qoi}, $\mathbf{r} \in \mathbb{C}^R$, is transformed to an equivalent real-valued \gls{qoi}, $\mathbf{r} \in \mathbb{R}^{2R-1}$, by considering separately the real and imaginary parts of the frequency components (see section~\ref{sec:dimension-reduction}), for $R=11$ (see section~\ref{sec:results-dimension-reduction}). 

For training and validation, an initial dataset $\mathcal{D} = \left\{\mathbf{p}^{(m)}, \boldsymbol{\tau}^{(m)} = \boldsymbol{\tau}\left(\mathbf{p}^{(m)}\right) \right\}_{m=1}^M$, $M=2000$, is partitioned into a training dataset $\mathcal{D}_{\text{t}}$ and a validation dataset $\mathcal{D}_{\text{v}}$, where $\mathcal{D}_{\text{t}} \cap \mathcal{D}_{\text{v}} = \emptyset$.
As previously discussed in section~\ref{sec:results-dimension-reduction}, the dataset $\mathcal{D}$ is generated by means of uniform random sampling of the design parameters.
The training dataset size, $M_\text{t}$, is progressively increased such that $M_\text{t} \in \left\{600, 1200, 1800\right\}$, which allows to examine the effect of training data availability on surrogate modeling accuracy.
Contrarily, the validation dataset remains fixed with size $M_\text{v}=200$.
Given a validation sample $\left( \mathbf{p}, \boldsymbol{\tau} \right) \in \mathcal{D}_{\text{v}}$, the \gls{ape} for a specific rotation angle $\beta$ with corresponding torque value $\tau_\beta \in \boldsymbol{\tau}$ is computed as
\begin{equation}
\label{eq:ape}
\mathrm{APE} = \frac{\left| \tau_\beta - \hat{\tau}_\beta \right|}{\tau_\beta},
\end{equation}
where $\hat{\tau}_\beta$ denotes the surrogate-based torque estimate.
Aggregating the \gls{ape} over the validation dataset, the error mean (\acrshort{mape}) and standard deviation (\acrshort{sdape}) can be computed, where the latter indicates the robustness of the surrogate with respect to its predictive accuracy.
Doing the same for all rotation angles $\beta_n$, $n=0, \dots, N-1$, the \acrshort{mape} and \acrshort{sdape} are computed over a full signal period.

\begin{figure}[t!]
    \centering
    \begin{subfigure}[b]{0.5\textwidth}
         \centering
         \fbox{\includegraphics[width=1\textwidth]{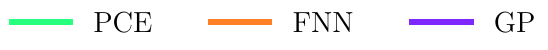}}
         \label{fig:mape_leg}
     \end{subfigure}
     \\
     \begin{subfigure}[b]{0.32\textwidth}
         \centering
         \includegraphics[width=\textwidth]{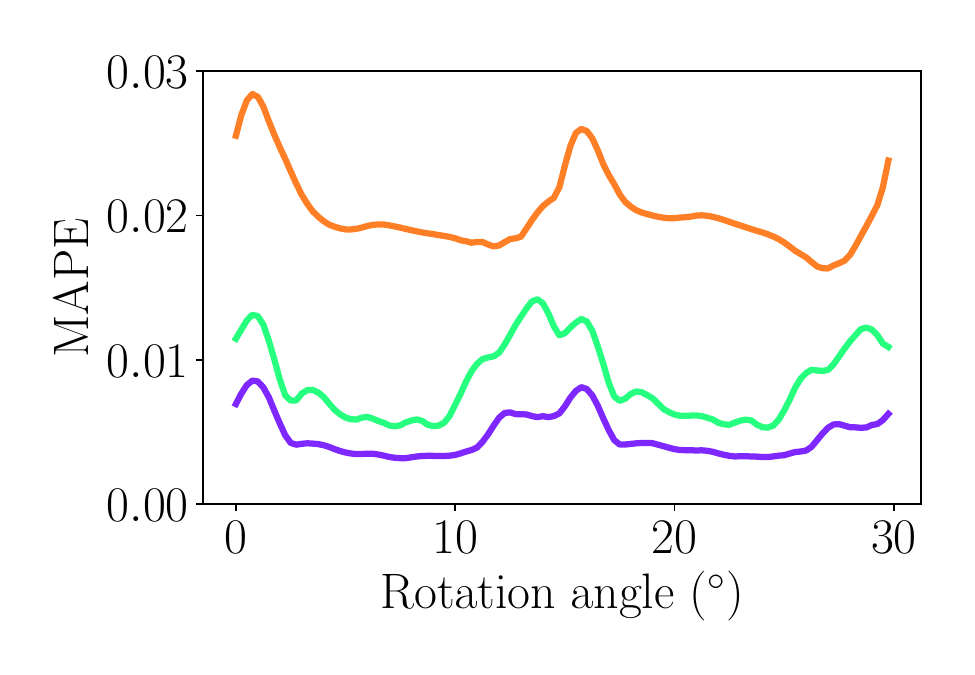}
         \caption{$M_\text{t}=600$ and \gls{dft}.}
         \label{fig:mape_fft_600}
     \end{subfigure}
     \hfill
     \begin{subfigure}[b]{0.32\textwidth}
         \centering
         \includegraphics[width=\textwidth]{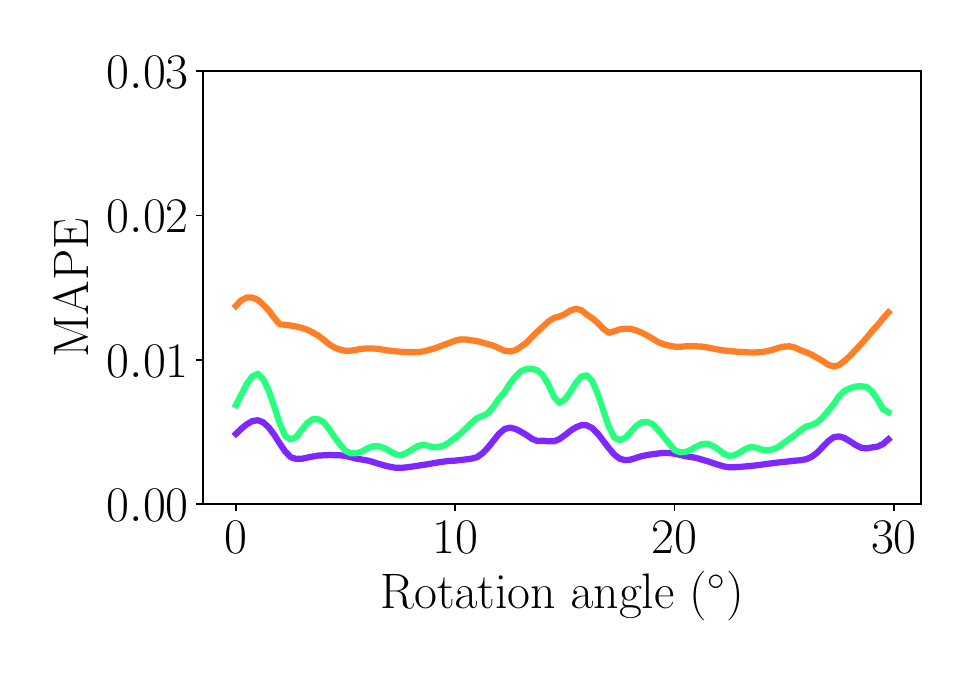}
         \caption{$M_\text{t}=1200$ and \gls{dft}.}
         \label{fig:mape_fft_1200}
     \end{subfigure}
     \hfill
     \begin{subfigure}[b]{0.32\textwidth}
         \centering
         \includegraphics[width=\textwidth]{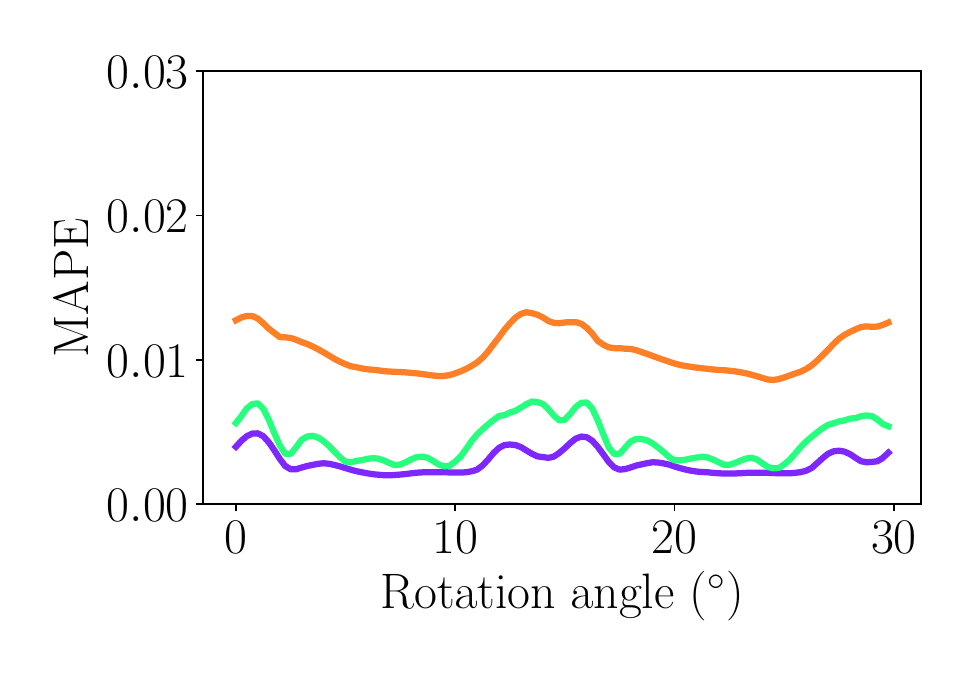}
         \caption{$M_\text{t}=1800$ and \gls{dft}.}
         \label{fig:mape_fft_1800}
     \end{subfigure}
    \begin{subfigure}[b]{0.32\textwidth}
         \centering
         \includegraphics[width=\textwidth]{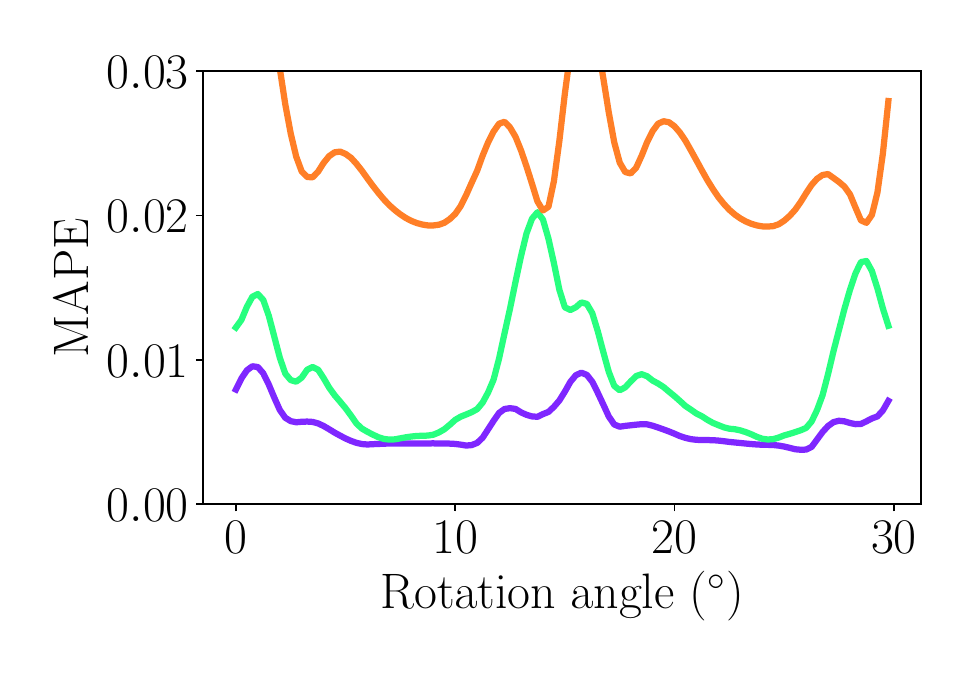}
         \caption{$M_\text{t}=600$ and \gls{pca}.}
         \label{fig:mape_pca_600}
     \end{subfigure}
     \hfill
     \begin{subfigure}[b]{0.32\textwidth}
         \centering
         \includegraphics[width=\textwidth]{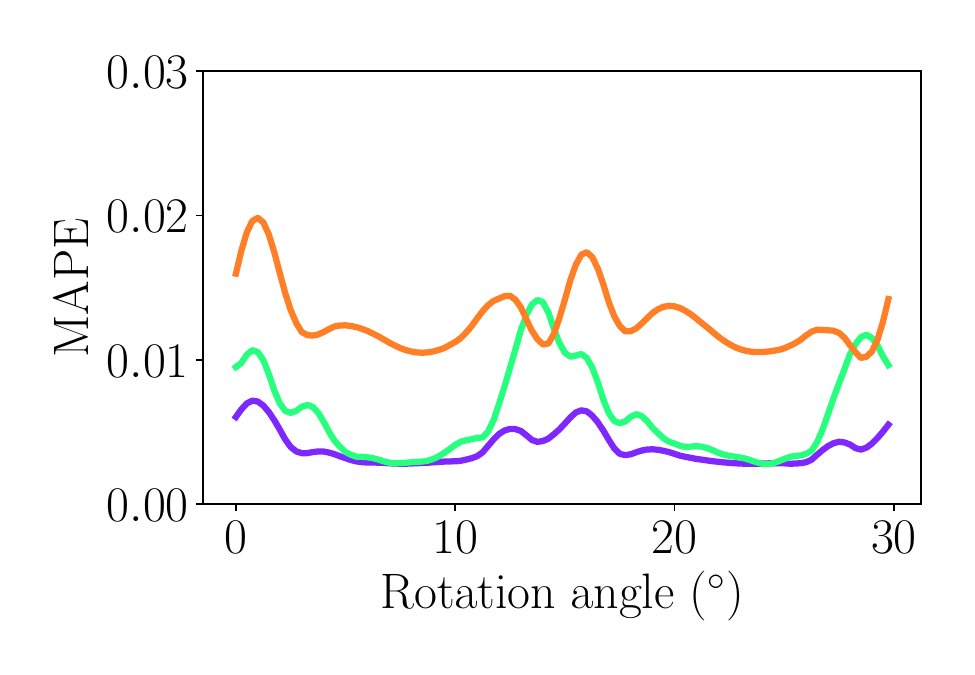}
         \caption{$M_\text{t}=1200$ and \gls{pca}.}
         \label{fig:mape_pca_1200}
     \end{subfigure}
     \hfill
     \begin{subfigure}[b]{0.32\textwidth}
         \centering
         \includegraphics[width=\textwidth]{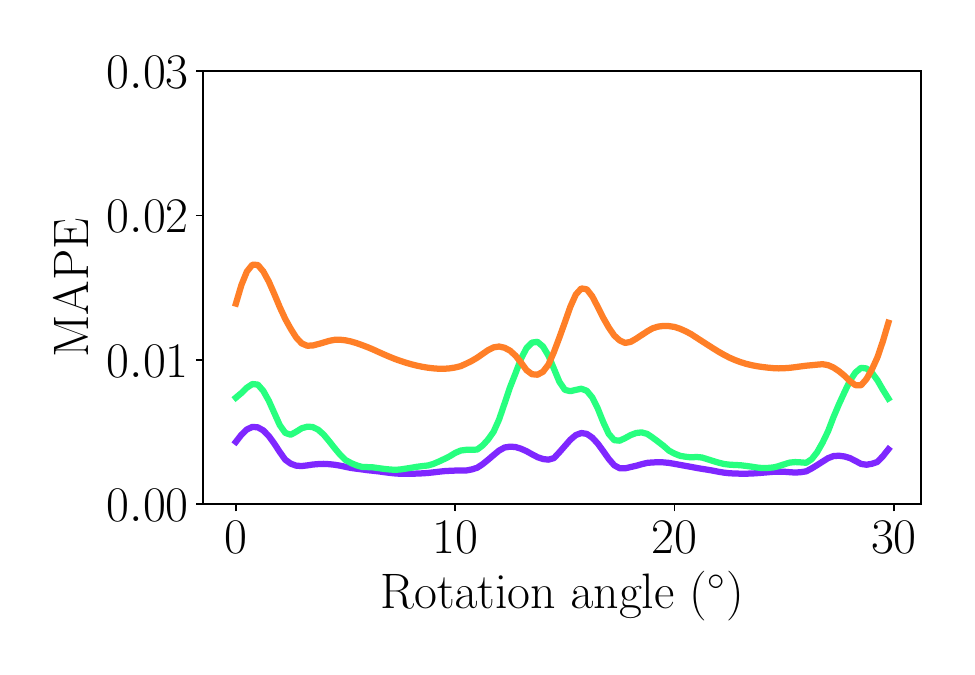}
         \caption{$M_\text{t}=1800$ and \gls{pca}.}
         \label{fig:mape_pca_1800}
     \end{subfigure}
    \begin{subfigure}[b]{0.32\textwidth}
         \centering
         \includegraphics[width=\textwidth]{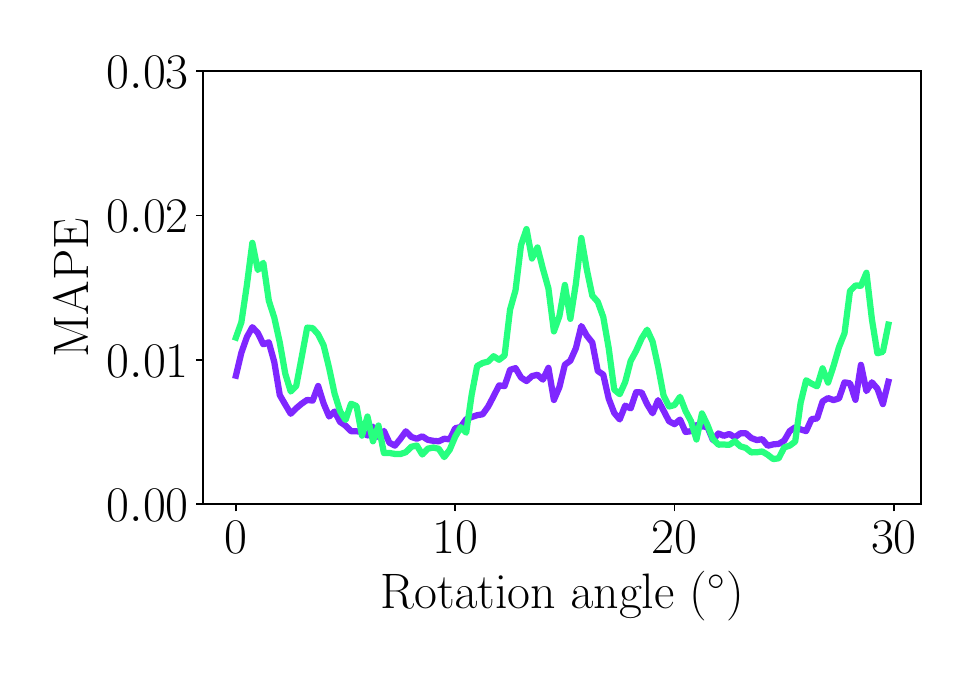}
         \caption{$M_\text{t}=600$ and no reduction.}
         \label{fig:mape_time_600}
     \end{subfigure}
     \hfill
     \begin{subfigure}[b]{0.32\textwidth}
         \centering
         \includegraphics[width=\textwidth]{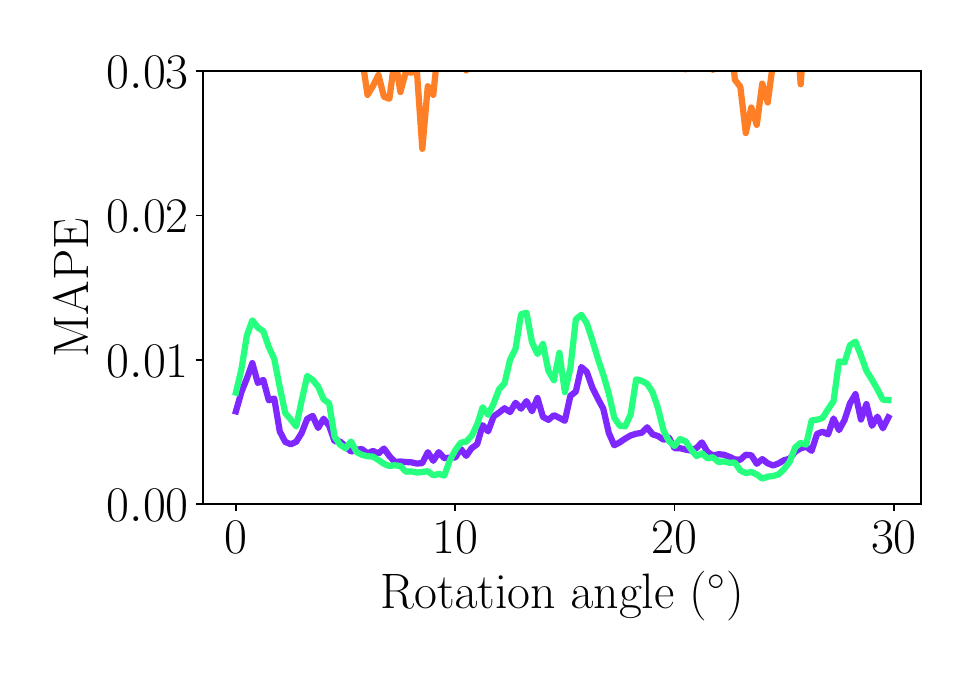}
         \caption{$M_\text{t}=1200$ and no reduction.}
         \label{fig:mape_time_1200}
     \end{subfigure}
     \hfill
     \begin{subfigure}[b]{0.32\textwidth}
         \centering
         \includegraphics[width=\textwidth]{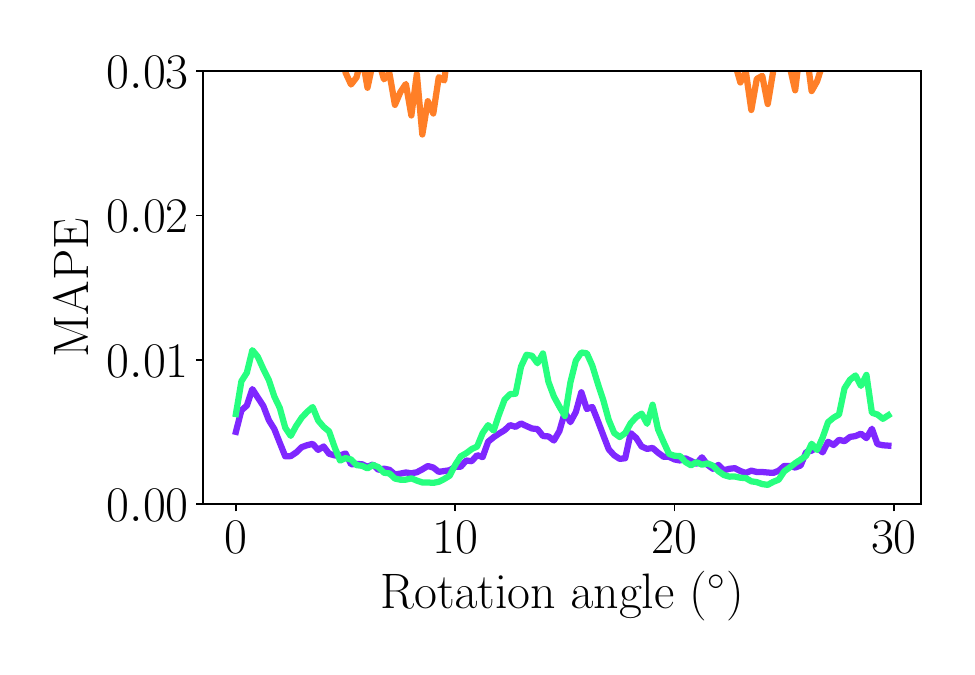}
         \caption{$M_\text{t}=1800$ and no reduction.}
         \label{fig:mape_time_1800}
     \end{subfigure}  
     \caption{\Gls{mape} for surrogates computed with different combinations of training dataset size, \gls{rsm}, and dimension reduction approach.}
    \label{fig:9_mape_sur}
\end{figure}

\begin{figure}[t!]
    \centering
    \begin{subfigure}[b]{0.5\textwidth}
         \centering
         \fbox{\includegraphics[width=1\textwidth]{figures/UQ/plot_legenduq.pdf}}
     \end{subfigure}
     \\
     \begin{subfigure}[b]{0.32\textwidth}
         \centering
         \includegraphics[width=\textwidth]{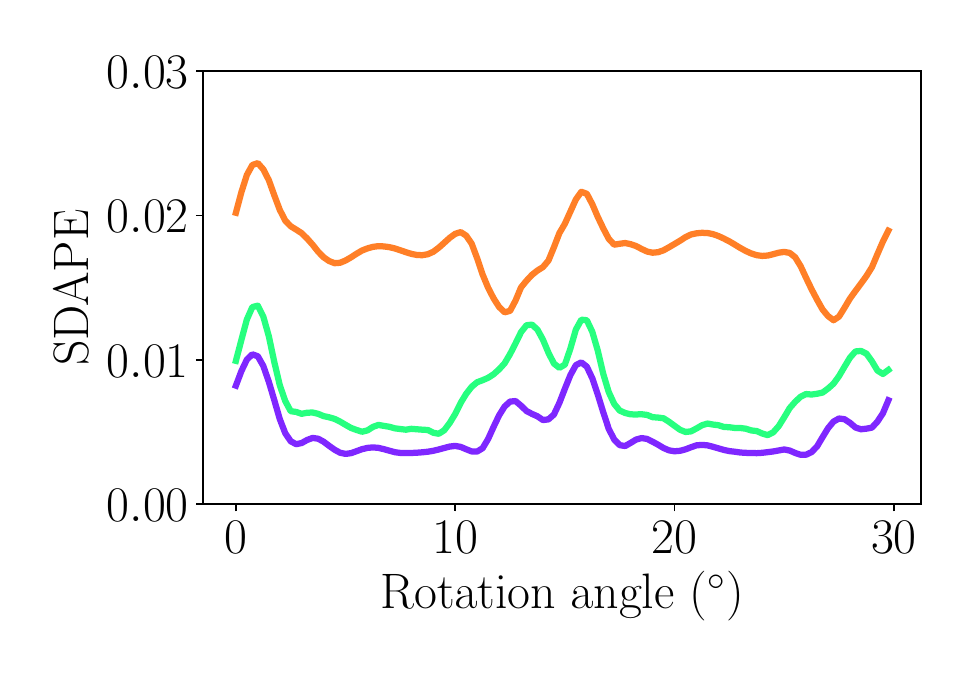}
         \caption{$M_\text{t}=600$ and \gls{dft}.}
         \label{fig:sdape_fft_600}
     \end{subfigure}
     \hfill
     \begin{subfigure}[b]{0.32\textwidth}
         \centering
         \includegraphics[width=\textwidth]{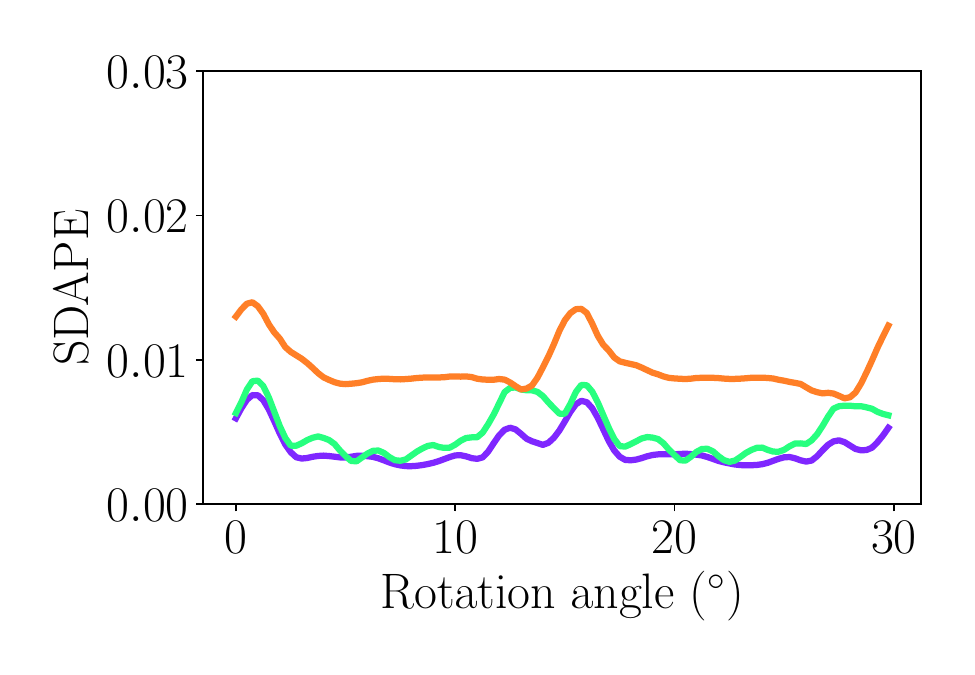}
         \caption{$M_\text{t}=1200$ and \gls{dft}.}
         \label{fig:sdape_fft_1200}
     \end{subfigure}
     \hfill
     \begin{subfigure}[b]{0.32\textwidth}
         \centering
         \includegraphics[width=\textwidth]{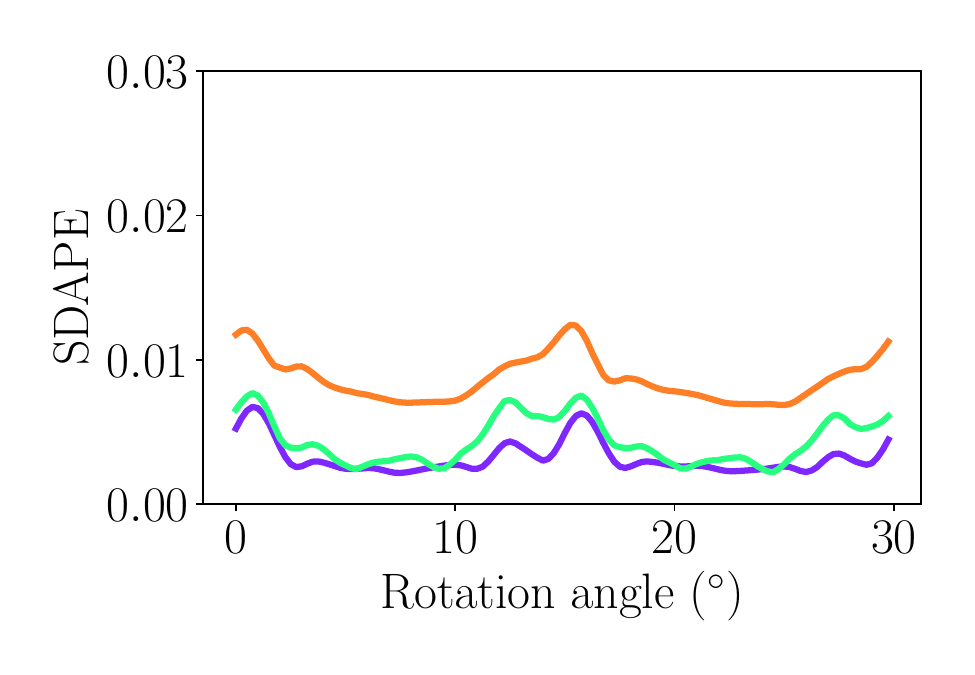}
         \caption{$M_\text{t}=1800$ and \gls{dft}.}
         \label{fig:sdape_fft_1800}
     \end{subfigure}
    \begin{subfigure}[b]{0.32\textwidth}
         \centering
         \includegraphics[width=\textwidth]{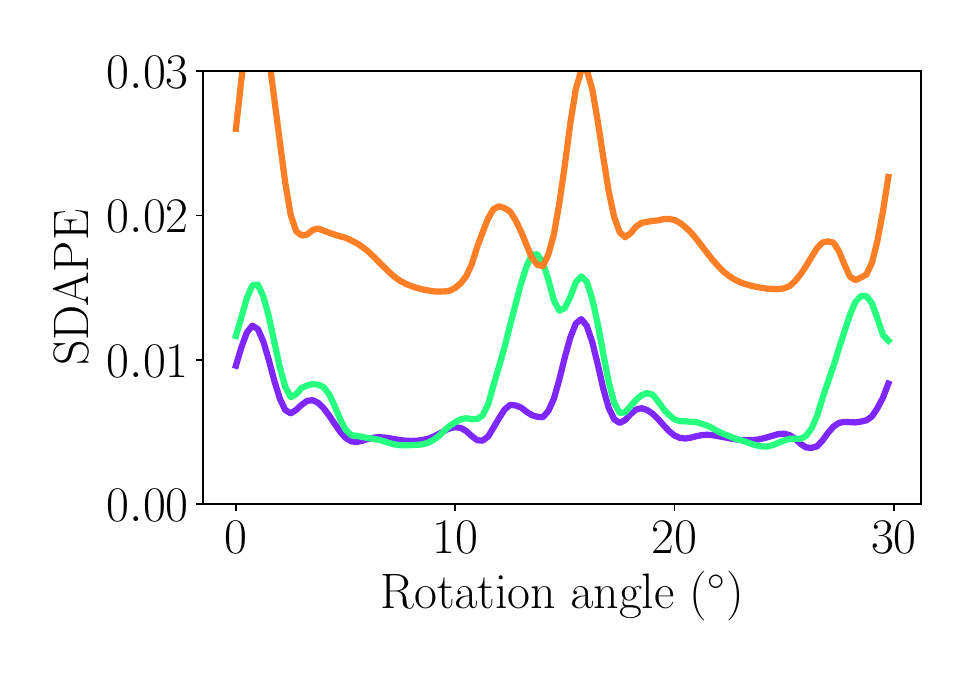}
         \caption{$M_\text{t}=600$ and \gls{pca}.}
         \label{fig:sdape_pca_600}
     \end{subfigure}
     \hfill
     \begin{subfigure}[b]{0.32\textwidth}
         \centering
         \includegraphics[width=\textwidth]{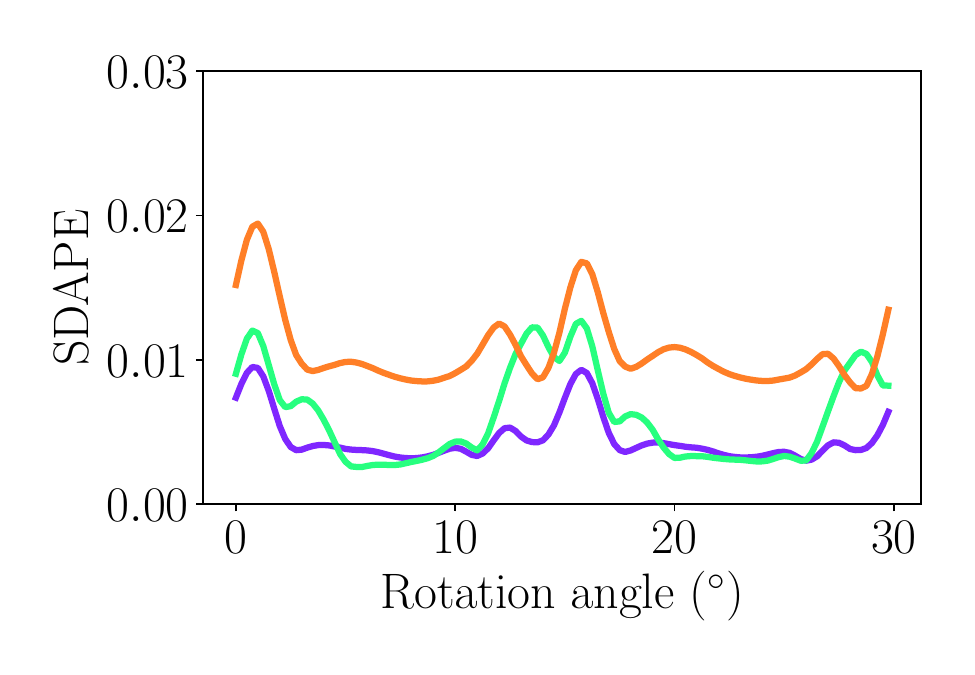}
         \caption{$M_\text{t}=1200$ and \gls{pca}.}
         \label{fig:sdape_pca_1200}
     \end{subfigure}
     \hfill
     \begin{subfigure}[b]{0.32\textwidth}
         \centering
         \includegraphics[width=\textwidth]{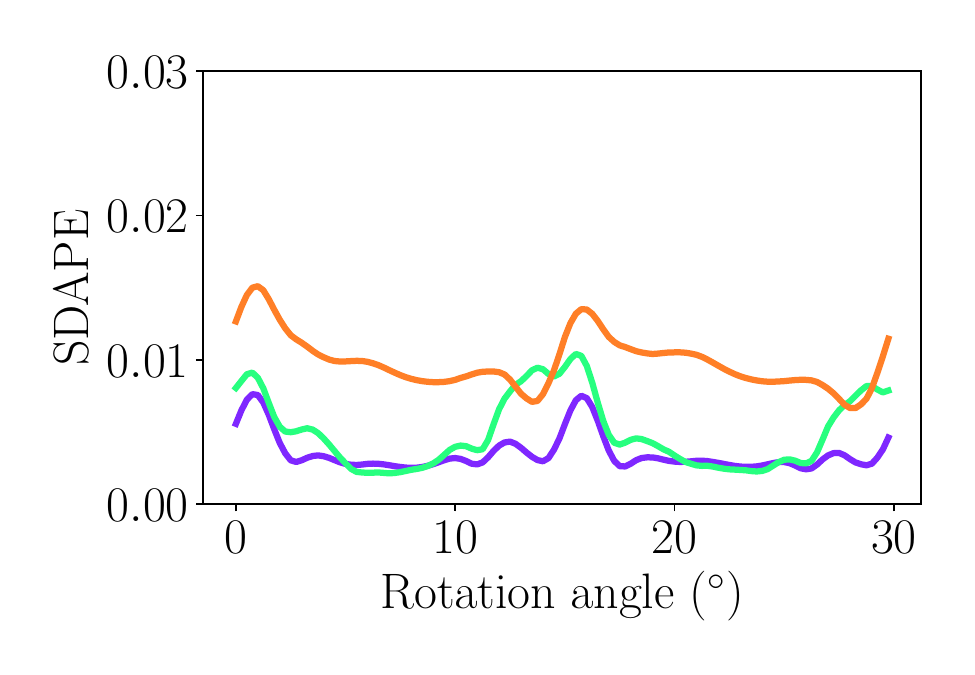}
         \caption{$M_\text{t}=1800$ and \gls{pca}.}
         \label{fig:sdape_pca_1800}
     \end{subfigure}
    \begin{subfigure}[b]{0.32\textwidth}
         \centering
         \includegraphics[width=\textwidth]{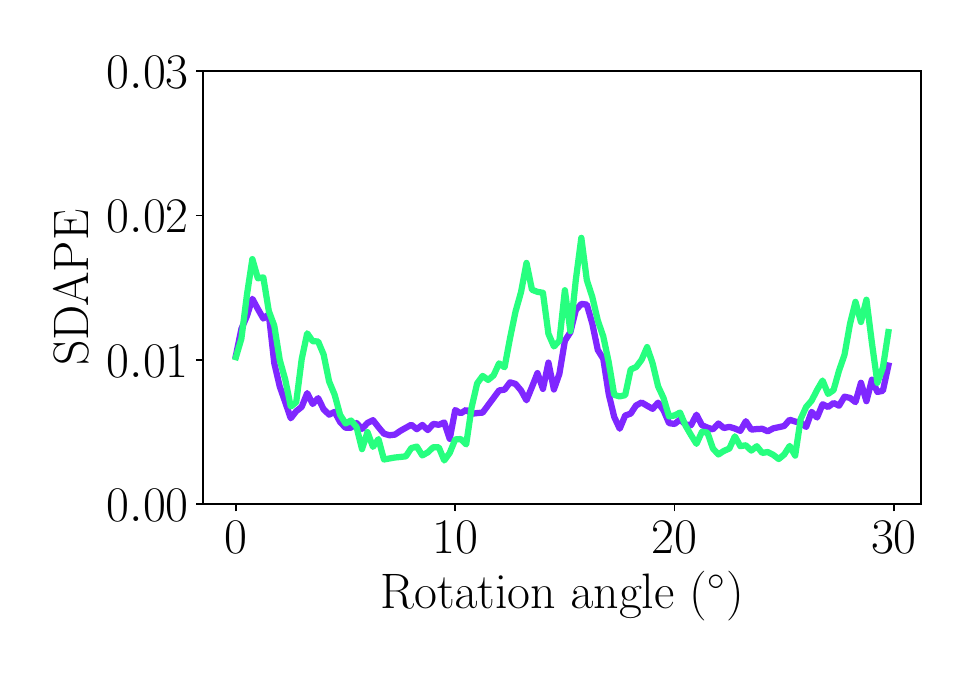}
         \caption{$M_\text{t}=600$ and no reduction.}
         \label{fig:sdape_time_600}
     \end{subfigure}
     \hfill
     \begin{subfigure}[b]{0.32\textwidth}
         \centering
         \includegraphics[width=\textwidth]{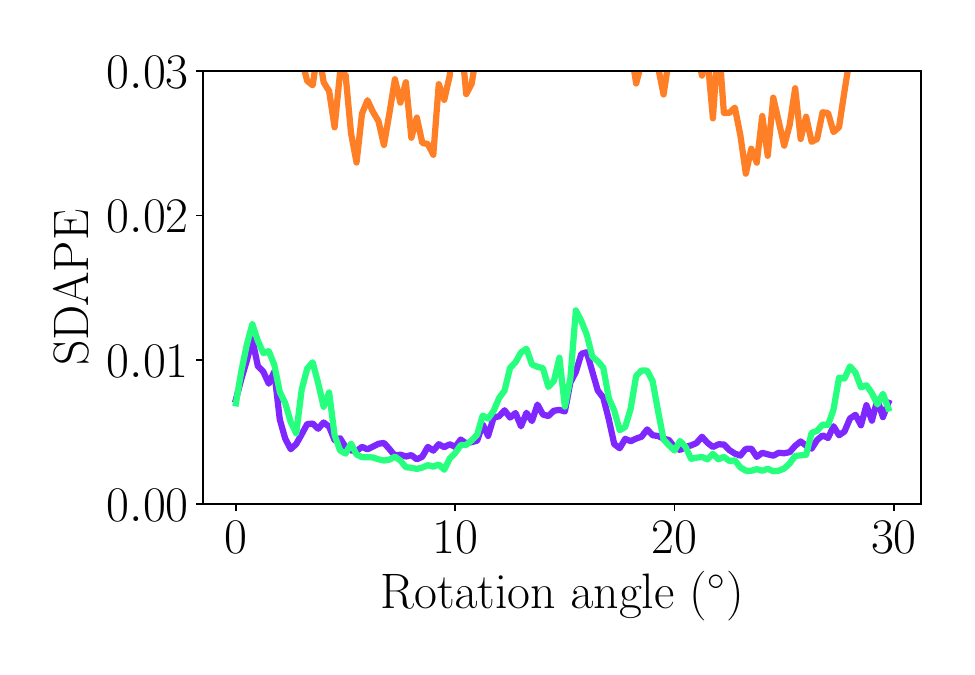}
         \caption{$M_\text{t}=1200$ and no reduction.}
         \label{fig:sdape_time_1200}
     \end{subfigure}
     \hfill
     \begin{subfigure}[b]{0.32\textwidth}
         \centering
         \includegraphics[width=\textwidth]{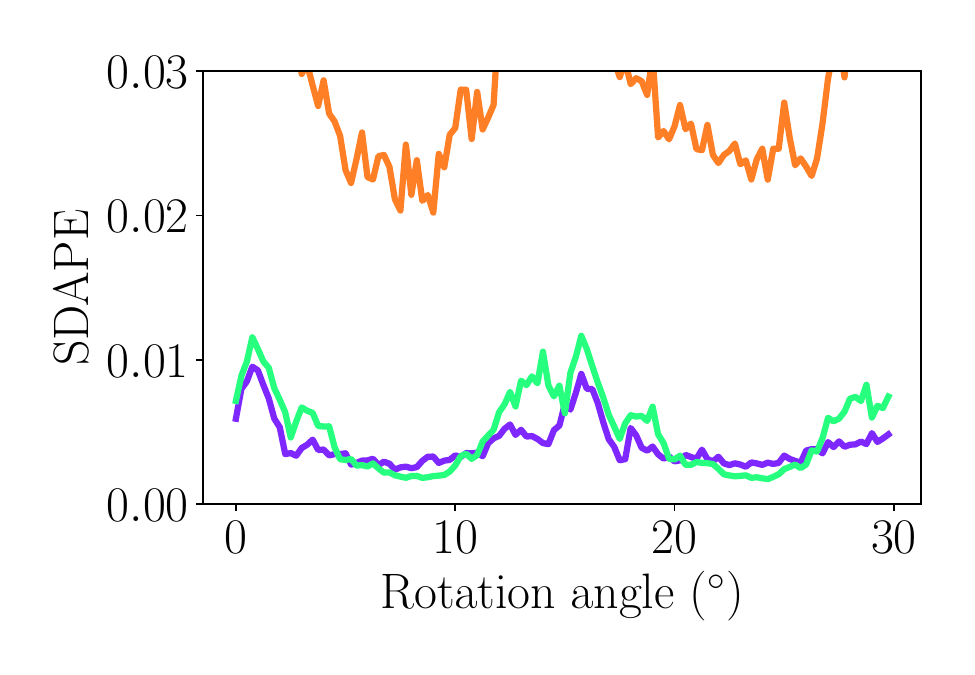}
         \caption{$M_\text{t}=1800$ and no reduction.}
         \label{fig:sdape_time_1800}
     \end{subfigure}  
     \caption{\Gls{sdape} for surrogates computed with different combinations of training dataset size, \gls{rsm}, and dimension reduction approach.}
    \label{fig:9_sdape_sur}
\end{figure}

The \acrshort{mape} and \acrshort{sdape} results are shown in Figures~\ref{fig:9_mape_sur} and \ref{fig:9_sdape_sur}, respectively, for all combinations of training dataset size, \glspl{rsm}, and dimension reduction approach (or lack thereof).
As would be expected, the accuracy and robustness of all surrogates is improved with more training data.
The improvement is significant when increasing the training dataset size from $M_{\text{t}}=600$ to $M_{\text{t}}=1200$, but only minor gains are obtained for a further increase to $M_{\text{t}}=1800$.
Irrespective of training dataset size or dimension reduction approach, \glspl{gp} are generally the best \gls{rsm} option in terms of surrogate modeling accuracy and robustness, especially if combined with dimension reduction.
On the opposite side, \gls{fnn} is consistently the worst-performing \gls{rsm} option.  
Using \gls{dft} for dimension reduction yields overall more accurate and robust surrogates compared to \gls{pca} or no dimension reduction, especially given smaller training datasets. 
The combination of \gls{dft}-based dimension reduction with a \gls{gp}-based \gls{rsm} yields the best-in-class surrogate model.
These observations are further supported by the results presented in Table~\ref{tab:mape-surrogates}, where the \acrshort{mape} is averaged over the torque signal.

\begin{table*}[b!]
\caption{Signal-averaged \acrshort{mape} for surrogates computed with different combinations of training dataset size, \gls{rsm}, and dimension reduction approach. The best-in-class error values for the same training dataset size are marked as bold.}
\label{tab:mape-surrogates}
\tabcolsep=0pt
\begin{threeparttable}
\begin{tabular*}{\textwidth}{@{\extracolsep{\fill}}lccccccccc@{\extracolsep{\fill}}}
\toprule
 & \multicolumn{3}{c}{DFT}& \multicolumn{3}{c}{PCA} & \multicolumn{3}{c}{No reduction}\\
\cmidrule(lr){2-4}\cmidrule(lr){5-7}\cmidrule(lr){8-10}
$M_\text{t}$ & PCE & FNN & GP & PCE & FNN & GP & PCE & FNN & GP\\
\midrule
600 & 0.0085 & 0.0203 & \textbf{0.0047} & 0.0091 & 0.02407 & 0.0055 & 0.0091 & 0.0526 & 0.0068 \\
1200 & 0.0057 & 0.0114 & \textbf{0.0037} & 0.0063 & 0.0127 & 0.0039 & 0.0063 & 0.0379 & 0.0049\\
1800 & 0.0044 & 0.0106 & \textbf{0.0029} & 0.0051 & 0.0110 & 0.0030 & 0.0050 & 0.0371 & 0.0038\\
\bottomrule
\end{tabular*}
\end{threeparttable}
\end{table*}

\begin{figure}[t!]
    \centering
    \begin{subfigure}[b]{0.7\textwidth}
         \centering
         \fbox{\includegraphics[width=1\textwidth]{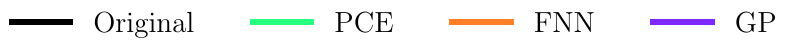}}
     \end{subfigure}
     \\
     \begin{subfigure}[b]{0.32\textwidth}
         \centering
         \includegraphics[width=\textwidth]{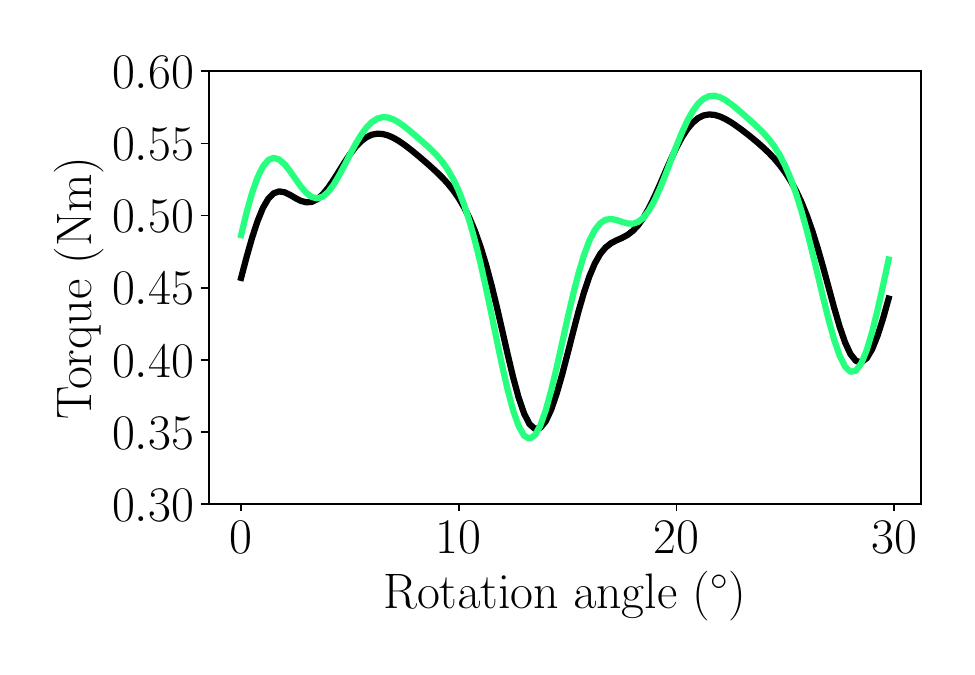}
         \caption{\gls{dft} and \gls{pce}.}
         \label{fig:fft_pce_1800}
     \end{subfigure}
     \hfill
     \begin{subfigure}[b]{0.32\textwidth}
         \centering
         \includegraphics[width=\textwidth]{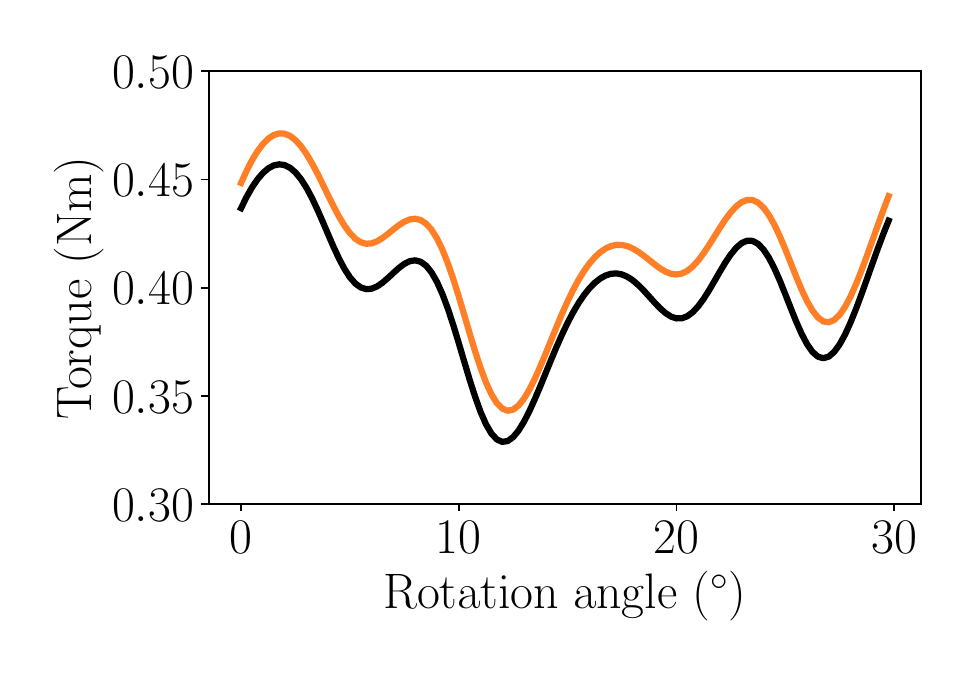}
         \caption{\gls{dft} and \gls{fnn}.}
         \label{fig:fft_nn_1800}
     \end{subfigure}
     \hfill
     \begin{subfigure}[b]{0.32\textwidth}
         \centering
         \includegraphics[width=\textwidth]{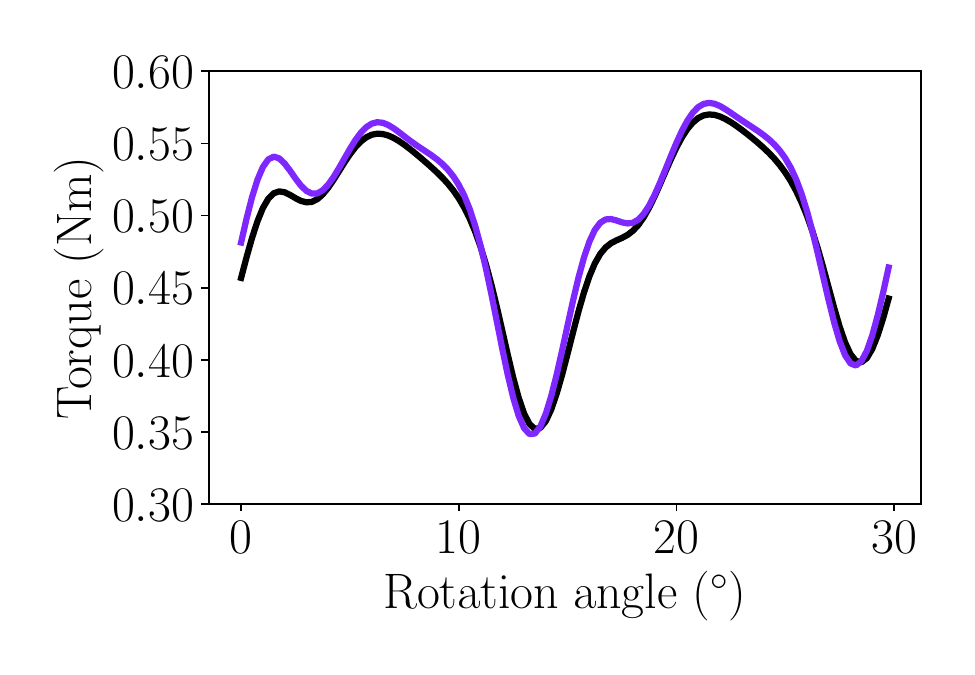}
         \caption{\gls{dft} and \gls{gp}.}
         \label{fig:fft_krig_1800}
     \end{subfigure}
     \\
     \begin{subfigure}[b]{0.32\textwidth}
         \centering
         \includegraphics[width=\textwidth]{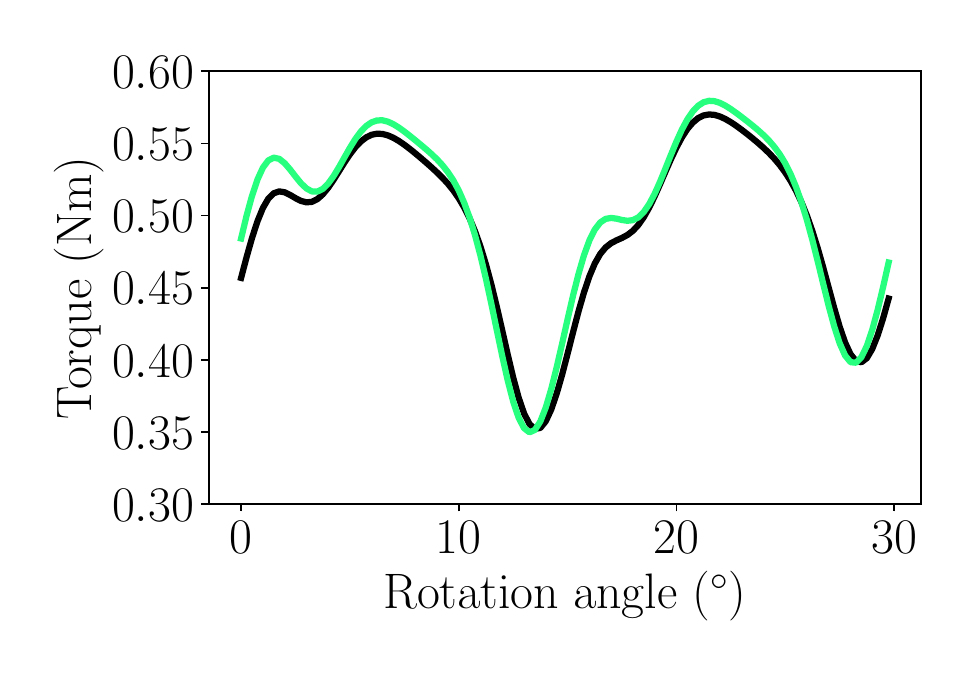}
         \caption{\gls{pca} and \gls{pce}.}
         \label{fig:pca_pce_1800}
     \end{subfigure}
     \hfill
     \begin{subfigure}[b]{0.32\textwidth}
         \centering
         \includegraphics[width=\textwidth]{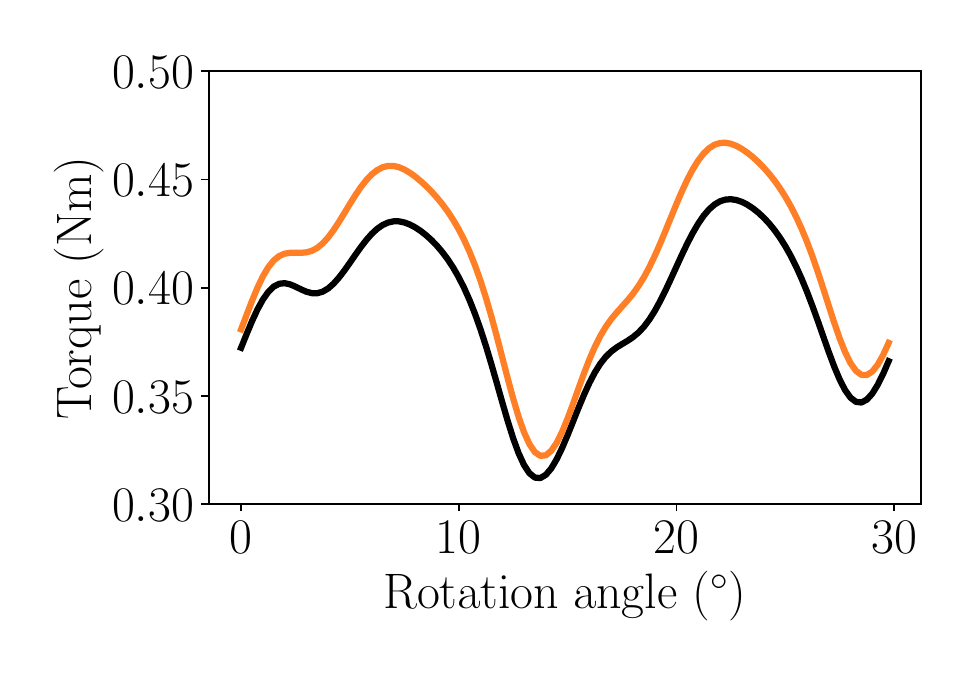}
         \caption{\gls{pca} and \gls{fnn}.}
         \label{fig:pca_nn_1800}
     \end{subfigure}
     \hfill
     \begin{subfigure}[b]{0.32\textwidth}
         \centering
         \includegraphics[width=\textwidth]{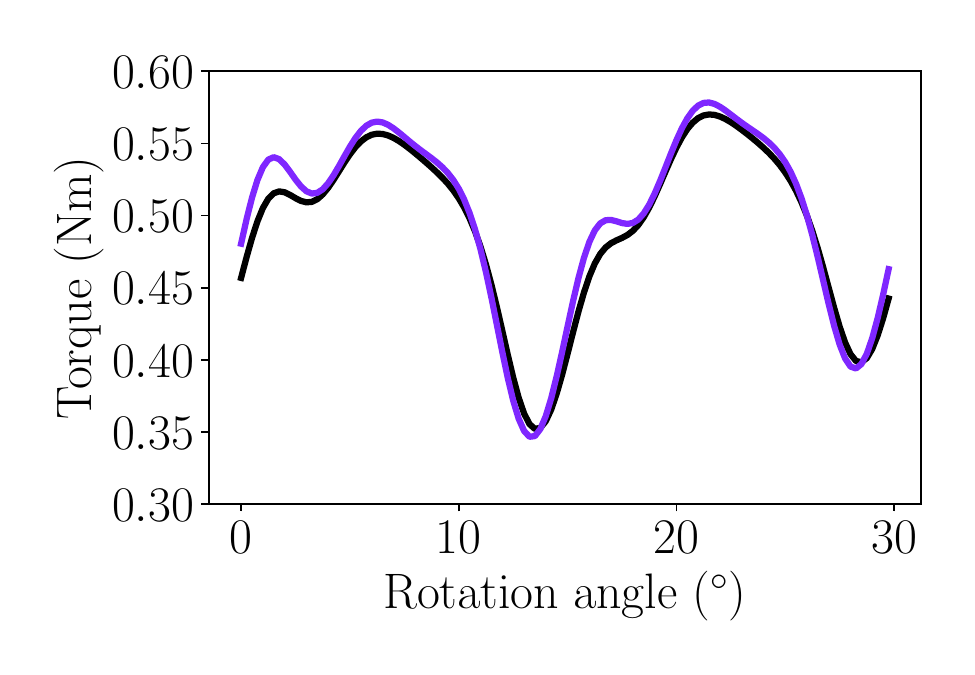}
         \caption{\gls{pca} and \gls{gp}.}
         \label{fig:pca_krig_1800}
     \end{subfigure}
     \\
     \begin{subfigure}[b]{0.32\textwidth}
         \centering
         \includegraphics[width=\textwidth]{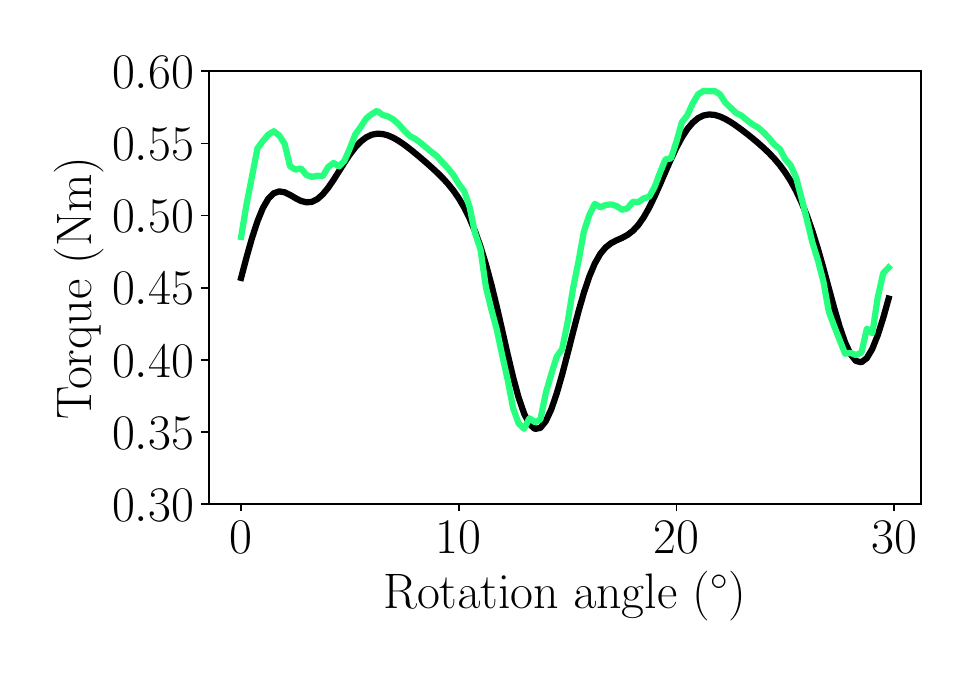}
         \caption{No reduction and \gls{pce}.}
         \label{fig:time_pce_1800}
     \end{subfigure}
     \hfill
     \begin{subfigure}[b]{0.32\textwidth}
         \centering
         \includegraphics[width=\textwidth]{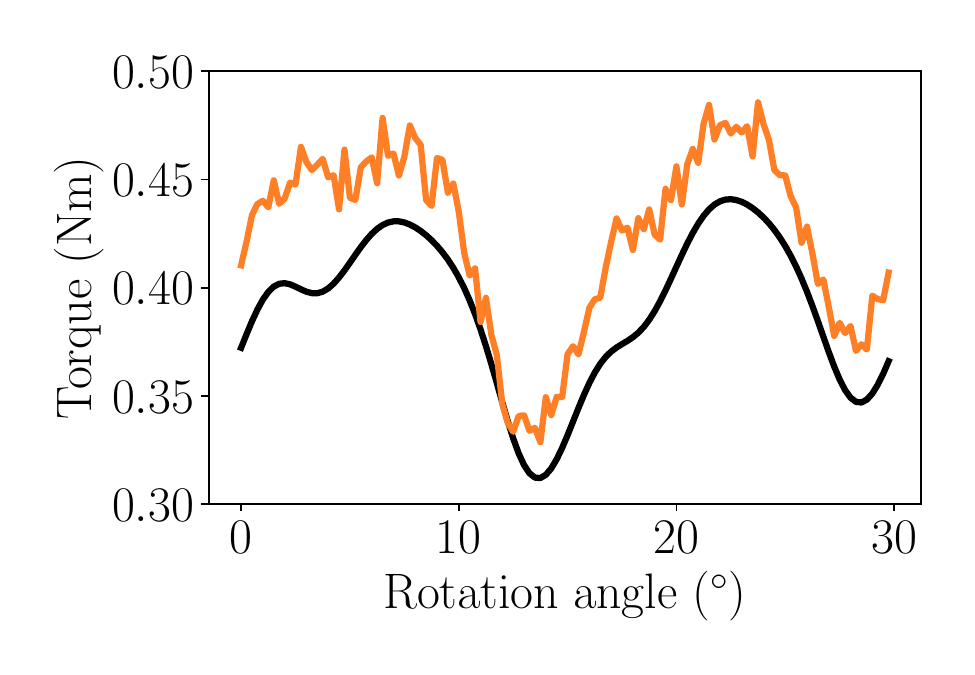}
         \caption{No reduction and \gls{fnn}.}
         \label{fig:time_nn_1800}
     \end{subfigure}
     \hfill
     \begin{subfigure}[b]{0.32\textwidth}
         \centering
         \includegraphics[width=\textwidth]{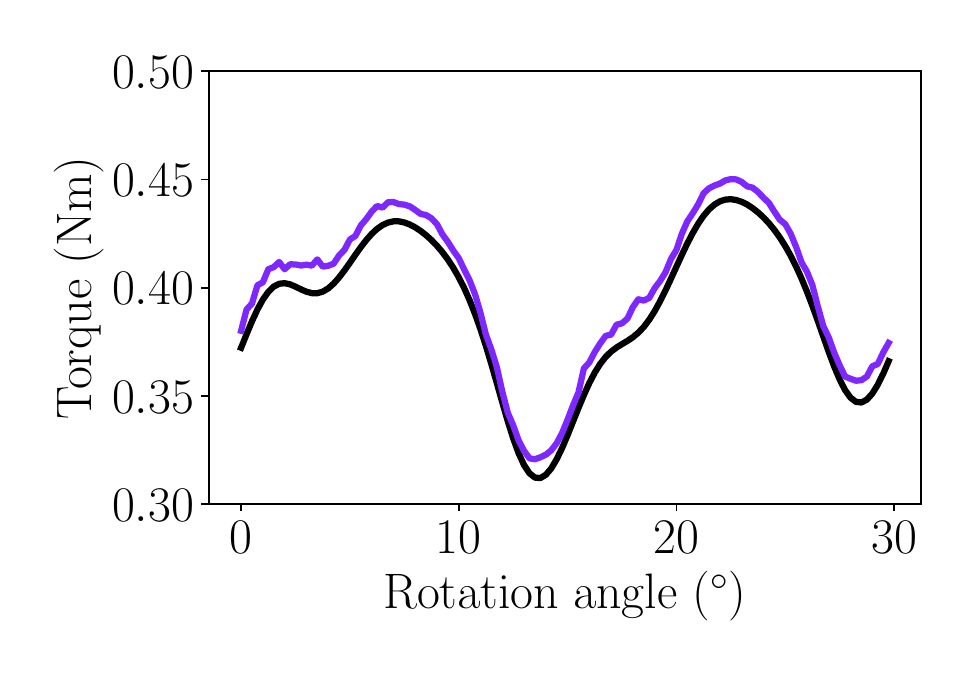}
         \caption{No reduction and \gls{gp}.}
         \label{fig:time_krig_1800}
     \end{subfigure}
     \caption{Worst-case surrogate-based torque signal predictions for training dataset size $M_\text{t}=1800$ and different combinations of \gls{rsm} and dimension reduction approach.}
    \label{fig:9-sur-1800}
\end{figure}

Figure~\ref{fig:9-sur-1800} shows the worst-case torque signal predictions for training dataset size $M_\text{t}=1800$ and each combination of \gls{rsm} and dimension reduction approach. 
The results for $M_\text{t}=600$ and $M_\text{t}=1200$ are available in the Appendix (Figures~\ref{fig:9-sur-600} and \ref{fig:9-sur-1200}).
In all cases, it is clear that surrogates based on \glspl{fnn} perform much worse than \gls{pce} and \gls{gp}-based ones.
The latter two \glspl{rsm} do not result in significant differences and only a slight edge can be claimed for \gls{gp}.
Note that, in the case of either \gls{dft} or \gls{pca}-based dimension reduction, the observed deviations between original and predicted torque signal can be attributed solely to training data availability and \gls{rsm} approximation capability (see section~\ref{sec:results-dimension-reduction}).

Last but not least, it is worth noting that surrogate modeling accuracy and robustness clearly benefit from dimension reduction, both on average and in worst-case. 
This is particularly evident for the \gls{fnn}, but actually applies to all \glspl{rsm}.
The enhancing effect of dimension reduction to the predictive accuracy of surrogate models is in fact known, e.g., considering time-series prediction \cite{verleysen2005curse} or deep operator learning \cite{kontolati2024learning}. 
For the use-case considered in this work, it is obvious both \gls{dft} and \gls{pca} provide this benefit, with a slight edge given to the former.

\subsection{Surrogate-based uncertainty quantification}
\label{sec:results-uq} 
In this last numerical study, surrogates obtained with the suggested framework are used for \gls{uq}. 
In particular, they replace the original, high-fidelity \gls{pmsm} model in Monte Carlo sampling, which aims at estimating torque statistics, specifically means and standard deviations of torque signals. 
Additionally, surrogate-based estimates are compared against reference statistics obtained by sampling the original model.
The model, either original or surrogate, is sampled $10^4$ times based on the considered parameter distributions.
The sample mean and standard deviation of a torque value $\tau_\beta$ are denoted with $\hat{\mu}\left[\tau_\beta\right]$ and $\hat{\sigma}\left[\tau_\beta\right]$, respectively.
In the following, we consider two distinct \gls{uq} scenarios concerning the parameter distributions, namely, uniform distributions in section~\ref{sec:uq-uniform}, and Gaussian distributions in section~\ref{sec:uq-gaussian}. 
Section~\ref{sec:costs-and-gains} discusses the computational costs and gains of the proposed method in the context of the aforementioned \gls{uq} studies.

\paragraph{Remark on modeling assumptions and \gls{uq}} As previously noted in section~\ref{sec:problem-setting}, only one-fourth of a two-dimensional cross-section of the \gls{pmsm} is actually modelled, by assumption of rotational symmetry and translational invariance along the $z$-axis.
In the \gls{uq} studies performed in the following, this assumption still holds, so that the \gls{pmsm} model described in sections \ref{sec:geometry-representation}, \ref{sec:magnetostatic-model}, and \ref{sec:numerical-approx} can be used. 
Nonetheless, we should note that this cannot be true in practice. For example, in a real-world setting, geometrical variations due to manufacturing tolerances will not be symmetrical.

\subsubsection{Uniform parameter distributions}
\label{sec:uq-uniform}

\begin{figure}[t!]
    \centering
    \begin{subfigure}[b]{0.7\textwidth}
         \centering
         \fbox{\includegraphics[width=1\textwidth]{figures/plot_legendsur1.pdf}}
     \end{subfigure}
     \\
     \begin{subfigure}[b]{0.32\textwidth}
         \centering
         \includegraphics[width=\textwidth]{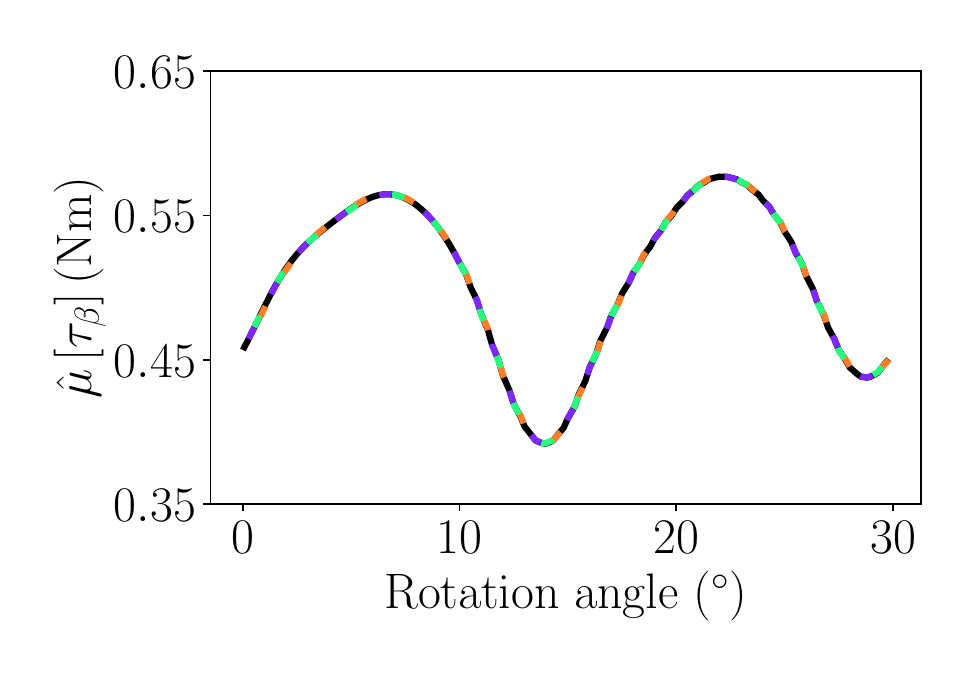}
         \caption{Mean, $M_\text{t}=600$.}
     \end{subfigure}
     \hfill
     \begin{subfigure}[b]{0.32\textwidth}
         \centering
         \includegraphics[width=\textwidth]{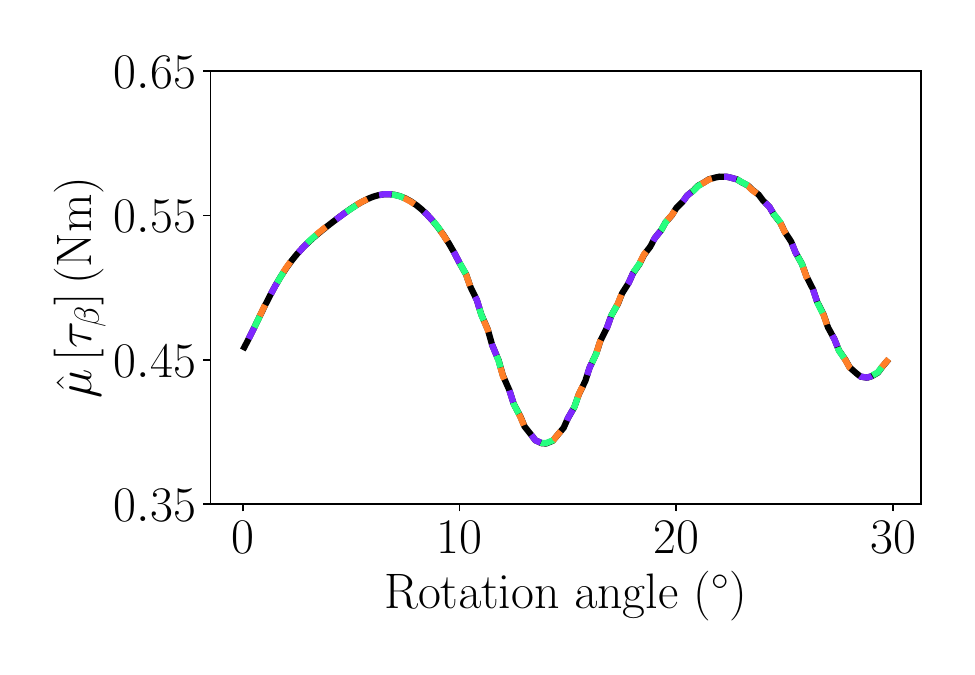}
         \caption{Mean, $M_\text{t}=1200$.}
     \end{subfigure}
     \hfill
     \begin{subfigure}[b]{0.32\textwidth}
         \centering
         \includegraphics[width=\textwidth]{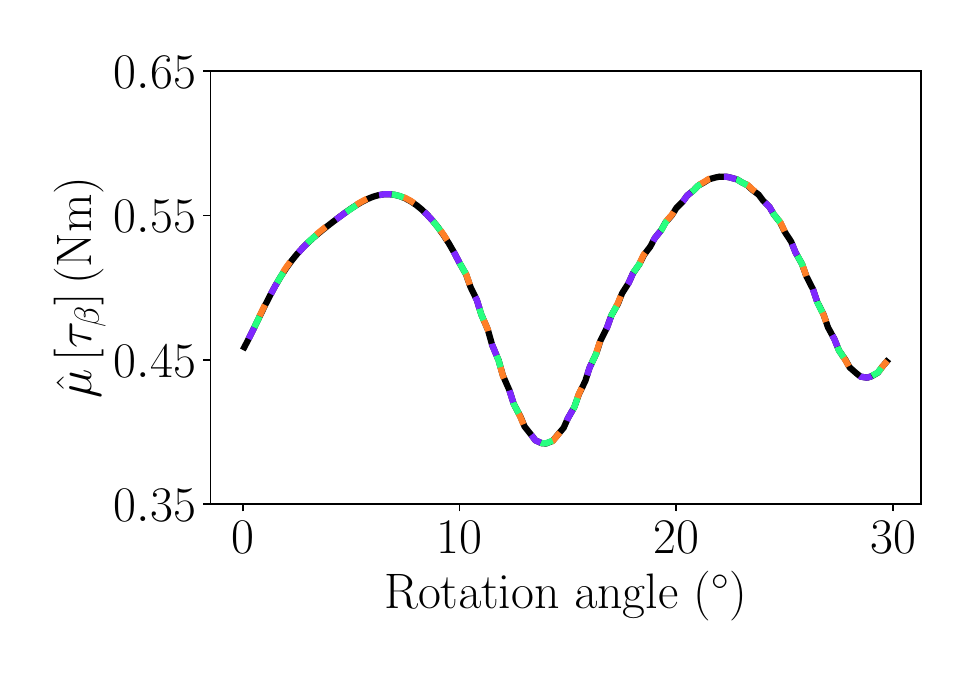}
         \caption{Mean, $M_\text{t}=1800$.}
     \end{subfigure} 
     \\
     \begin{subfigure}[b]{0.32\textwidth}
         \centering
         \includegraphics[width=\textwidth]{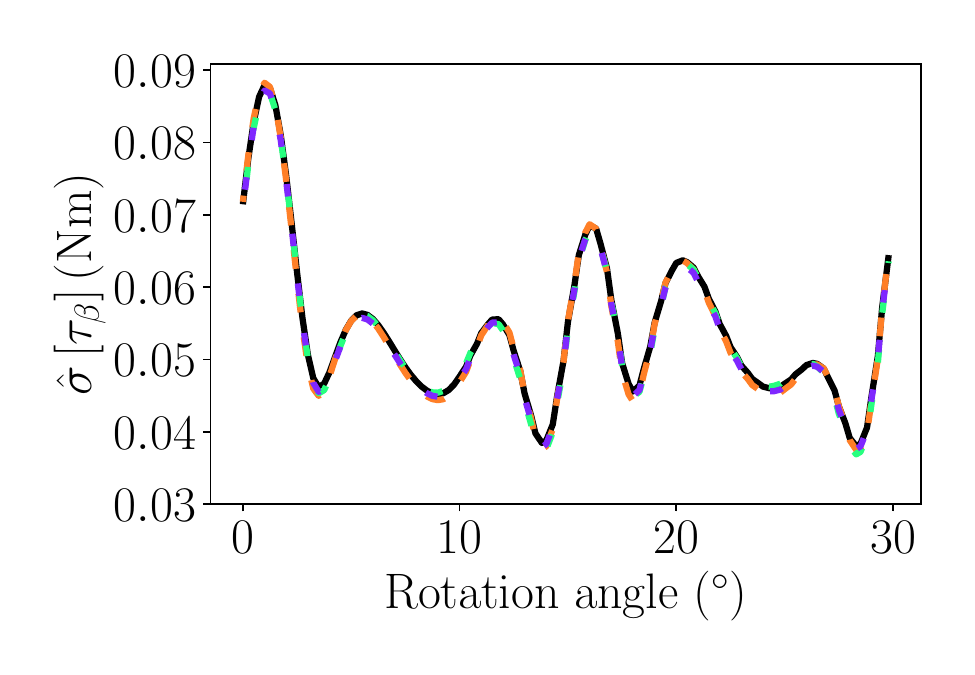}
         \caption{Standard deviation, $M_\text{t}=600$.}
     \end{subfigure}
     \hfill
     \begin{subfigure}[b]{0.32\textwidth}
         \centering
         \includegraphics[width=\textwidth]{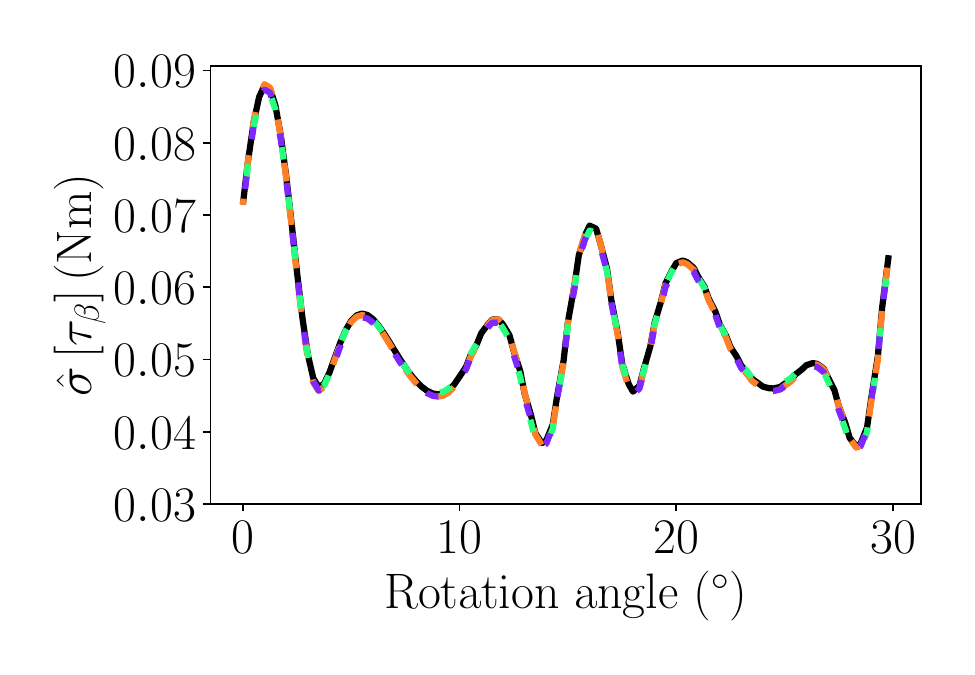}
         \caption{Standard deviation, $M_\text{t}=1200$.}
     \end{subfigure}
     \hfill
     \begin{subfigure}[b]{0.32\textwidth}
         \centering
         \includegraphics[width=\textwidth]{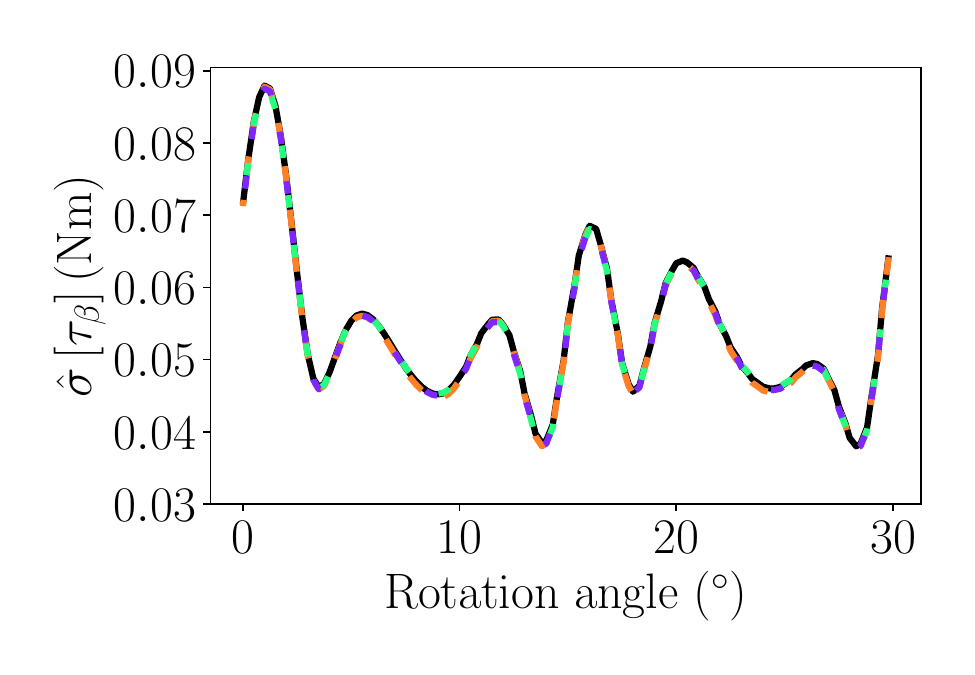}
         \caption{Standard deviation, $M_\text{t}=1800$.}
     \end{subfigure}   
     \caption{Torque mean and standard deviation estimates for uniform parameter distributions, obtained by sampling the original, high-fidelity model and the framework-based surrogates, where the latter are based on different combinations of training dataset size and \gls{rsm}.}
    \label{fig:statistics-uniform}
\end{figure}

\begin{figure}[t!]
    \centering
    \begin{subfigure}[b]{0.5\textwidth}
         \centering
         \fbox{\includegraphics[width=1\textwidth]{figures/UQ/plot_legenduq.pdf}}
     \end{subfigure}
     \\
     \begin{subfigure}[b]{0.32\textwidth}
         \centering
         \includegraphics[width=\textwidth]{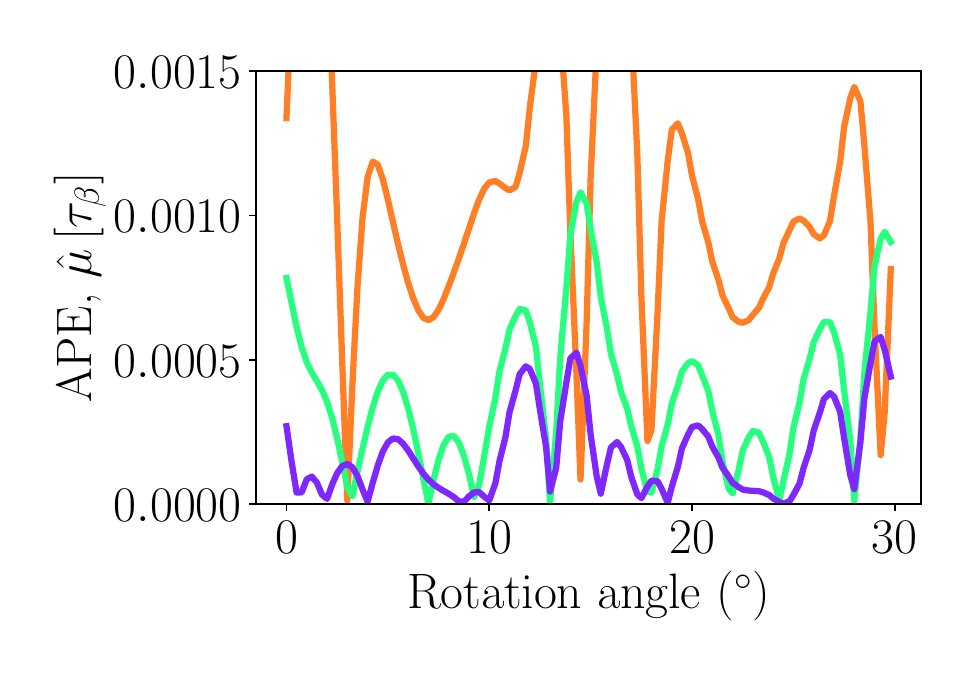}
         \caption{Mean, $M_\text{t}=600$.}
     \end{subfigure}
     \hfill
     \begin{subfigure}[b]{0.32\textwidth}
         \centering
         \includegraphics[width=\textwidth]{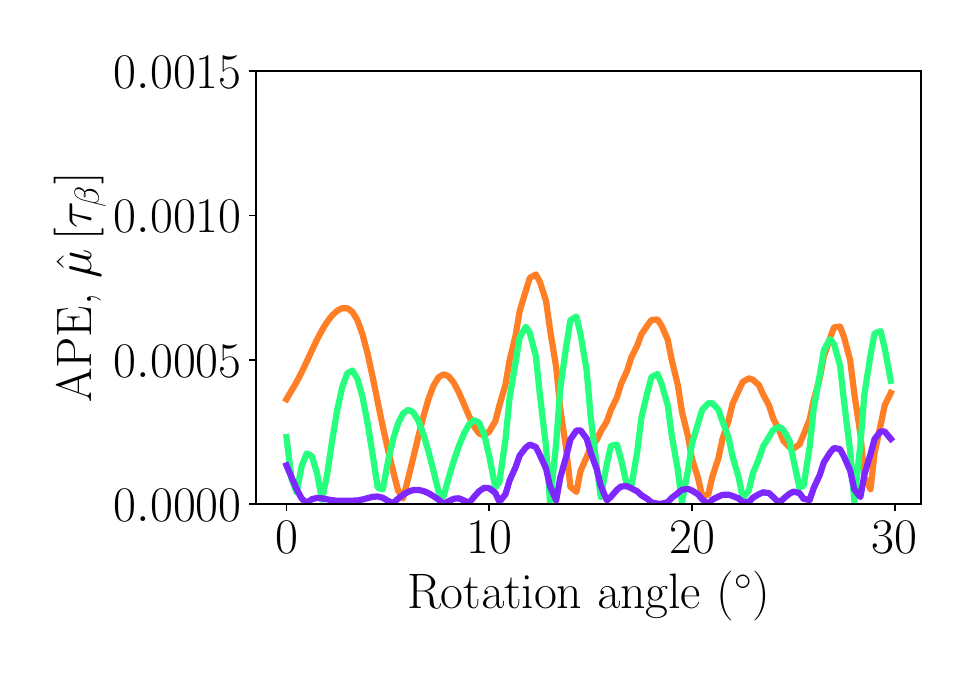}
         \caption{Mean, $M_\text{t}=1200$.}
     \end{subfigure}
     \hfill
     \begin{subfigure}[b]{0.32\textwidth}
         \centering
         \includegraphics[width=\textwidth]{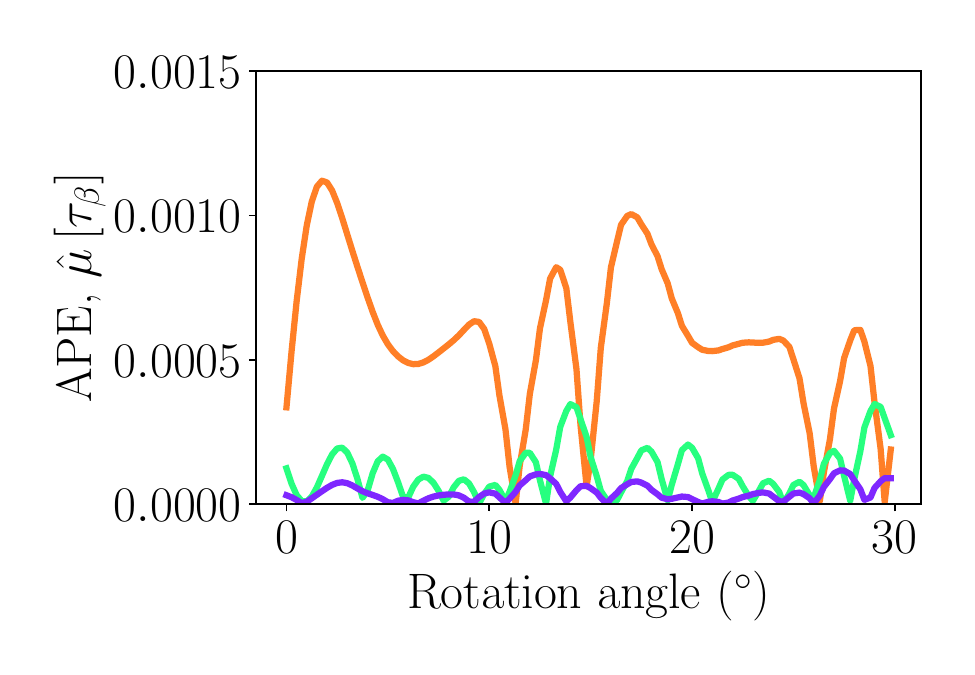}
         \caption{Mean, $M_\text{t}=1800$.}
     \end{subfigure} 
     \\
     \begin{subfigure}[b]{0.32\textwidth}
         \centering
         \includegraphics[width=\textwidth]{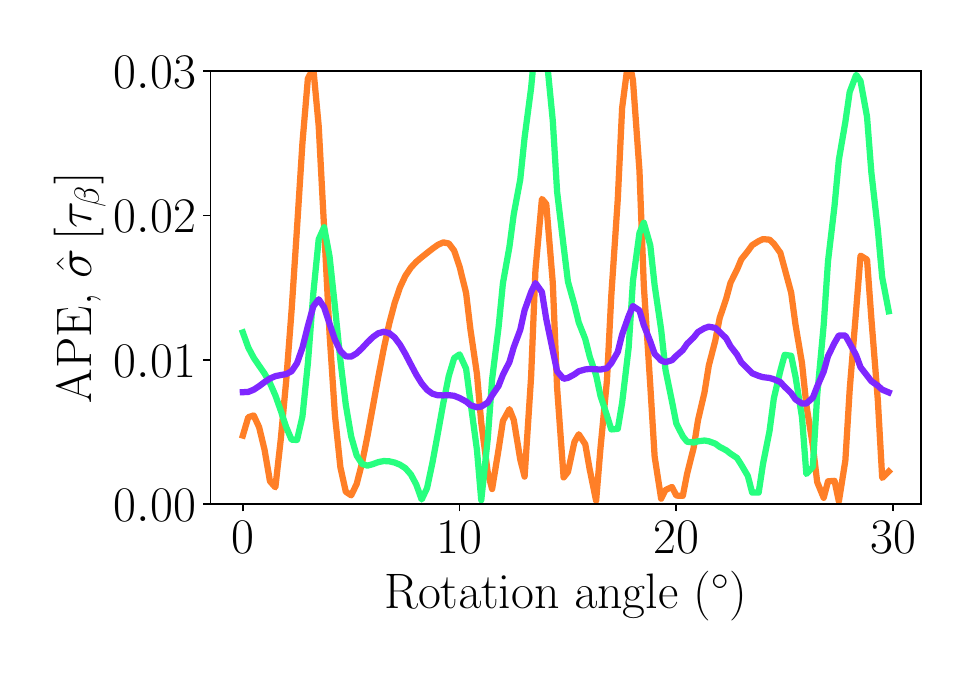}
         \caption{Standard deviation, $M_\text{t}=600$.}
     \end{subfigure}
     \hfill
     \begin{subfigure}[b]{0.32\textwidth}
         \centering
         \includegraphics[width=\textwidth]{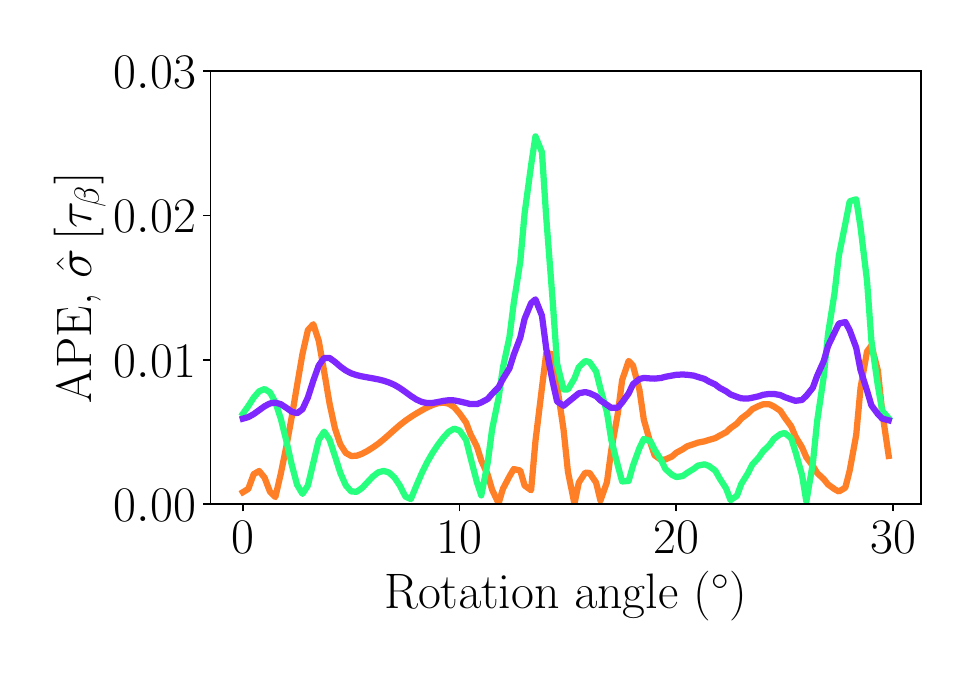}
         \caption{Standard deviation, $M_\text{t}=1200$.}
     \end{subfigure}
     \hfill
     \begin{subfigure}[b]{0.32\textwidth}
         \centering
         \includegraphics[width=\textwidth]{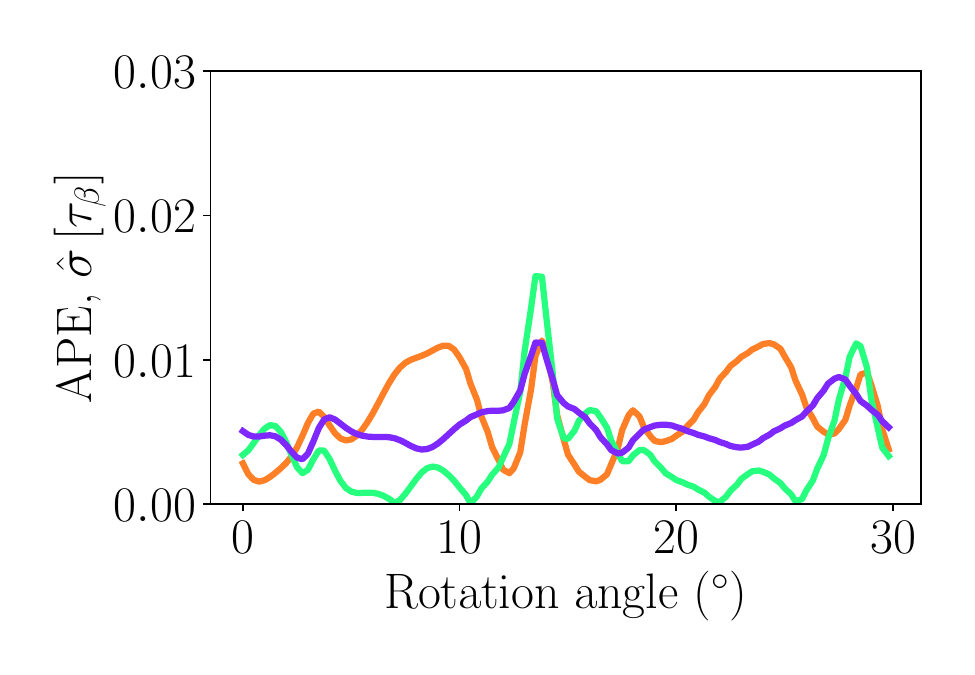}
         \caption{Standard deviation, $M_\text{t}=1800$.}
     \end{subfigure}   
     \caption{\Gls{ape} of torque mean and standard deviation estimates for uniform parameter distributions, obtained with framework-based surrogates computed with different combinations of training dataset size and \gls{rsm}.}
    \label{fig:statistics-uniform-APE}
\end{figure}

In the first \gls{uq} study, uniform distributions within the limits given in Table~\ref{tab:pmsm-parameters} are assumed for all design parameters.
Figure~\ref{fig:statistics-uniform} presents the sample mean and standard deviation of the torque signal for both the original and the surrogate models. 
Figure \ref{fig:statistics-uniform-APE} shows the \gls{ape} for the mean and standard deviation estimates relative to the reference statistics, i.e., the ones obtained by sampling the high-fidelity model. 
Additionally, signal-averaged \glspl{ape} per training dataset size and \gls{rsm} are reported in Table~\ref{tab:mape-uq-uniform}.

\begin{table*}[b!]
\caption{Signal-averaged \acrshort{ape} of surrogate-based mean and standard deviation estimates for uniform parameter distributions. The surrogates are computed using the suggested framework, for different training dataset sizes and \glspl{rsm}. The best-in-class error values for the same training dataset size are marked as bold.}
\label{tab:mape-uq-uniform}
\tabcolsep=0pt
\begin{threeparttable}
\begin{tabular*}{\textwidth}{@{\extracolsep{\fill}}ccccccc@{\extracolsep{\fill}}}
\toprule
& \multicolumn{3}{c}{Mean} & \multicolumn{3}{c}{Standard Deviation} \\
\cmidrule(lr){2-4}\cmidrule(lr){5-7}
$M_{\text{t}}$ & PCE & FNN & GP& PCE & FNN & GP\\
\midrule
600 & 0.0004 & 0.0011 & \textbf{0.0002}  & 0.0106 & 0.0105 & \textbf{0.0100}\\
1200 & 0.0003 & 0.0004 & \textbf{0.0001} & 0.0060 & \textbf{0.0048} & 0.0083  \\
1800 & 0.0001 & 0.0006 & $\mathbf{4 \cdot 10^{-5}}$ & \textbf{0.0035} & 0.0063 & 0.0056 \\
\bottomrule
\end{tabular*}
\end{threeparttable}
\end{table*}

With respect to mean estimates, it is observed that \gls{gp} is consistently the best \gls{rsm} option, followed by \glspl{pce} and \glspl{fnn}, irrespective of training dataset size. 
This aligns with our findings in section~\ref{sec:results-rom} regarding surrogate modeling accuracy.
In the case of standard deviation, however, no \gls{rsm} choice consistently outperforms the others. 
One important advantage of \gls{gp} is the comparatively smaller error fluctuation and the typically smaller maximum error values, with the only exception being the case for $M_{\text{t}}=1200$ shown in Figure~\ref{fig:statistics-uniform-APE}(e).

\subsubsection{Gaussian parameter distributions}
\label{sec:uq-gaussian}
\begin{figure}[t!]
    \centering
    \begin{subfigure}[b]{0.7\textwidth}
         \centering
         \fbox{\includegraphics[width=1\textwidth]{figures/plot_legendsur1.pdf}}
     \end{subfigure}
     \\
     \begin{subfigure}[b]{0.32\textwidth}
         \centering
         \includegraphics[width=\textwidth]{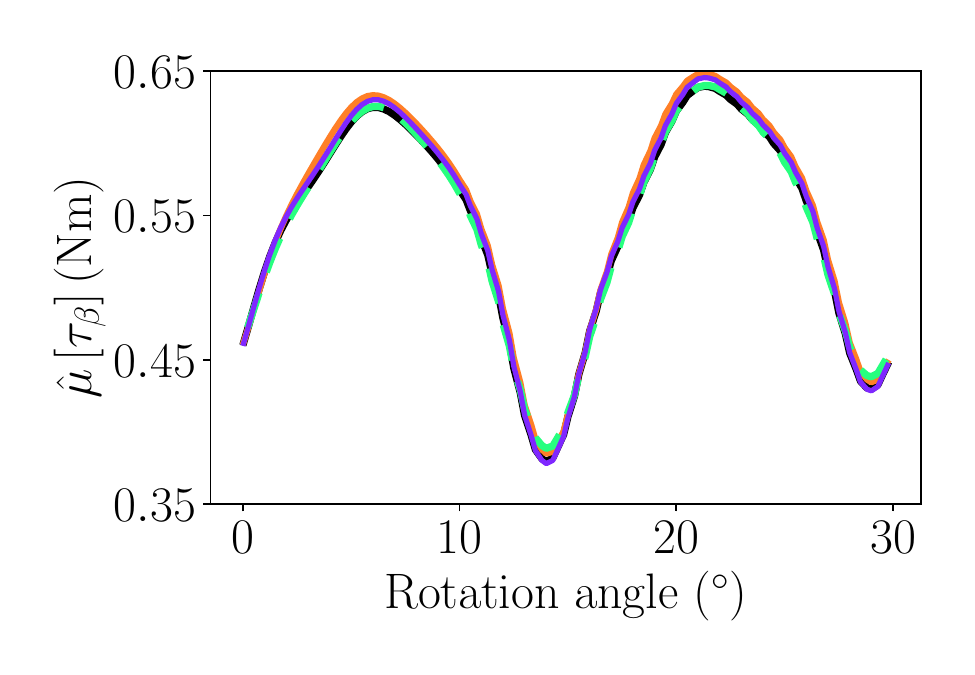}
         \caption{Mean, $M_\text{t}=600$.}
     \end{subfigure}
     \hfill
     \begin{subfigure}[b]{0.32\textwidth}
         \centering
         \includegraphics[width=\textwidth]{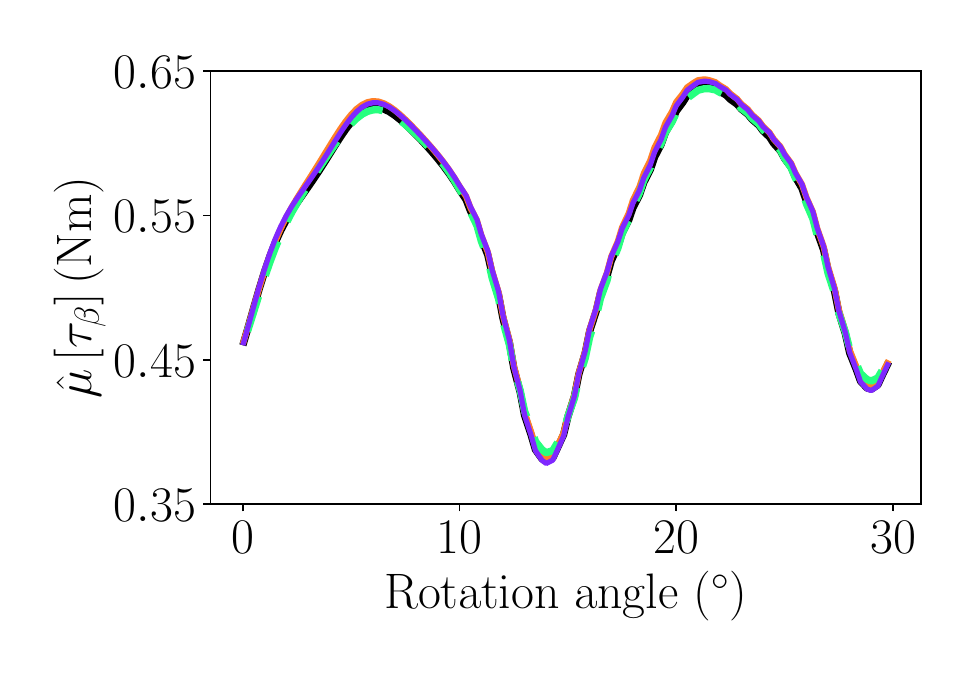}
         \caption{Mean, $M_\text{t}=1200$.}
     \end{subfigure}
     \hfill
     \begin{subfigure}[b]{0.32\textwidth}
         \centering
         \includegraphics[width=\textwidth]{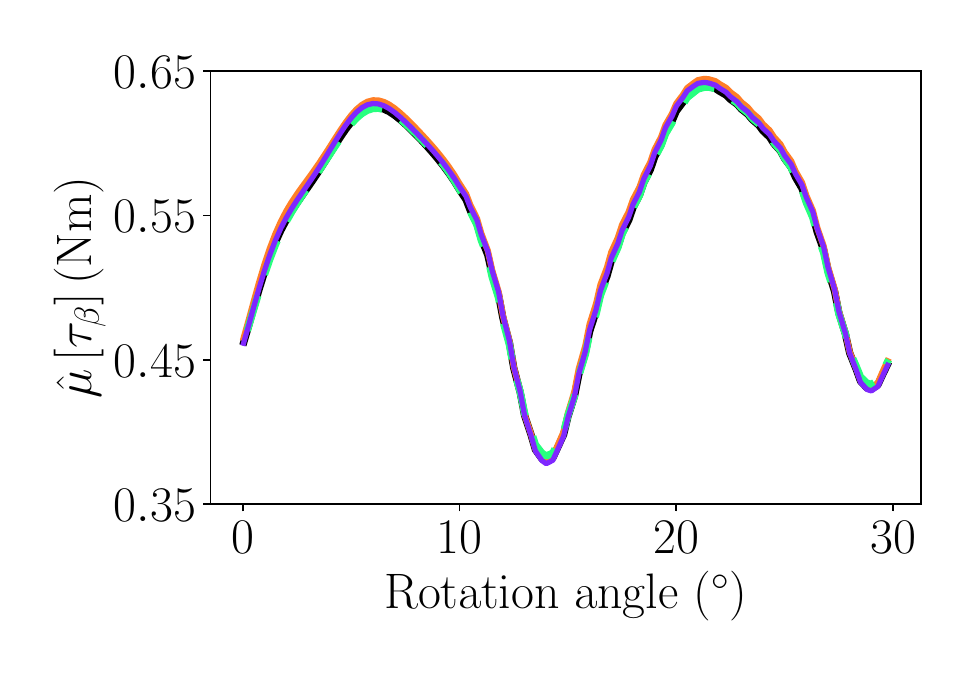}
         \caption{Mean, $M_\text{t}=1800$.}
     \end{subfigure} 
     \\
     \begin{subfigure}[b]{0.32\textwidth}
         \centering
         \includegraphics[width=\textwidth]{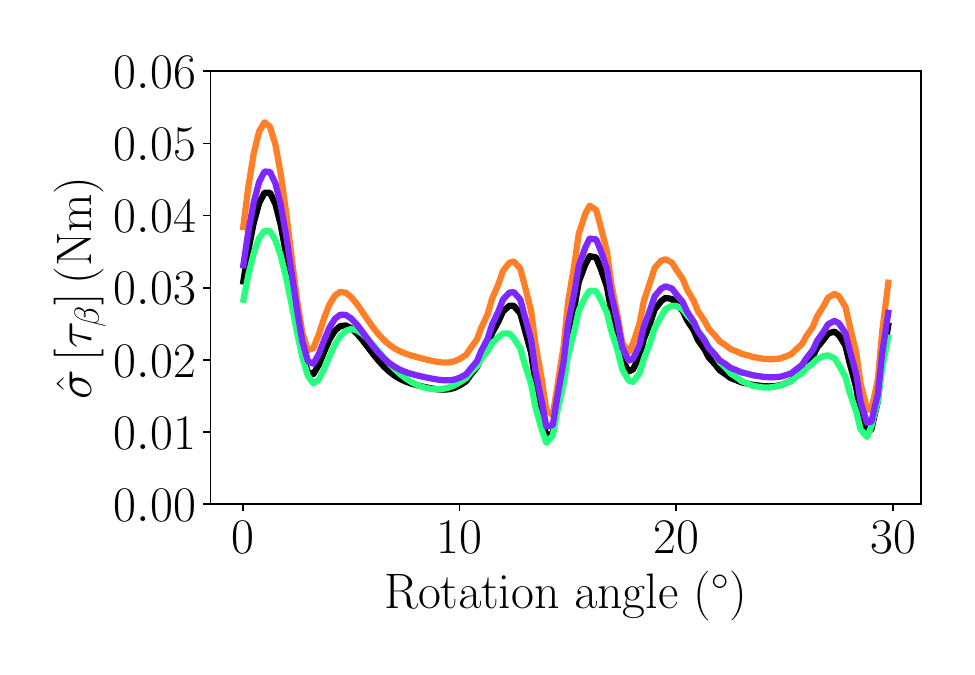}
         \caption{Standard deviation, $M_\text{t}=600$.}
     \end{subfigure}
     \hfill
     \begin{subfigure}[b]{0.32\textwidth}
         \centering
         \includegraphics[width=\textwidth]{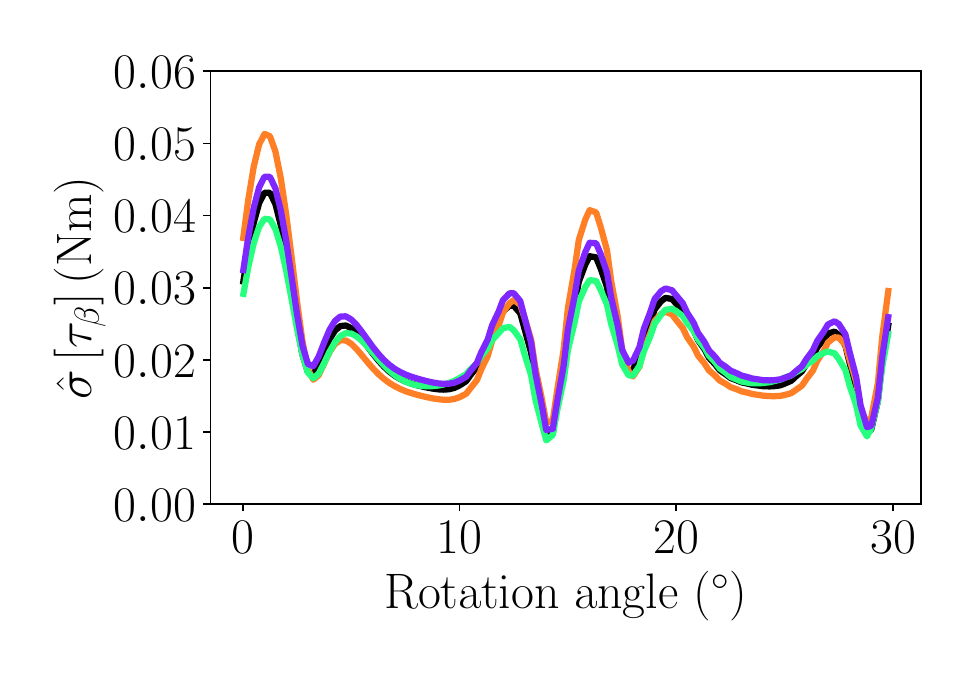}
         \caption{Standard deviation, $M_\text{t}=1200$.}
     \end{subfigure}
     \hfill
     \begin{subfigure}[b]{0.32\textwidth}
         \centering
         \includegraphics[width=\textwidth]{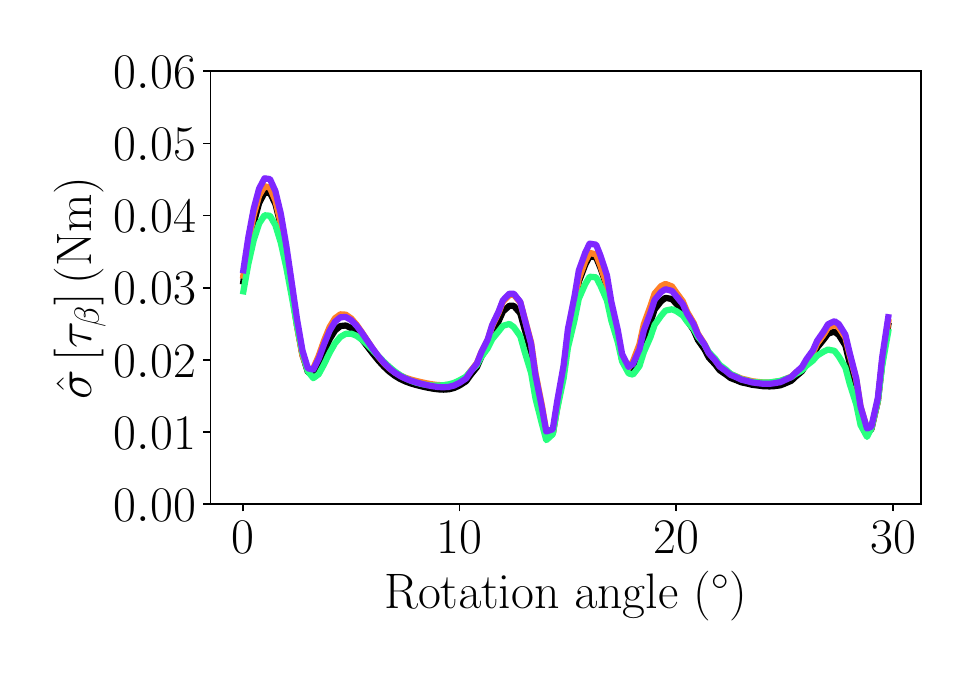}
         \caption{Standard deviation, $M_\text{t}=1800$.}
     \end{subfigure}   
     \caption{Torque mean and standard deviation estimates for Gaussian parameter distributions, obtained by sampling the original, high-fidelity model and the framework-based surrogates, where the latter are based on different combinations of training dataset size and \gls{rsm}.}
    \label{fig:statistics-gaussian}
\end{figure}

\begin{figure}[t!]
    \centering
    \begin{subfigure}[b]{0.5\textwidth}
         \centering
         \fbox{\includegraphics[width=1\textwidth]{figures/UQ/plot_legenduq.pdf}}
     \end{subfigure}
     \\
     \begin{subfigure}[b]{0.32\textwidth}
         \centering
         \includegraphics[width=\textwidth]{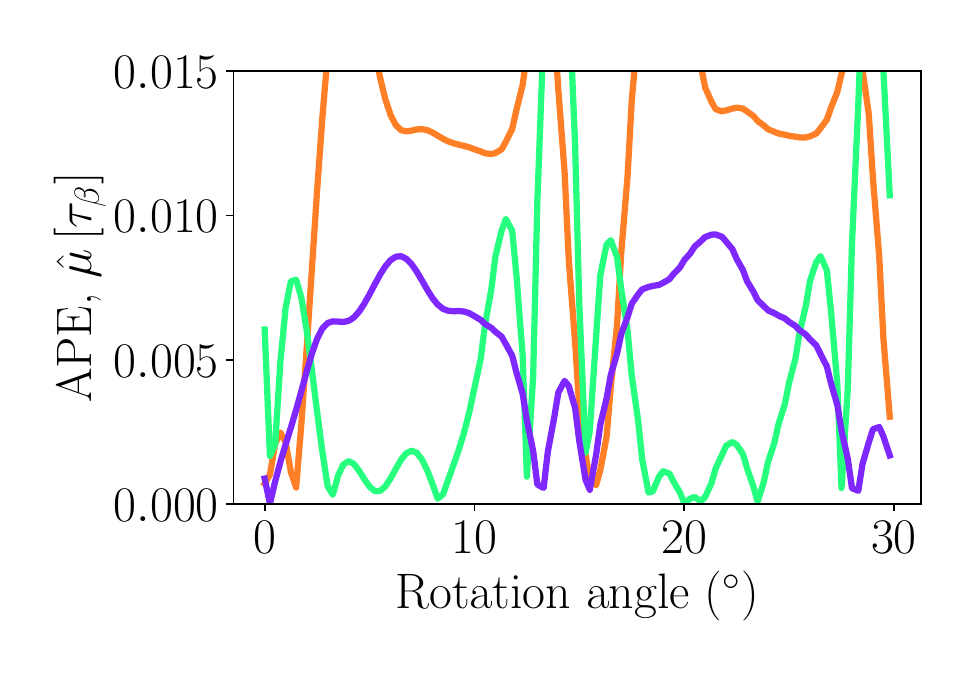}
         \caption{Mean, $M_\text{t}=600$.}
     \end{subfigure}
     \hfill
     \begin{subfigure}[b]{0.32\textwidth}
         \centering
         \includegraphics[width=\textwidth]{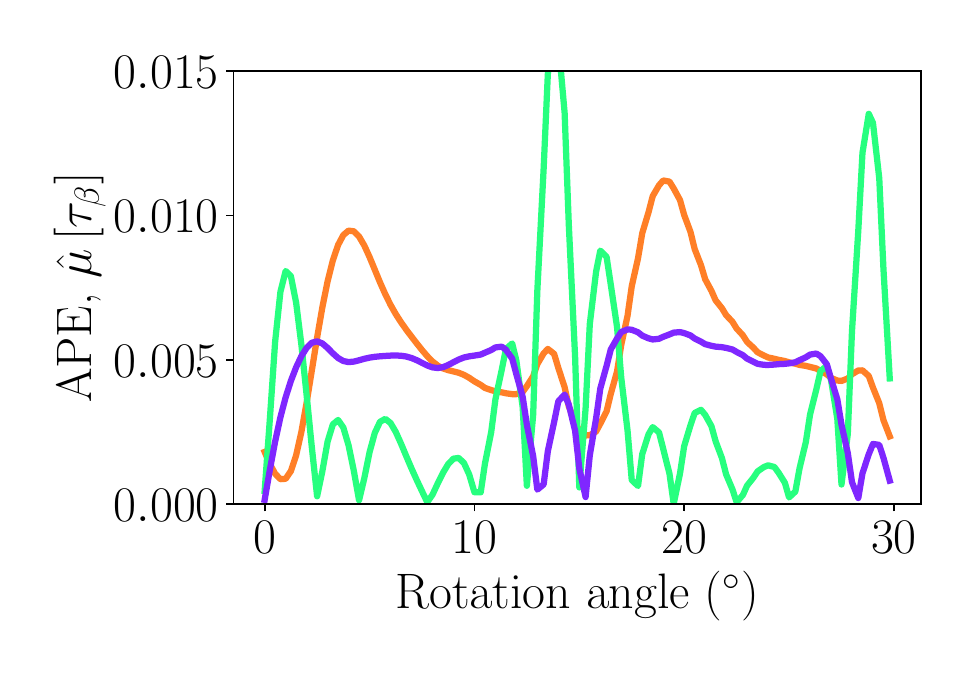}
         \caption{Mean, $M_\text{t}=1200$.}
     \end{subfigure}
     \hfill
     \begin{subfigure}[b]{0.32\textwidth}
         \centering
         \includegraphics[width=\textwidth]{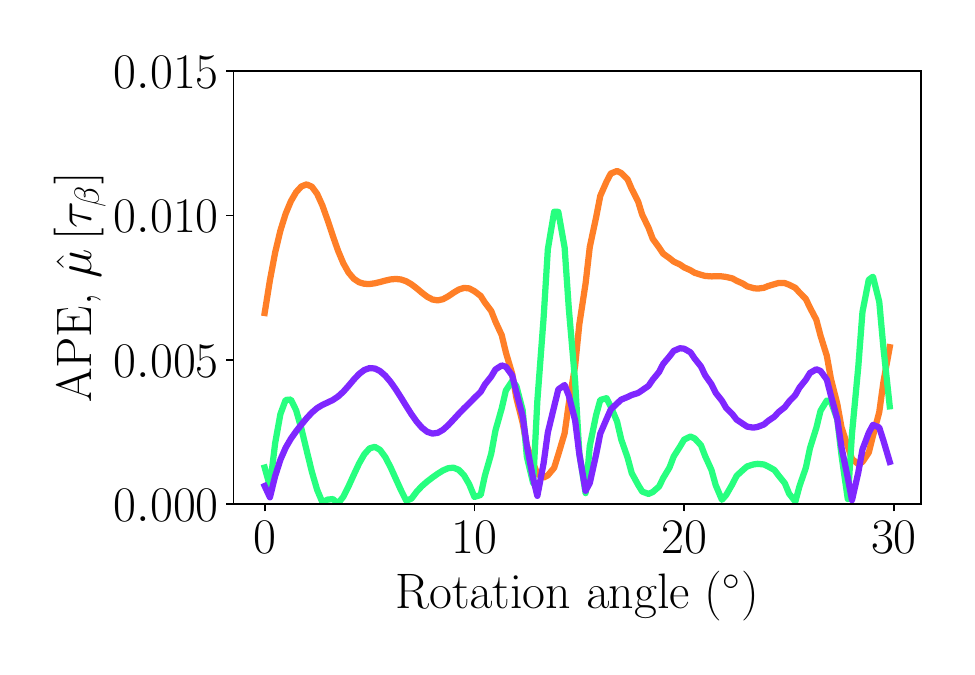}
         \caption{Mean, $M_\text{t}=1800$.}
     \end{subfigure} 
     \\
     \begin{subfigure}[b]{0.32\textwidth}
         \centering
         \includegraphics[width=\textwidth]{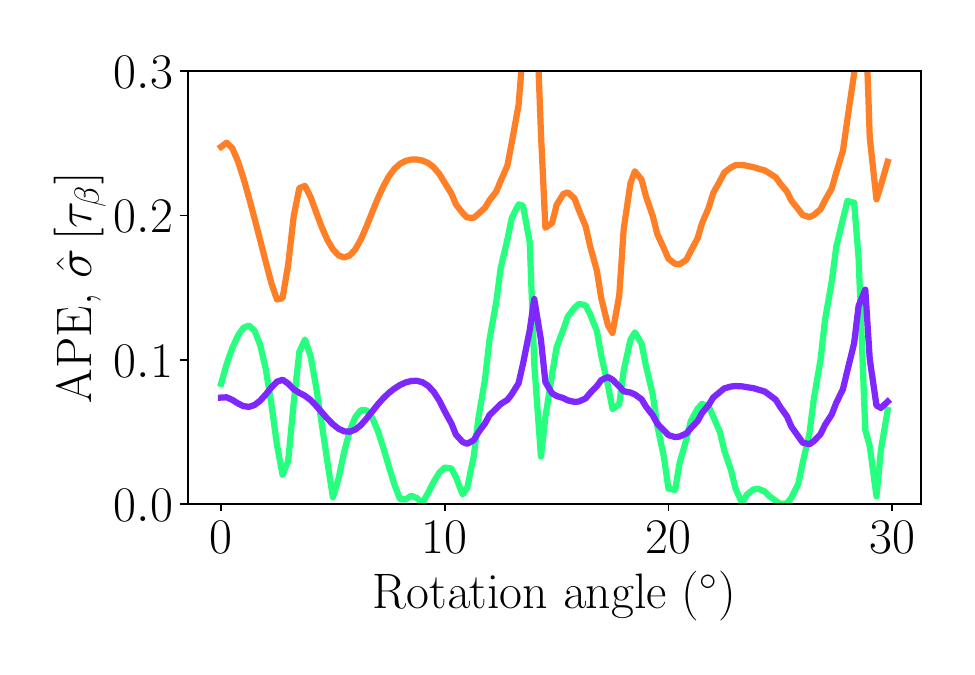}
         \caption{Standard deviation, $M_\text{t}=600$.}
     \end{subfigure}
     \hfill
     \begin{subfigure}[b]{0.32\textwidth}
         \centering
         \includegraphics[width=\textwidth]{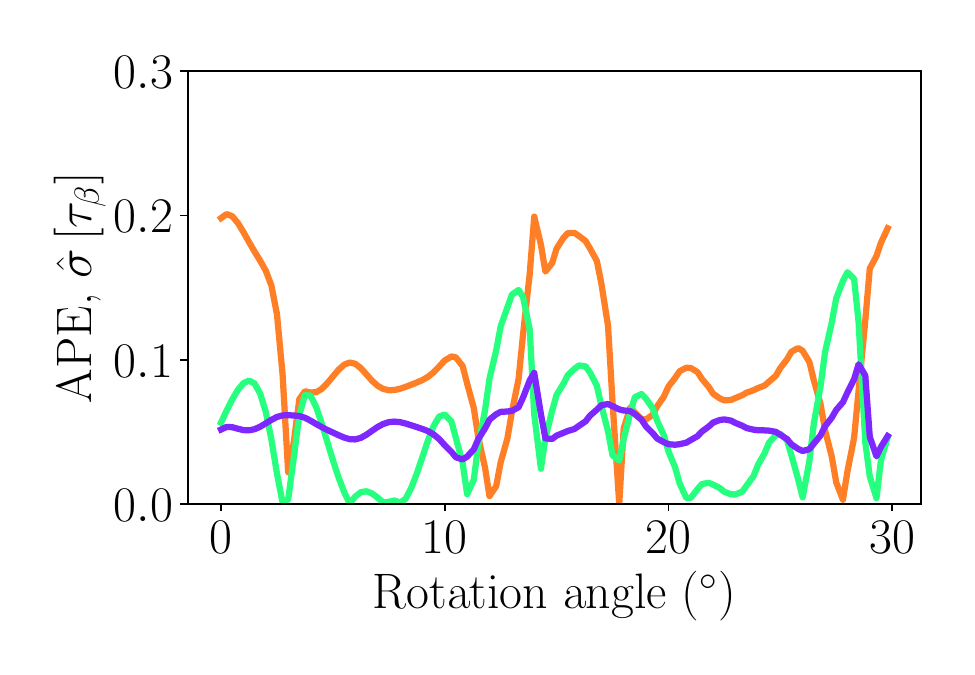}
         \caption{Standard deviation, $M_\text{t}=1200$.}
     \end{subfigure}
     \hfill
     \begin{subfigure}[b]{0.32\textwidth}
         \centering
         \includegraphics[width=\textwidth]{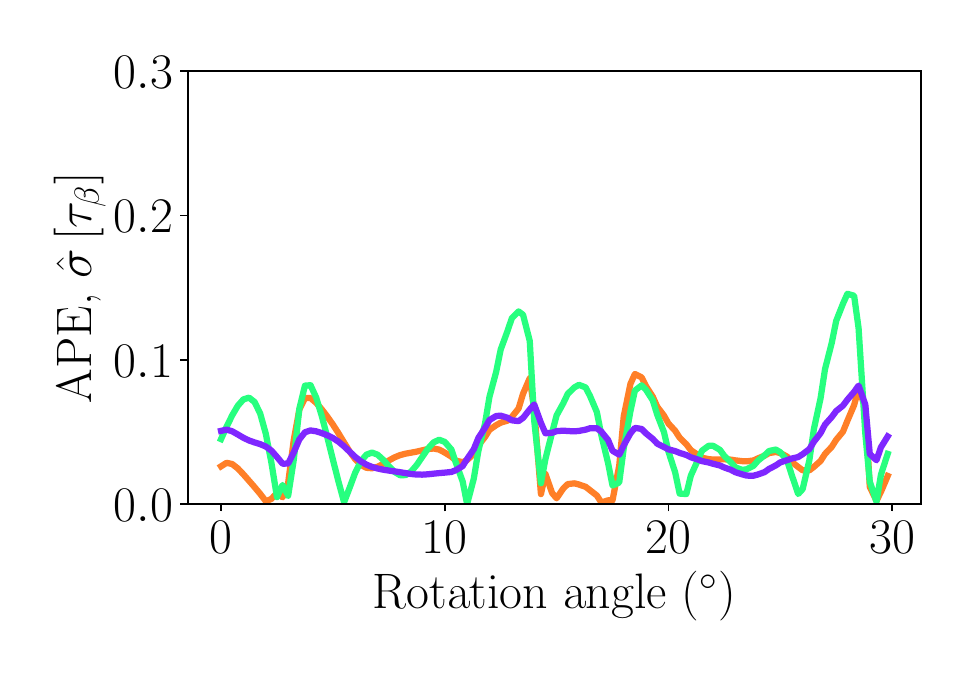}
         \caption{Standard deviation, $M_\text{t}=1800$.}
     \end{subfigure}   
     \caption{\Gls{ape} of torque mean and standard deviation estimates for Gaussian parameter distributions, obtained with framework-based surrogates computed with different combinations of training dataset size and \gls{rsm}.}
    \label{fig:statistics-gaussian-APE}
\end{figure}

In the second \gls{uq} study, we assume a fixed nominal value for each parameter, which is equal to the middle value of its range as given in Table~\ref{tab:pmsm-parameters}. 
Additionally, we consider an additive Gaussian noise term per parameter, with zero mean and standard deviation equal to $1\%$ of the nominal parameter value.
Similar to the case of uniform parameter distributions discussed in section~\ref{sec:uq-uniform}, we report sample means and standard deviations  for both the original and the surrogate models in Figure~\ref{fig:statistics-gaussian}, while the corresponding \glspl{ape} of the surrogate-based statistics estimates are shown in  
Figure \ref{fig:statistics-gaussian-APE}.
Signal-averaged \glspl{ape} per training dataset size and \gls{rsm} are reported in Table~\ref{tab:mape-uq-gaussian}.

\begin{table*}[b!]
\caption{Signal-averaged \acrshort{ape} of surrogate-based mean and standard deviation estimates for Gaussian parameter distributions. The surrogates are computed using the suggested framework, for different training dataset sizes and \glspl{rsm}. The best-in-class error values for the same training dataset size are marked as bold.}
\label{tab:mape-uq-gaussian}
\tabcolsep=0pt
\begin{threeparttable}
\begin{tabular*}{\textwidth}{@{\extracolsep{\fill}}ccccccc@{\extracolsep{\fill}}}
\toprule
& \multicolumn{3}{c}{Mean} & \multicolumn{3}{c}{Standard Deviation} \\
\cmidrule(lr){2-4}\cmidrule(lr){5-7}
$M_{\text{t}}$ & PCE & FNN & GP& PCE & FNN & GP\\
\midrule
600 & \textbf{0.0052} & 0.0125 & 0.0055  & \textbf{0.071} & 0.2140 & 0.0721\\
1200 & \textbf{0.0037} & 0.0055 & 0.0044 & \textbf{0.0516} & 0.0998 & 0.0541  \\
1800 & \textbf{0.0022} & 0.0071 & 0.0033 & 0.0496 & \textbf{0.0365} & 0.0405 \\
\bottomrule
\end{tabular*}
\end{threeparttable}
\end{table*}

Compared to the results reported in previous section for the case of uniform parameter distributions, important differences can be observed. 
First, the mean and standard deviation estimates are now visually discernible from the corresponding references, which was not the case before. 
Moreover, the errors presented in Figure~\ref{fig:statistics-gaussian-APE} and in Table~\ref{tab:mape-uq-gaussian} are significantly higher, in fact by approximately one order of magnitude.
This deterioration in statistics estimation accuracy can be mainly attributed to the difference between the training and sampling distributions, since the surrogate models are trained on uniform data and are then sampled for Gaussian parameter distributions. 
This \emph{distributional shift}, as is the more common term in the context of \gls{ml}, is known to deteriorate the performance of data-driven models in terms of their predictive ability \cite{mougan2023explanation}. 
Naturally, the reduced surrogate modeling accuracy affects the quality of the statistics estimates as well.

Additionally, looking at Table~\ref{tab:mape-uq-gaussian}, the \gls{pce} seems to be the best \gls{rsm} option concerning statistics estimation, with the only exception being the standard deviation estimate for $M_{\text{t}} = 1800$ training data points, in which case the \gls{fnn} prevails. 
However, the signal-averaged errors can be quite misleading, as revealed by Figure~\ref{fig:statistics-gaussian-APE}. 
In fact, the better signal-averaged errors of the \gls{pce} can be explained by very low errors in some parts for the signal, however, they are accompanied by large error fluctuations and high maximum errors. 
Once again, the \gls{gp} yields consistently the lowest maximum errors and significantly less error fluctuation, and is clearly the best option in almost all cases. 
The only possible exception is the standard deviation estimate for $M_{\text{t}} = 1800$ training data points, in which case the \gls{fnn} could also be selected.

\subsubsection{Computational costs and gains}
\label{sec:costs-and-gains}

Last, we underline the computational gains achieved by using surrogate models for Monte Carlo-based statistics estimates.
First, we note that a single high-fidelity \gls{pmsm} model evaluation takes approximately $300$ seconds on an up-to-date CPU. 
Accordingly, a Monte Carlo-based \gls{uq} study based on $10^4$ random samples costs approximately $833$ CPU-hours.
In contrast, a surrogate-based \gls{uq} study using the same number of random samples can be performed in just $10$ seconds, as the surrogate model is evaluated in $10^{-3}$ seconds at most.

However, in the latter case, one must also consider the so-called offline costs due to data acquisition and \gls{rsm} training.
Generating training datasets with sizes $M_{\text{t}} \in \left\{600, 1200, 1800\right\}$ amounts to (approximately) $50$, $100$, and $150$ CPU-hours, respectively.
The cost due to \gls{ml} regression depends on the \gls{rsm} and the training dataset size, as shown in Table~\ref{tab:training-times}.
Even in the worst case, \gls{ml} regression accounts for less than $1\%$ of the total computational cost.
Then, depending on the training dataset size, the computational cost of the surrogate-based \gls{uq} study is $16.7\times$ -- $5.5\times$ smaller compared to using the high-fidelity model, with statistics errors that can be considered more than acceptable. 

Note that the computational gains could possibly be even greater, for example, considering scenarios with smaller training datasets or larger Monte Carlo samples. 
Moreover, it must be noted that the surrogate model is trained only once and can then be sampled for different distributions, as long as these distributions respect the given parameter ranges. 
That is, if new parameter distributions are to be considered, the data acquisition cost remains the same and only the marginal cost of sampling the surrogate model is added.
This comes at the cost of possible deterioration in surrogate modeling accuracy, as shown in section~\ref{sec:uq-gaussian}.
On the contrary, in case of utilizing the original model within the \gls{uq} study, each change in the parameters' distributions would necessitate new costly model evaluations.

\begin{table*}[b!]
\caption{Training times (in seconds) per \gls{rsm} and training dataset size.}
\label{tab:training-times}
\tabcolsep=0pt
\begin{threeparttable}
\begin{tabular*}{\textwidth}{@{\extracolsep{\fill}}cccc@{\extracolsep{\fill}}}
\toprule
$M_\text{t}$ & PCE & FNN & GP \\
\midrule
600 & 675 & 27 & 450 \\
1200 & 2030 & 41 & 1780 \\
1800 & 5100 & 51 & 4800 \\
\bottomrule
\end{tabular*}
\end{threeparttable}
\end{table*}

\section{Conclusions}
\label{sec:conclusions}
This work presented a novel, data-driven surrogate modeling framework, aiming at predicting the torque of a \gls{pmsm} under geometric design variations cost-effectively, and utilizing these predictions for cost-efficient \gls{uq} studies.
The framework consists of a reduced-order modeling and an inference part. 
The reduced-order modeling part combines \gls{dft}-based dimension reduction of torque signals with regression-based \glspl{rsm} that approximate the functional dependency between the \gls{pmsm}'s design parameters and the reduced \gls{qoi}. 
Three \glspl{rsm} are tested, namely, \gls{pce}, \gls{fnn}, and \gls{gp}.
In the inference part, the trained \gls{rsm} is first evaluated for a set of new design parameters and subsequently an inverse \gls{dft} is applied to yield the predicted torque signal.

Our framework is applied for torque inference and \gls{uq} for a \gls{pmsm} with $20$ geometric design parameters. 
The torque signals contain $120$ torque values evaluated upon equidistant rotation angles within a period. 
The numerical studies reveal a number of interesting observations.
First, at least $R=11$ frequency components must be kept in the \gls{rom} to ensure a sufficiently accurate torque signal reconstruction. 
Next, the  best-in-class surrogate model in terms of prediction accuracy and robustness is obtained by combining \gls{dft}-based dimension reduction and \gls{gp}-based \glspl{rsm}.
This result includes comparisons against non-framework-based surrogate models, which either replace \gls{dft} with \gls{pca} in the dimension reduction step or omit it altogether.  
Importantly, omitting the dimension reduction step leads to severe deterioration in prediction accuracy.
In surrogate-based \gls{uq} by means of Monte Carlo sampling, the \gls{gp} remains the best \gls{rsm} choice concerning mean torque estimation accuracy.
For standard deviation estimation, no \gls{rsm} can claim best performance, however, \gls{gp} offers the advantages of smaller maximum errors and less error fluctuation.
Last, taking into account both the offline and online costs of the suggested framework, that is, due to data acquisition, \gls{ml} regression, and model evaluation, surrogate-based \gls{uq} studies by means of Monte Carlo are several times more cost-efficient compared to sampling the original, high-fidelity \gls{pmsm} model.

Overall, the suggested framework results in surrogate models that can reliably replace the high-fidelity \gls{pmsm} model and enable otherwise computationally intractable parameter studies such as \gls{uq}, even outperforming commonly used approaches like direct \gls{qoi} approximation or dimension reduction via \gls{pca}.
The \gls{dft}-based dimension reduction is particularly suitable in this use case, given the physical characteristics of the \gls{qoi}, mainly its periodicity.
Periodicity is indeed present for a wide array of electric machine \glspl{qoi} besides the torque.
The method remains applicable for non-periodic \glspl{qoi} as well, however, the extent of this applicability should be investigated further. 
Other extensions could consider a richer selection of \gls{rsm} and dimension reduction possibilities, including nonlinear methods.

\paragraph{Acknowledgements}
The authors are partially supported by the joint DFG/FWF Collaborative Research Centre CREATOR (DFG: Project-ID 492661287/TRR 361; FWF: 10.55776/F90) at TU Darmstadt, TU Graz, and JKU Linz.
The authors would like to thank Prof. Dr.-Ing. Herbert De Gersem for the fruitful exchanges during the revision of this work.

\paragraph{Author contributions}
A.P. implemented the software and performed all simulations and numerical studies. A.P. and D.L. analysed the numerical results. D.L. and S.Sc. conceived the original idea, acquired funding, and provided supervision. All authors contributed to writing and reviewing the manuscript. 

\printbibliography

\section*{Appendix}

\begin{figure}[h!]
    \centering
    \begin{subfigure}[b]{0.7\textwidth}
         \centering
         \fbox{\includegraphics[width=1\textwidth]{figures/plot_legendsur1.pdf}}
     \end{subfigure}
     \\
     \begin{subfigure}[b]{0.32\textwidth}
         \centering
         \includegraphics[width=\textwidth]{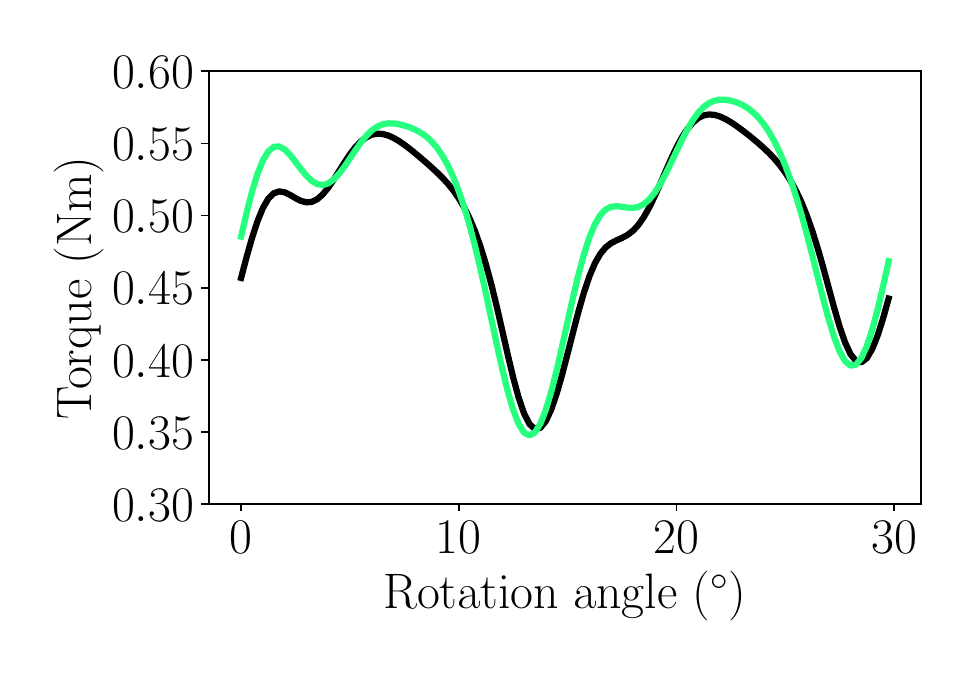}
         \caption{\gls{dft} and \gls{pce}.}
         \label{fig:fft_pce_600}
     \end{subfigure}
     \hfill
     \begin{subfigure}[b]{0.32\textwidth}
         \centering
         \includegraphics[width=\textwidth]{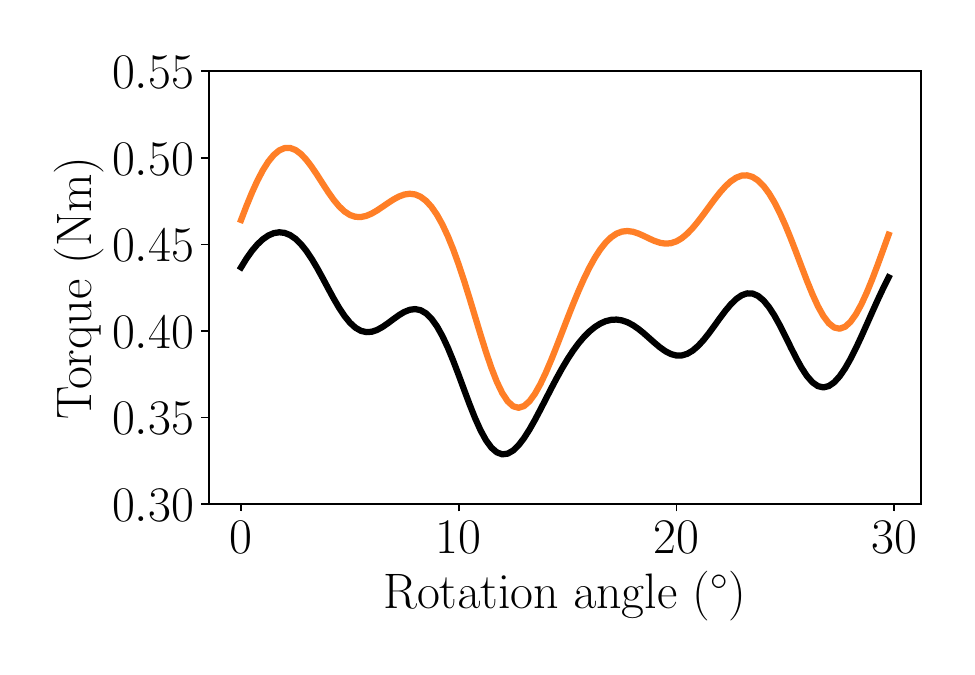}
         \caption{\gls{dft} and \gls{fnn}.}
         \label{fig:fft_nn_600}
     \end{subfigure}
     \hfill
     \begin{subfigure}[b]{0.32\textwidth}
         \centering
         \includegraphics[width=\textwidth]{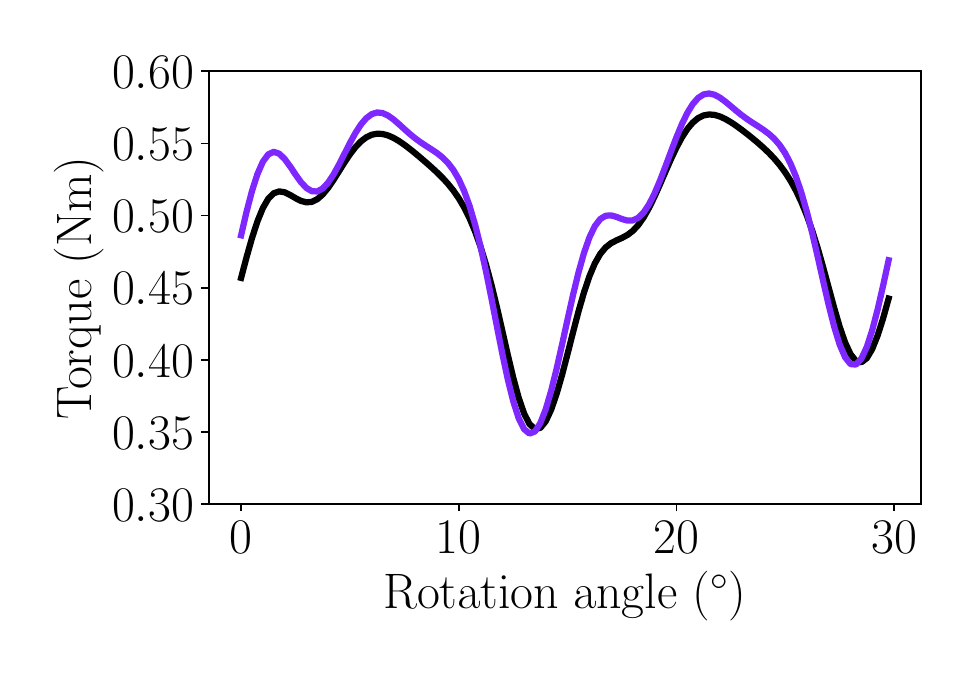}
         \caption{\gls{dft} and \gls{gp}.}
         \label{fig:fft_krig_600}
     \end{subfigure}
     \\
     \begin{subfigure}[b]{0.32\textwidth}
         \centering
         \includegraphics[width=\textwidth]{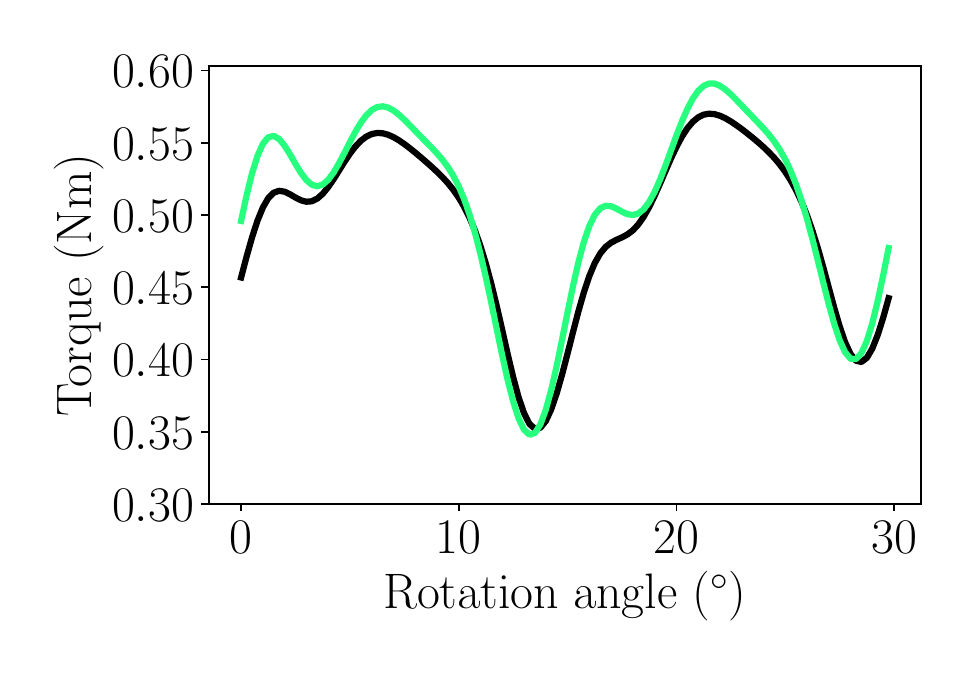}
         \caption{\gls{pca} and \gls{pce}.}
         \label{fig:pca_pce_600}
     \end{subfigure}
     \hfill
     \begin{subfigure}[b]{0.32\textwidth}
         \centering
         \includegraphics[width=\textwidth]{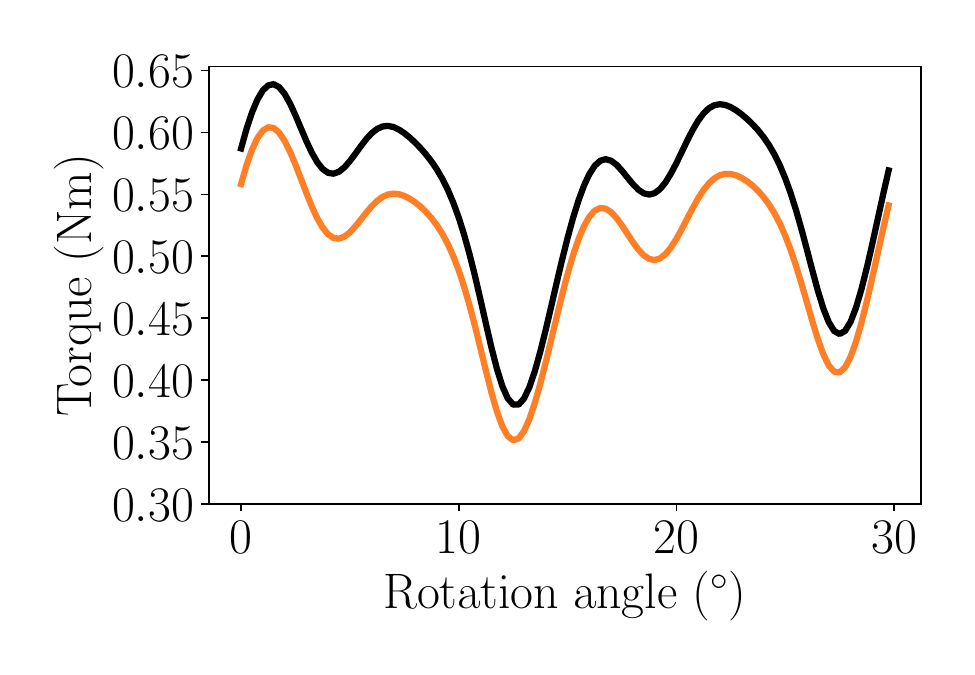}
         \caption{\gls{pca} and \gls{fnn}.}
         \label{fig:pca_nn_600}
     \end{subfigure}
     \hfill
     \begin{subfigure}[b]{0.32\textwidth}
         \centering
         \includegraphics[width=\textwidth]{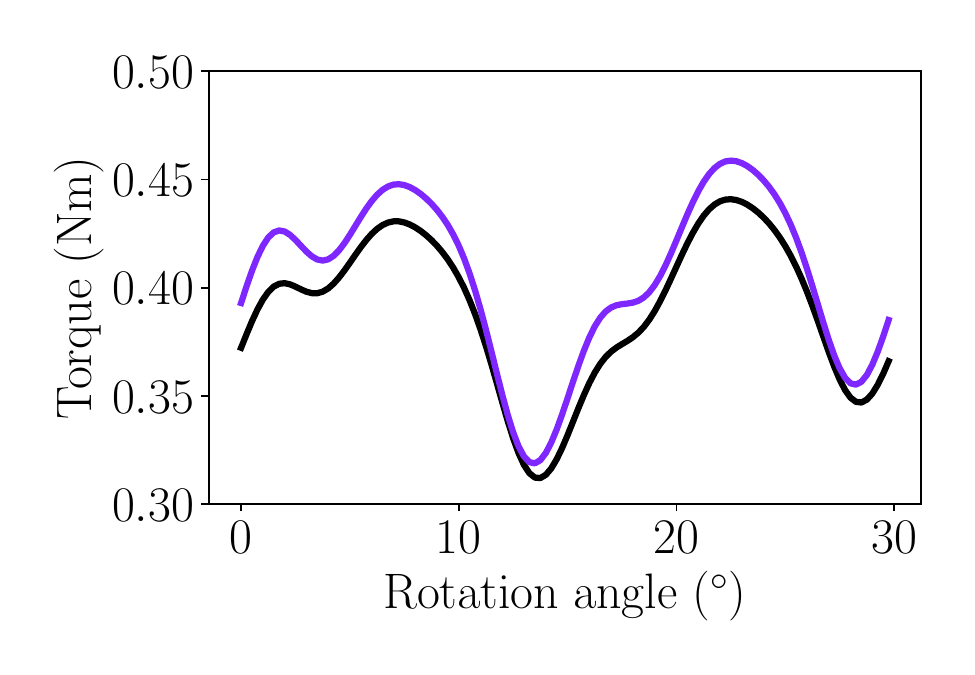}
         \caption{\gls{pca} and \gls{gp}.}
         \label{fig:pca_krig_600}
     \end{subfigure}
     \\
     \begin{subfigure}[b]{0.32\textwidth}
         \centering
         \includegraphics[width=\textwidth]{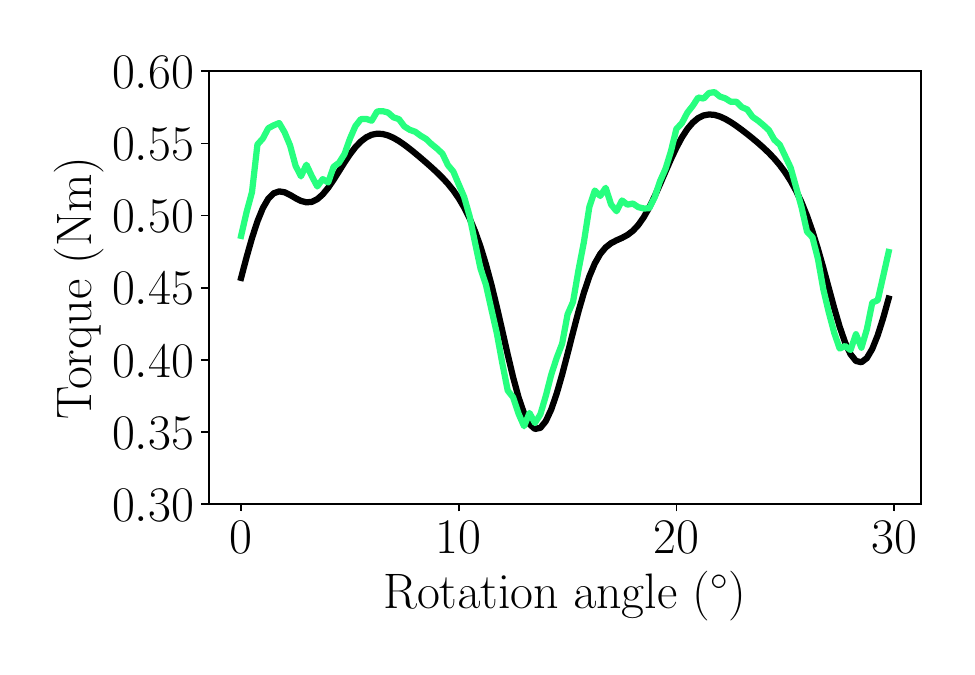}
         \caption{No reduction and \gls{pce}.}
         \label{fig:time_pce_600}
     \end{subfigure}
     \hfill
     \begin{subfigure}[b]{0.32\textwidth}
         \centering
         \includegraphics[width=\textwidth]{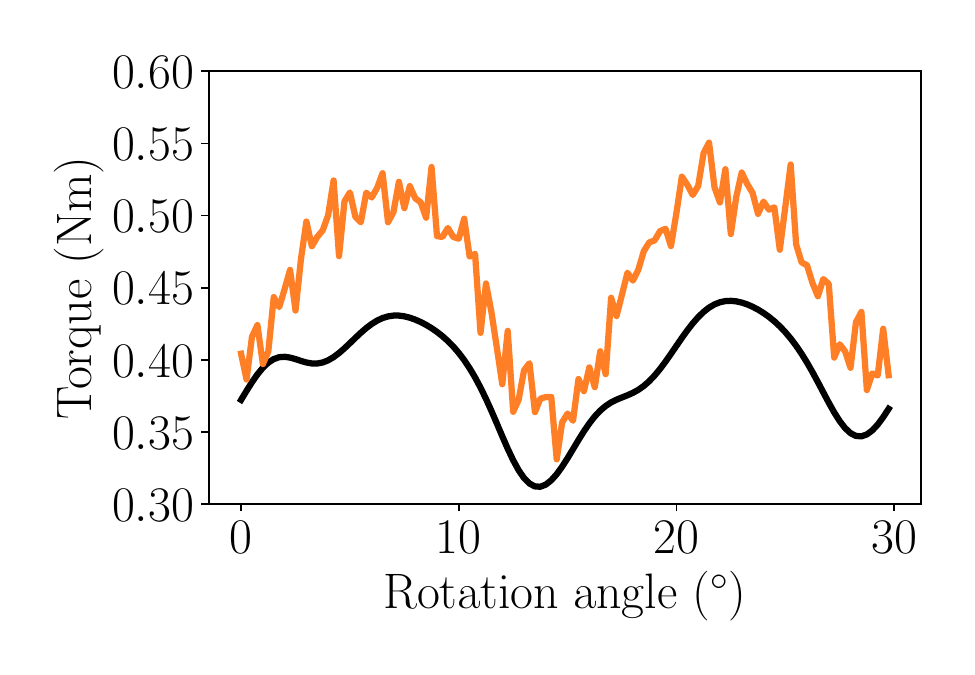}
         \caption{No reduction and \gls{fnn}.}
         \label{fig:time_nn_600}
     \end{subfigure}
     \hfill
     \begin{subfigure}[b]{0.32\textwidth}
         \centering
         \includegraphics[width=\textwidth]{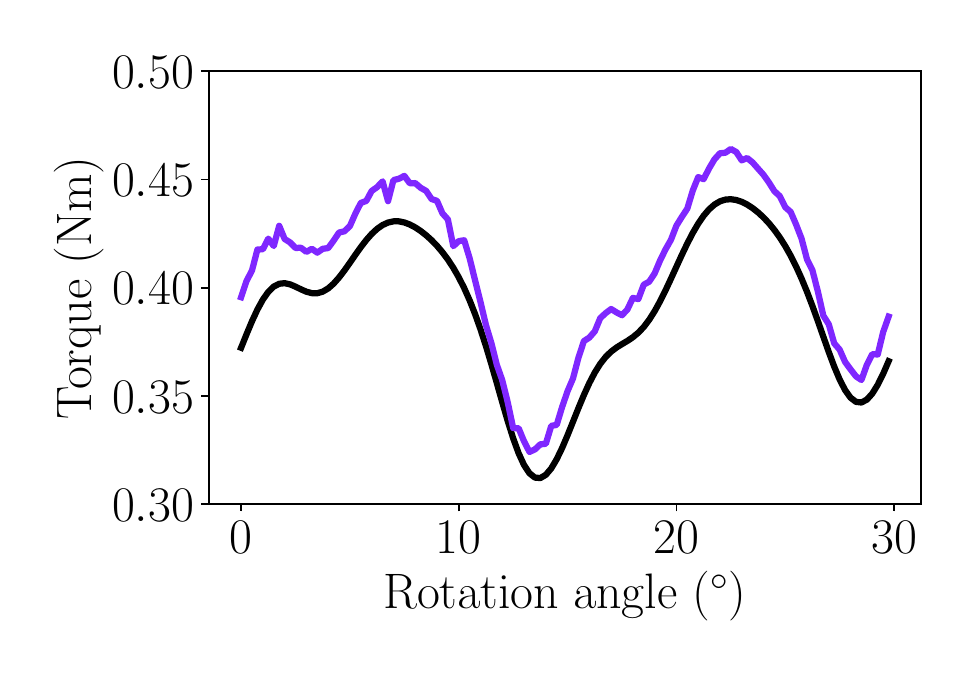}
         \caption{No reduction and \gls{gp}.}
         \label{fig:time_krig_600}
     \end{subfigure}
     \caption{Worst-case surrogate-based torque signal signal predictions for training dataset size $M_\text{t}=600$ and different combinations of \gls{rsm} and dimension reduction approach.}
    \label{fig:9-sur-600}
\end{figure}

\begin{figure}[h!]
    \centering
    \begin{subfigure}[b]{0.7\textwidth}
         \centering
         \fbox{\includegraphics[width=1\textwidth]{figures/plot_legendsur1.pdf}}
     \end{subfigure}
     \\
     \begin{subfigure}[b]{0.32\textwidth}
         \centering
         \includegraphics[width=\textwidth]{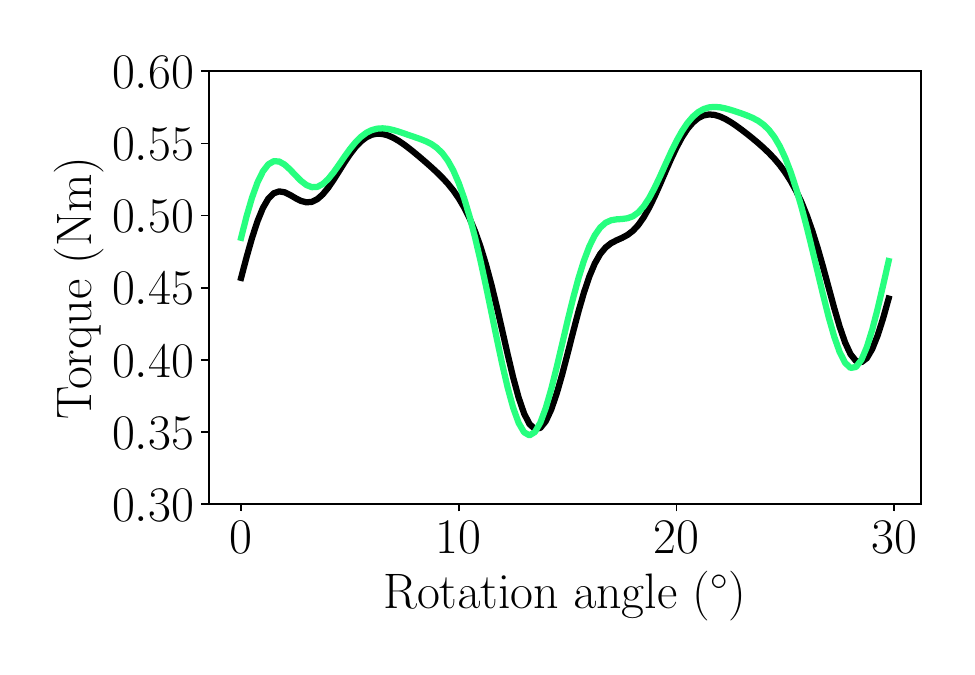}
         \caption{\gls{dft} and \gls{pce}.}
         \label{fig:fft_pce_1200}
     \end{subfigure}
     \hfill
     \begin{subfigure}[b]{0.32\textwidth}
         \centering
         \includegraphics[width=\textwidth]{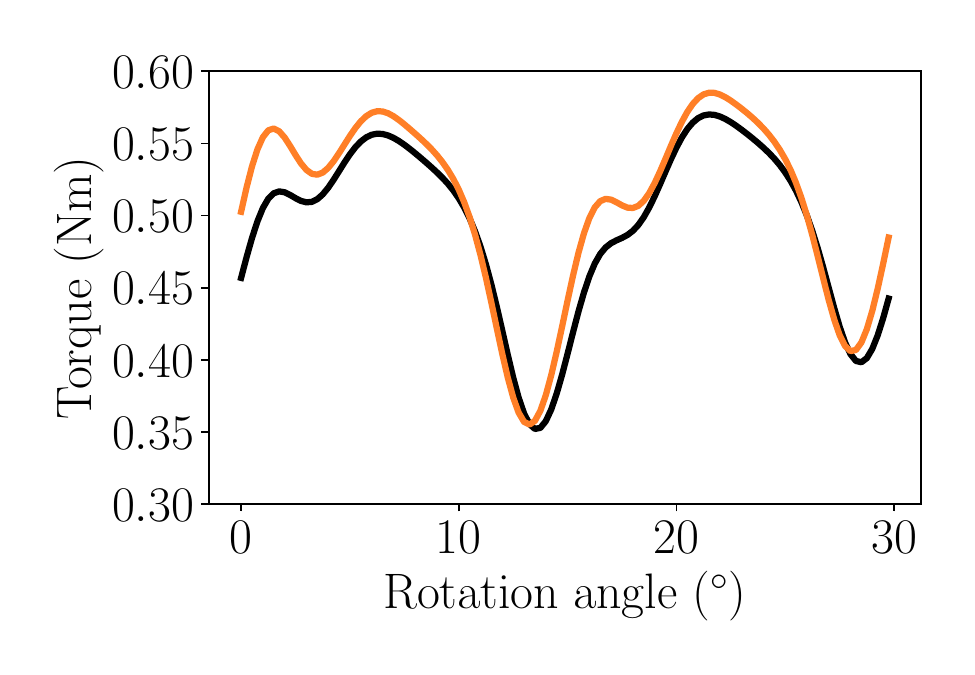}
         \caption{\gls{dft} and \gls{fnn}.}
         \label{fig:fft_nn_1200}
     \end{subfigure}
     \hfill
     \begin{subfigure}[b]{0.32\textwidth}
         \centering
         \includegraphics[width=\textwidth]{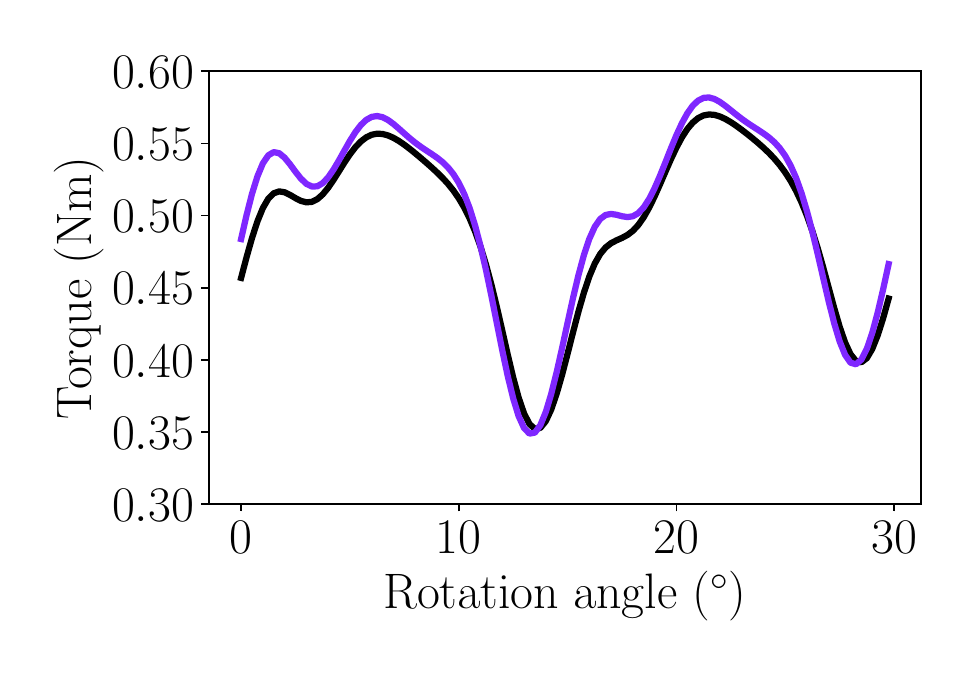}
         \caption{\gls{dft} and \gls{gp}.}
         \label{fig:fft_krig_1200}
     \end{subfigure}
     \\
     \begin{subfigure}[b]{0.32\textwidth}
         \centering
         \includegraphics[width=\textwidth]{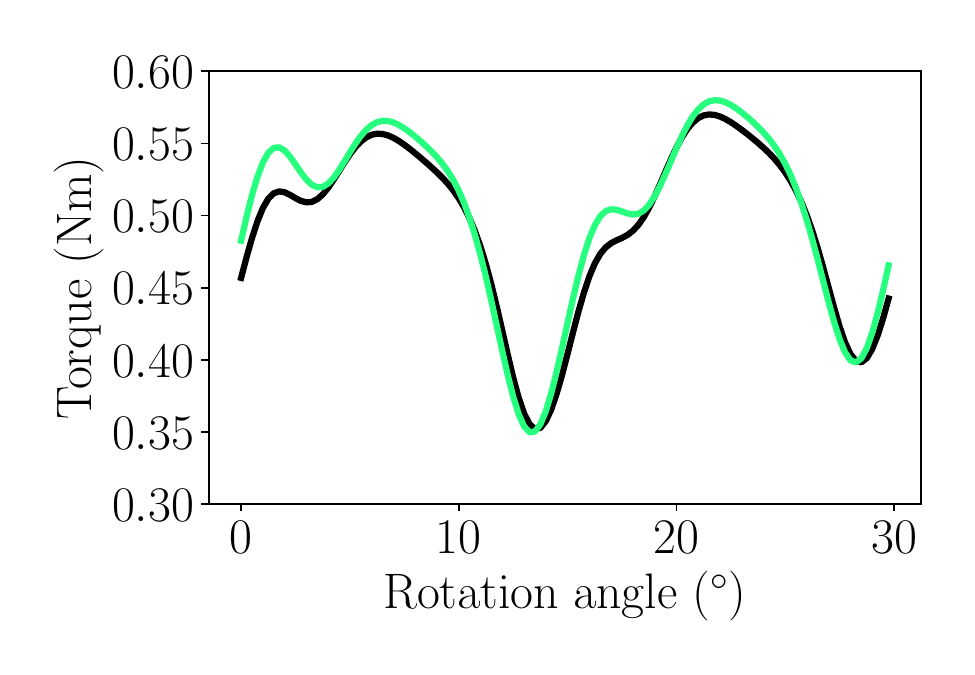}
         \caption{\gls{pca} and \gls{pce}.}
         \label{fig:pca_pce_1200}
     \end{subfigure}
     \hfill
     \begin{subfigure}[b]{0.32\textwidth}
         \centering
         \includegraphics[width=\textwidth]{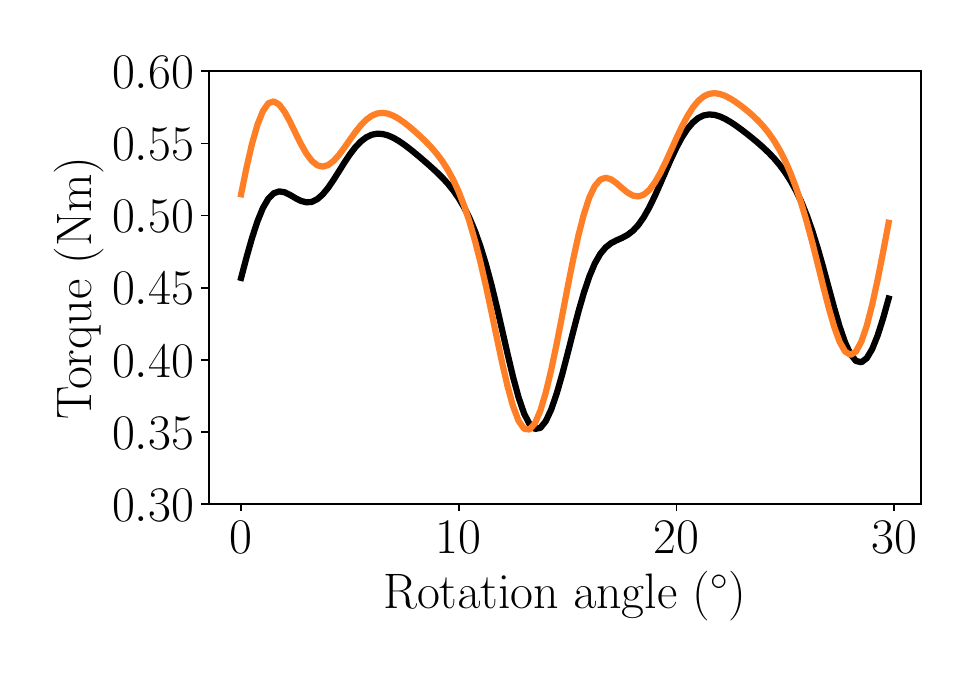}
         \caption{\gls{pca} and \gls{fnn}.}
         \label{fig:pca_nn_1200}
     \end{subfigure}
     \hfill
     \begin{subfigure}[b]{0.32\textwidth}
         \centering
         \includegraphics[width=\textwidth]{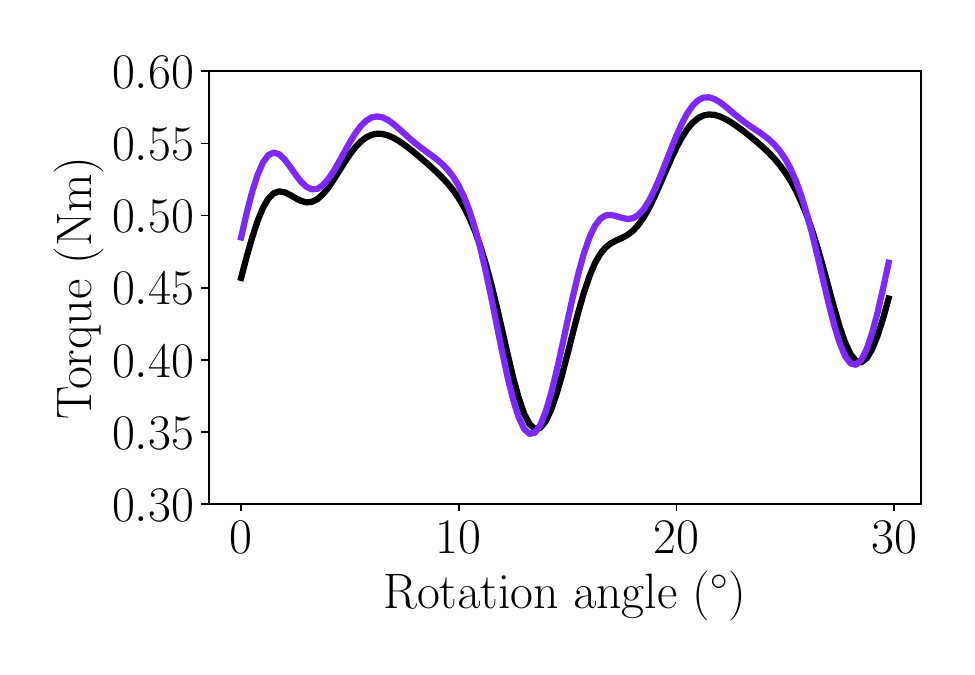}
         \caption{\gls{pca} and \gls{gp}.}
         \label{fig:pca_krig_1200}
     \end{subfigure}
     \\
     \begin{subfigure}[b]{0.32\textwidth}
         \centering
         \includegraphics[width=\textwidth]{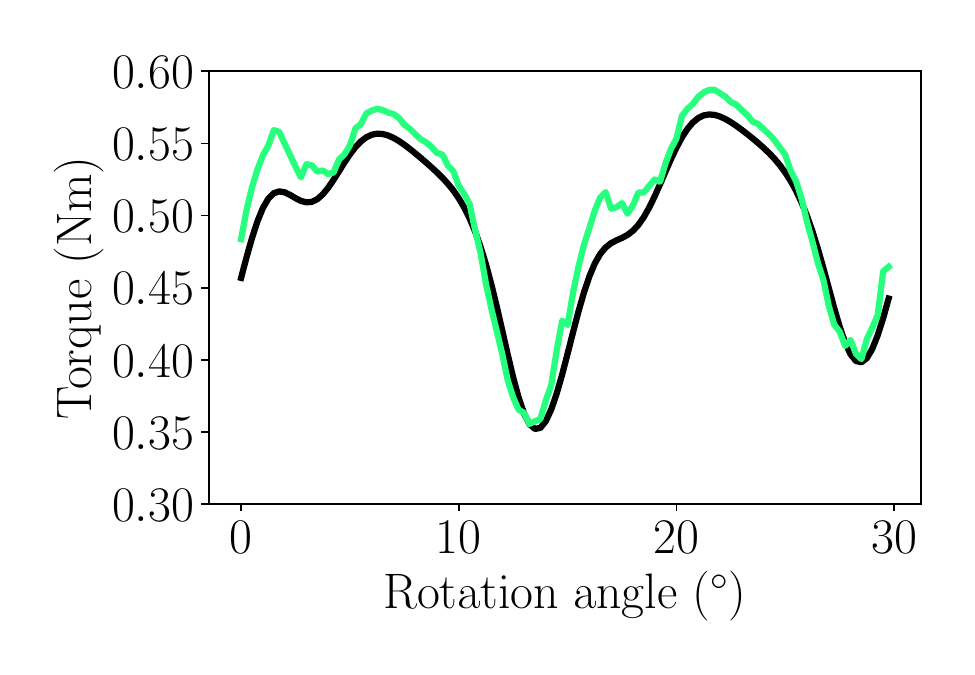}
         \caption{No reduction and \gls{pce}.}
         \label{fig:time_pce_1200}
     \end{subfigure}
     \hfill
     \begin{subfigure}[b]{0.32\textwidth}
         \centering
         \includegraphics[width=\textwidth]{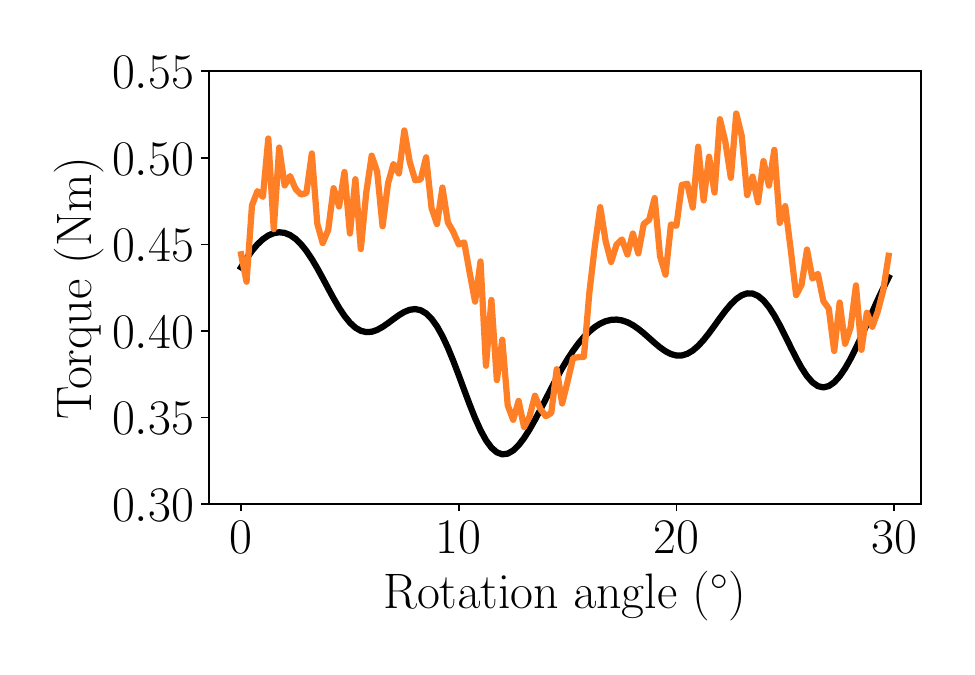}
         \caption{No reduction and \gls{fnn}.}
         \label{fig:time_nn_1200}
     \end{subfigure}
     \hfill
     \begin{subfigure}[b]{0.32\textwidth}
         \centering
         \includegraphics[width=\textwidth]{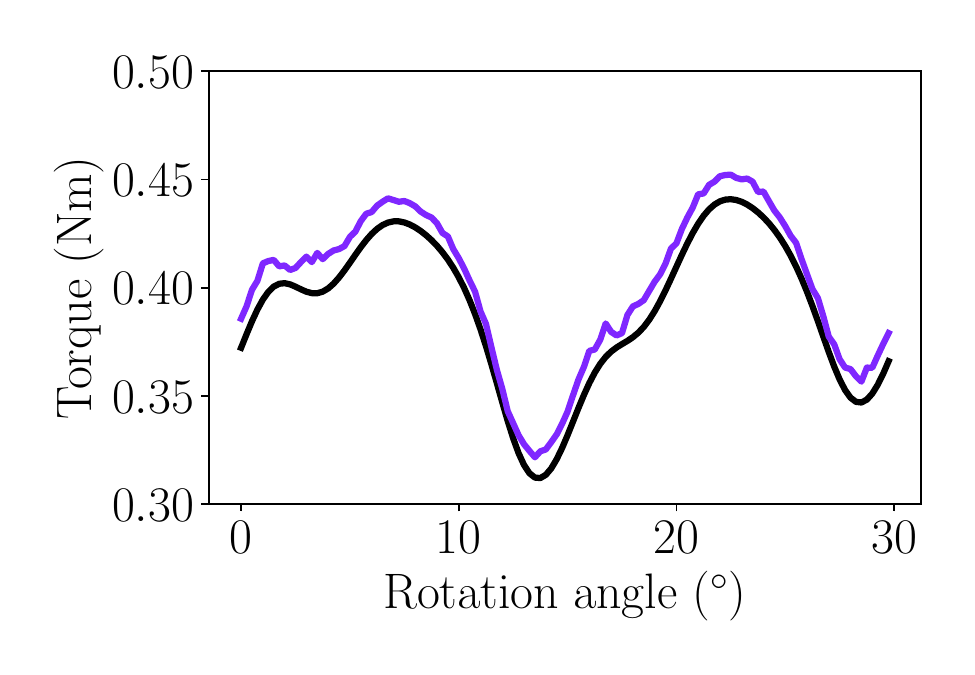}
         \caption{No reduction and \gls{gp}.}
         \label{fig:time_krig_1200}
     \end{subfigure}
     \caption{Worst-case surrogate-based torque signal predictions for training dataset size $M_\text{t}=1200$ and different combinations of \gls{rsm} and dimension reduction approach.}
    \label{fig:9-sur-1200}
\end{figure}

\end{document}